\documentclass[11pt]{article}
\usepackage[english]{babel}
\usepackage{amssymb,amsfonts,amsmath}

\newcommand{\spa}{\hspace{-2mm}}
\newcommand{\intr}{\int_{\mathbb{R}}}
\newcommand{\intrr}{\int_{\mathbb{R}\times\mathbb{R}}}
\newcommand{\dqdp}{\;\mathrm{d}q\,\mathrm{d}p\hspace{0.4mm}}
\newcommand{\dens}{\hat{\rho}}
\newcommand{\opa}{\hat{A}}
\newcommand{\opb}{\hat{B}}
\newcommand{\tr}{{\hspace{0.3mm}\mathsf{tr}\hspace{0.3mm}}}
\newcommand{\qua}{\mathcal{Q}}
\newcommand{\lr}{\mathrm{L}^2(\mathbb{R})}
\newcommand{\lrr}{\mathrm{L}^2(\mathbb{R}\times\mathbb{R})}
\newcommand{\lurr}{\mathrm{L}^1(\mathbb{R}\times\mathbb{R})}
\newcommand{\hpsi}{{\hat{\psi}}}
\newcommand{\psipsi}{{|\psi\rangle\langle\psi |}}
\newcommand{\quapsi}{\qua_\hpsi}
\newcommand{\rr}{{\mathbb{R}\times\mathbb{R}}}
\newcommand{\fac}{{\frac{1}{2\pi}}}
\newcommand{\dx}{\;\mathrm{d}x\hspace{0.4mm}}
\newcommand{\ima}{{\hspace{0.2mm}\mathrm{i}\hspace{0.4mm}}}

\newcommand{\lru}{\mathrm{L}^1(\mathbb{R})}
\newcommand{\cfvi}{\mathrm{C}_{0}}
\newcommand{\dep}{\;\mathrm{d}p\hspace{0.4mm}}
\newcommand{\deq}{\;\mathrm{d}q\hspace{0.4mm}}
\newcommand{\hq}{{\hat{q}}}
\newcommand{\hp}{{\hat{p}}}

\newcommand{\phipsi}{{|\phi\rangle\langle\psi |}}
\newcommand{\hfp}{{\widehat{\phi\psi}}}
\newcommand{\quafp}{\qua_\hfp}
\newcommand{\quapf}{\qua_{\widehat{\psi\phi}}}
\newcommand{\fwfp}{\mathcal{V}_{\widehat{\phi\psi}}}
\newcommand{\fwfpu}{\mathcal{V}_{\widehat{\phi_1\psi_1}}}
\newcommand{\fwfpd}{\mathcal{V}_{\widehat{\phi_2\psi_2}}}
\newcommand{\hfpa}{{\widehat{\phi_1\psi_1}}}
\newcommand{\hfpb}{{\widehat{\phi_2\psi_2}}}
\newcommand{\quafpa}{\qua_\hfpa}
\newcommand{\quafpb}{\qua_\hfpb}
\newcommand{\qp}{{(q,p)}}
\newcommand{\disp}{\exp\!\left(\ima(p\hspace{0.5mm}\hq-q\hspace{0.3mm}\hp)\right)}
\newcommand{\unig}{\mathcal{U}(\mathcal{H})}
\newcommand{\gi}{g_1}
\newcommand{\hi}{g_2}
\newcommand{\op}{\hat{T}}
\newcommand{\hs}{\mathcal{B}_2(\mathcal{H})}
\newcommand{\tgh}{\hat{T}(\gi,\hi)}
\newcommand{\txx}{\hat{T}(\cdot,\cdot)}
\newcommand{\ranglehs}{\rangle_{\hs}}
\newcommand{\norhs}{\|_{\hs}}
\newcommand{\ug}{U(\gi)}
\newcommand{\uh}{U(\hi)}
\newcommand{\lgg}{\mathrm{L}^2(G\times G)}
\newcommand{\lggi}{\mathrm{L}^2(G\times G,\mu_G\otimes\mu_G;\mathbb{C})}
\newcommand{\lugg}{\mathrm{L}^1(G\times G)}
\newcommand{\lug}{\mathrm{L}^1(G)}
\newcommand{\intg}{\int_{G}}
\newcommand{\intgg}{\int_{G\times G}}
\newcommand{\degh}{\;\mathrm{d}\mu_G\otimes\mu_G\hspace{0.3mm}(\gi,\hi)\hspace{0.3mm}}
\newcommand{\deggi}{\;\mathrm{d}\mu_G(g)\hspace{0.3mm}}
\newcommand{\degi}{\;\mathrm{d}\mu_G(\gi)\hspace{0.3mm}}
\newcommand{\deh}{\;\mathrm{d}\mu_G(\hi)\hspace{0.3mm}}
\newcommand{\enne}{{\mbox{\tiny $\mathsf{N}$}}}
\newcommand{\mme}{{\mbox{\tiny $\mathsf{M}$}}}
\newcommand{\dequa}{{\hspace{0.3mm}\boldsymbol{\mathfrak{D}\hspace{0.3mm}}}}
\newcommand{\quant}{{\hspace{0.3mm}\boldsymbol{\mathfrak{Q}\hspace{0.3mm}}}}
\newcommand{\quat}{{\hspace{0.3mm}\boldsymbol{\mathfrak{Q}}_{\op}}}
\newcommand{\dequat}{{\hspace{0.3mm}\boldsymbol{\mathfrak{D}}_{\op}}}
\newcommand{\dequatt}{{\hspace{0.3mm}\boldsymbol{\mathfrak{D}}_{\op^\ast}}}
\newcommand{\dequatu}{{\hspace{0.3mm}\boldsymbol{\mathfrak{D}}_{\op_1}}}
\newcommand{\dequatd}{{\hspace{0.3mm}\boldsymbol{\mathfrak{D}}_{\op_2}}}
\newcommand{\dequas}{{\hspace{0.3mm}\boldsymbol{\mathfrak{D}}_{\hspace{-0.5mm}\hat{S}}}}
\newcommand{\tdequat}{{\hspace{0.3mm}\widetilde{\boldsymbol{\mathfrak{D}}}_{\op^\prime}}}
\newcommand{\rep}{{U\hspace{-0.5mm}\vee\hspace{-0.5mm} U}}
\newcommand{\mm}{{\hspace{0.3mm}\mathtt{m}\hspace{0.5mm}}}
\newcommand{\MM}{{\hspace{0.3mm}\mathtt{M}\hspace{0.5mm}}}
\newcommand{\repr}{\mathcal{L}_{\MM^{\phantom{\hat{I}}}}\hspace{-2mm}}
\newcommand{\ccc}{\mathbb{C}}
\newcommand{\rrr}{\mathbb{R}}
\newcommand{\Hilb}{\mathcal{S}}
\newcommand{\mes}{\mathcal{X}}
\newcommand{\mesy}{\mathcal{Y}}
\newcommand{\fra}{{\Hilb_X}}
\newcommand{\psix}{{\psi_x}}
\newcommand{\psic}{{\psi_{(\cdot)}}}
\newcommand{\lx}{{\mathrm{L}^2(X)}}
\newcommand{\ly}{{\mathrm{L}^2(Y)}}
\newcommand{\lxc}{{\mathrm{L}^2(X,\mu; \mathbb{C})}}
\newcommand{\lyc}{{\mathrm{L}^2(Y,\nu; \mathbb{C})}}
\newcommand{\frop}{{\mathfrak{F}}}
\newcommand{\infrop}{{\mathfrak{F}^{\mbox{\tiny $\leftarrow$}}}}
\newcommand{\met}{{\hat{M}}}
\newcommand{\frad}{{\Hilb^X}}
\newcommand{\psixd}{{\psi^x}}
\newcommand{\tpsix}{{\tilde{\psi}_x}}

\newcommand{\pr}{{\hat{\mathrm{P}}_{\hspace{-1.2mm}\mbox{\tiny $\mathrm{Ran}(\frop)$}}}}

\newcommand{\deyu}{\mathrm{d}\nu\hspace{0.3mm}(y_1)\hspace{0.3mm}}
\newcommand{\deyd}{\mathrm{d}\nu\hspace{0.3mm}(y_2)\hspace{0.3mm}}
\newcommand{\deyt}{\mathrm{d}\nu\hspace{0.3mm}(y_3)\hspace{0.3mm}}
\newcommand{\tx}{\hat{T}_y}
\newcommand{\txd}{\hat{T}^y}
\newcommand{\txdu}{\hat{T}^{y_1}}
\newcommand{\txdd}{\hat{T}^{y_2}}
\newcommand{\fropd}{{\widetilde{\frop}}}

\newcommand{\ut}{{\widetilde{U}}}
\newcommand{\mmt}{{\hspace{0.3mm}\widetilde{\mathtt{m}}\hspace{0.5mm}}}
\newcommand{\MMt}{{\hspace{0.3mm}\widetilde{\mathtt{M}}\hspace{0.5mm}}}
\newcommand{\rept}{{\ut\hspace{-0.5mm}\vee\hspace{-0.5mm} \ut}}
\newcommand{\reprt}{\mathcal{L}_\MMt\hspace{-0.4mm}}
\newcommand{\wt}{\boldsymbol{\mathfrak{W}}_\psi^{\phantom{\ast}}}
\newcommand{\wta}{\boldsymbol{\mathfrak{W}}_\psi^{\hspace{0.3mm}\ast}}
\newcommand{\fr}{\mathsf{FR}\hspace{0.3mm}(\mathcal{H})}
\newcommand{\fri}{\mathsf{FR}\hspace{0.3mm}(\mathcal{H};U)}
\newcommand{\fru}{\breve{\mathsf{FR}}\hspace{0.3mm}(\mathcal{H};U)}
\newcommand{\du}{\hat{D}_U}
\newcommand{\duu}{\hat{D}_U^{\phantom{1}}}

\newcommand{\Kuu}{\hat{K}_U^{\phantom{1}}}
\newcommand{\ku}{\mathfrak{K}_U^{\phantom{x}}}
\newcommand{\preku}{\breve{\mathfrak{K}}_U^{\phantom{x}}}
\newcommand{\lef}{\mathsf{L}_{\hspace{-0.5mm}\opa}}
\newcommand{\rig}{\mathsf{R}_{\hspace{-0.5mm}\opa}}
\newcommand{\hrho}{\hat{\rho}}
\newcommand{\chil}{\chi^{\mbox {\tiny ${\mathsf{L}}$}}}
\newcommand{\chir}{\chi^{\mbox {\tiny ${\mathsf{R}}$}}}
\newcommand{\chilt}{\chi^{\mbox {\tiny ${\mathsf{L}}$}}_{\hat{T}}}
\newcommand{\chirt}{\chi^{\mbox {\tiny ${\mathsf{R}}$}}_{\hat{T}}}
\newcommand{\prd}{{\hat{\mathrm{P}}_{\hspace{-1.2mm}\mbox{\tiny $\mathrm{Ran}(\mathfrak{D})$}}}}
\newcommand{\selfb}{\mathcal{B}(\mathcal{H})_{\mathbb{R}}}
\newcommand{\fs}{\mathcal{F}_{\hspace{-0.6mm}\mbox{\tiny sp}}^{\phantom{x}}}
\newcommand{\fsy}{\mathcal{F}_{\hspace{-0.6mm}\mbox{\tiny sp}}}
\newcommand{\wig}{\mathfrak{S}_U^{\phantom{x}}}
\newcommand{\wigg}{\mathfrak{S}_U}
\newcommand{\ldg}{\mathrm{L}^2(G)}

\newcommand{\two}{\mathcal{T}_{\mm}\hspace{-0.3mm}}
\newcommand{\mmm}{\overset{\leftrightarrow}{\mm}}
\newcommand{\ru}{\mathcal{R}_U}
\newcommand{\proi}{{\hat{\mathrm{P}}_{\hspace{-1mm}\mbox{\tiny $\ru$}}}}
\newcommand{\ddu}{d_U^{\phantom{x}}}
\newcommand{\GG}{\boldsymbol{G}}
\newcommand{\gigi}{\boldsymbol{g}}
\newcommand{\diag}{\overset{\frown}{g}}
\newcommand{\mugg}{\mu_{\GG}}
\newcommand{\ldgg}{\mathrm{L}^2(\GG)}
\newcommand{\kerst}{{\hspace{0.6mm}\kappa_{\op}\big(\gigi,\gigi^{\prime}\hspace{-1mm},\gigi^{\prime\prime}\big)}}
\newcommand{\repker}{{\hspace{0.6mm}\varkappa_{\op}\big(\gigi,\gigi^{\prime}\big)}}
\newcommand{\repkert}{{\hspace{0.6mm}\varkappa_{\op}}}
\newcommand{\gam}{{\hspace{0.6mm}\gamma_{\op\hspace{-0.8mm},\hspace{0.3mm}\hpsi}\big(g,\gigi\big)}}
\newcommand{\gams}{{\hspace{0.6mm}\gamma_{\op\hspace{-0.8mm},\hspace{0.3mm}\hat{S}}\big(g,\gigi\big)}}
\newcommand{\gamd}{{\hspace{0.6mm}\gamma_{\op\hspace{-0.8mm},\hspace{0.3mm}\hat{S}}(\cdot,\boldsymbol{\cdot})}}
\newcommand{\invo}{\mathfrak{J}}
\newcommand{\wigt}{\mathfrak{T}}
\newcommand{\gigigia}{\gigi^{\prime}\hspace{-0.7mm},\gigi^{\prime\prime}}
\newcommand{\gigigib}{\gigi^{\prime}\hspace{-0.7mm},\gigi}
\newcommand{\gigigic}{\gigi,\gigi^{\prime\prime}}
\newcommand{\rpsi}{\mathcal{R}_\psi}
\newcommand{\rpsip}{\mathcal{R}_\psi^\perp}
\newcommand{\rande}{\mathrm{Ran}(\boldsymbol{\mathfrak{D}})}
\newcommand{\re}{\Re\mathrm{e}}
\newcommand{\ops}{\hat{T}_s}
\newcommand{\zz}{\boldsymbol{z}}
\newcommand{\tzz}{\tilde{\zz}}
\newcommand{\zu}{z_1^{\phantom{x}}}
\newcommand{\zd}{z_2^{\phantom{x}}}
\newcommand{\tzu}{\tilde{z}_1^{\phantom{x}}}
\newcommand{\tzd}{\tilde{z}_2^{\phantom{x}}}
\newcommand{\diagz}{\overset{\frown}{z}}
\newcommand{\BB}{\mathfrak{B}_2}
\newcommand{\ZZ}{\mathfrak{L}^2}
\newcommand{\aas}{\mathsf{A}_{-s}}
\newcommand{\as}{A_s}
\newcommand{\Apsi}{\widehat{A_{\psi}}}
\newcommand{\randet}{\mathrm{Ran}\big(\dequat\big)}
\newcommand{\doma}{G}

\newtheorem{proposition}{Proposition}

\newtheorem{theorem}{Theorem}

\newtheorem{remark}{Remark}

\newcommand{\de}{\mathrm{d}}

\newcommand{\mG}{\mu_G}

\newcommand{\sdp}{\hspace{-0.2mm}\rtimes}

\begin{document}

\title{Frame transforms, star products and quantum mechanics on phase space}

\author{
P. Aniello$^{\ast\,\ddagger}$, V.I. Man'ko$^\dagger$, G. Marmo$^\ast$  \vspace{2mm}\\
\small \it $^\ast$ Dipartimento di Scienze Fisiche dell'Universit\`a
di Napoli `Federico
II'\\ \small \it and Istituto Nazionale di Fisica Nucleare (INFN) -- Sezione di Napoli, \\
\small \it Complesso Universitario di Monte S.\ Angelo, via Cintia,
80126 Napoli, Italy \\ \small \it $^\ddagger$ Facolt\`a di Scienze
Biotecnologiche, Universit\`a di Napoli `Federico II' \\
\small \it $^\dagger$ P.N. Lebedev Physical Institute, Leninskii
Prospect 53, Moscow 119991, Russia\\ {\footnotesize E-mail:
paolo.aniello@na.infn.it, manko@na.infn.it, marmo@na.infn.it} }

\maketitle

\begin{abstract} \noindent Using the notions of frame transform and
of square integrable projective representation of a locally compact
group $G$, we introduce a class of isometries (tight frame
transforms) from the space of Hilbert-Schmidt operators in the
carrier Hilbert space of the representation into the space of square
integrable functions on the direct product group $G\times G$. These
transforms have remarkable properties. In particular, their ranges
are reproducing kernel Hilbert spaces endowed with a suitable `star
product' which mimics, at the level of functions, the original
product of operators. A `phase space formulation' of quantum
mechanics relying on the frame transforms introduced in the present
paper, and the link of these maps with both the Wigner transform and
the wavelet transform are discussed.
\end{abstract}

%------------------------------------------------------------------------------
\section{Introduction}
\label{intro}
%------------------------------------------------------------------------------

The formulation of quantum mechanics `on phase space' dates back to
the early stages of development of quantum theory. As it is well
known, the foundations of this elegant formulation have been laid by
E.~Wigner in his 1932 celebrated paper~{\cite{Wigner}}, with the aim
of exploring the quantum corrections to classical statistical
mechanics. Strictly related to Wigner's work are the pioneering
studies of H.~Weyl on quantization~\cite{Weyl}. On one hand, Wigner
was interested in associating with a quantum state a suitable phase
space `quasi-probability distribution' (association that leads to
the Wigner transform). On the other hand, Weyl aimed at associating
with a classical observable
--- a function on phase space --- a quantum observable in such a way to
overcome the ambiguities related to the `operator ordering'
(association that leads to the Weyl map). These procedures can be
regarded as the two `arrows' of a unique theoretical framework that
we may call the `quantization-dequantization theory'. This subject
is a richly branched, old --- but still extremely vital --- tree.
Since it is really huge, we will not attempt at giving even a brief
overview; the reader may consult the collection of
papers~{\cite{Zachos}} (and the bibliography therein) as a general
reference on the subject.

It is also worth mentioning the fact that, quite recently, the
impressive progress of experimental techniques
--- as well as the need of gaining a deeper understanding of some
fundamental (and controversial) aspects of quantum mechanics ---
have motivated a renewed interest in the description of quantum
states by means of phase space functions, the so-called `quantum
state tomography' or simply `quantum tomography'; see e.g.\
refs.~{\cite{Manko1,Manko2,Manko3,Dariano1,Dariano2,Dariano3}}.

There is a deep link between the quantization-dequantization theory
(including the formalism of quantum tomography) and another huge
research area
--- mainly focused on applications to signal analysis --- which we
may globally call `(generalized) wavelet analysis'. The main
mathematical tool in wavelet analysis is that of
\emph{frame}~{\cite{Duffin}}, a notion that will play a central role
in the present paper. Again, we will make no attempt at providing an
overview on this vast and interesting subject; we will then refer
the reader to the excellent
references~{\cite{Ali,Ali2,Daubechies,Kaiser}}. It is a remarkable
fact that several issues, concepts and techniques can be translated
`from one language into another' --- from quantum theory into signal
analysis and vice versa --- opening the way to new insights (see,
e.g., ref.~{\cite{Manko-new}}). Several anticipations of the unified
framework encompassing the quantization-dequantization theory and
wavelet analysis were already present in the pioneering work of
Klauder (and his co-authors), who introduced a `continuous
representation theory'~{\cite{Klauder1,Klauder2}}, and of Cahill and
Glauber~\cite{Cahill}.

It turns out that, from the mathematical point of view, the main
\emph{trait d'union} between the two mentioned subjects is the
remarkable notion of \emph{square integrable
representation}~\cite{Duflo,Grossmann,Aniello-sdp,Aniello}. In fact,
using this invaluable mathematical tool, one is able to perform all
the fundamental tasks of the quantization-dequantization theory and
of generalized wavelet analysis:
\begin{itemize}

\item to define generalized families of coherent states (covariant
frames), see refs.~\cite{Ali,Ali2,Perelomov}; in particular, the
standard family of coherent states of
Schr\"odinger~{\cite{Schrodinger}}, Glauber~{\cite{Glauber}},
Klauder~{\cite{Klauder1}} and Sudarshan~{\cite{Sudarshan}};

\item to obtain `discretized frames' from the covariant frames;
see e.g.\ refs.~{\cite{Aniello-discr,Aniello-discr-bis}};

\item to define suitable --- \`a la Weyl-Wigner ---
quantization-dequantization maps; see e.g.\
refs.~{\cite{Ali,Ali2,Folland-bis,Wong-bis}}.

\end{itemize}

Aim of the present contribution is to reconsider the previously
mentioned link between the quantization-dequantization theory and
the generalized wavelet analysis. In fact, we believe that to a
renewed interest in this area of research should correspond a
renewed study of its conceptual and mathematical foundations. As we
will try to show, this study leads, in a quite natural way, to the
definition of a certain class of `frame transforms' associated with
square integrable representations. These transforms are isometries
mapping a space of Hilbert-Schmidt operators (which is, obviously, a
Hilbert space) onto a space of square integrable functions having
remarkable properties. More precisely, given a square integrable
projective representation $U$ of a locally compact group $G$ in a
Hilbert space $\mathcal{H}$ and a (suitable) Hilbert-Schmidt
operator $\op$ in $\mathcal{H}$, one can associate with $\op$ an
isometry $\dequat$ mapping $\hs$ (the space of Hilbert-Schmidt
operators in $\mathcal{H}$) into $\lgg$ (the Hilbert space of square
integrable $\ccc$-valued functions on the direct product group
$G\times G$, with respect to the left Haar measure). As it will be
shown, the isometry $\dequat$ has remarkable properties that can be
regarded as direct consequences of the fact that $\dequat$ is a
frame transform; in particular:
\begin{enumerate}

\item The range $\randet$ of the isometry $\dequat$ is a `reproducing kernel
Hilbert space' embedded in $\lgg$.

\item The image, through the isometry $\dequat$, of the product of
operators in $\hs$ is a `star product of functions' in $\randet$.

\item The standard expectation value formula of quantum mechanics
--- namely,
\begin{equation*}
\langle \opa\rangle_{\hrho}=\tr(\opa\,\hrho),
\end{equation*}
where $\opa$ and $\hrho$ are, respectively, a bounded selfadjoint
operator and a density operator (a positive trace class operator of
unit trace) in $\mathcal{H}$
--- admits, in this framework, a suitable expression in terms of
$\mathbb{C}$-valued functions `on phase space'.

\end{enumerate}
The adjoint $\quat$ of the isometry $\dequat$, like the Weyl map,
has a simple integral expression and can be regarded as a
`quantization map'.

The paper is organized as follows. In Sect.~\ref{frame}, we discuss
the notion of `frame transform' and its main consequences. In
Sect.~\ref{wigdis}, we briefly review the definition of the Wigner
distribution and its relation with projective representations. Next,
in Sect.~\ref{interlude}, we recall the basic properties of square
integrable projective representations, tools that are fundamental
for the definition of the (generalized) Wigner transform and of its
reverse arrow, the (generalized) Weyl map, see
Sect.~\ref{revisited}; we will also argue that the generalized
Wigner transform is not, in general, a frame transform. Our analysis
will culminate in the introduction of the class of frame transforms
mentioned before --- Sect.~\ref{transforms} --- and in the
discussion of the main consequences from the point of view of
quantum mechanics, see Sect.~\ref{quantum}. In Sect.~\ref{example},
we consider a remarkable example that allows to show the link of our
results with the formalism of $s$-parametrized quasi-distributions
developed by Cahill and Glauber~\cite{Cahill}. Eventually, in
Sect.~\ref{conclusions}, a few conclusions are drawn.

%------------------------------------------------------------------------------
\section{Frame transforms and star products}
\label{frame}
%------------------------------------------------------------------------------

In this section, we will introduce the mathematical notions of
`frame' and of `frame transform' that will be central in the
following. In particular --- in the present section and later, on
the base of our main results, in Sect.~\ref{quantum}
--- we will show that by means of these notions it is possible, in a natural way,
to define a class of `star products' of functions and to introduce a
formulation of quantum mechanics `on phase space'.

In the first part of the section, we will collect a few basic fact
on frames in Hilbert spaces, a subject which is discussed with
plenty of applications in several excellent references; see e.g.\
refs.~{\cite{Daubechies,Kaiser}}. In the second part of the section,
we will focus on the peculiar case of frames in Hilbert-Schmidt
spaces (of operators). As we will show, in this case the theory of
frames enjoys extra results reflecting the fact that a space of
Hilbert-Schmidt operators is not only an Hilbert space but is also
endowed with the structure of an algebra.

Let $\Hilb$ be a separable complex Hilbert space (we denote by
$\langle\cdot,\cdot\rangle$ the associated scalar product, which
will be assumed to be linear in the second argument) and $\mes =
(X,\mu)$ a measure space. A family of vectors $\fra$ in $\Hilb$,
labelled by points in $X$,
\begin{equation}
\fra = \{\psix\in\Hilb : \ x\in X\},
\end{equation}
is called a \emph{frame} (in $\Hilb$, based on the measure space
$\mes$) if it satisfies the following defining conditions:
\begin{itemize}
\item for every $\phi\in\Hilb$, the function
\begin{equation}
\Phi \colon X\ni x\mapsto\langle\psix,\phi\rangle\in\mathbb{C}
\end{equation}
is $\mu$-measurable and belongs to $\lx\equiv\lxc$;

\item the `stability condition' is verified, namely,
\begin{equation} \label{stability}
\alpha\,\|\phi\|_\Hilb^2\le \|\Phi\|_\lx^2=\int_X |\Phi (x)|^2 \;
\de\mu(x)\le \beta\,\|\phi\|_\Hilb^2,\ \ \ \forall\hspace{0.3mm}
\phi\in\Hilb,
\end{equation}
for some (fixed) $\alpha,\beta\in\rrr$ such that $0<\alpha\le\beta$.
\end{itemize}
A couple of strictly positive numbers $\alpha,\beta$ --- such the
the stability condition~{(\ref{stability})} is satisfied --- are
called (lower and upper) \emph{frame bounds} for the frame $\fra$;
in particular, the frame $\fra$ is said to be \emph{tight} if one
can set $\alpha=\beta$.

Therefore, a frame $\fra= \{\psix\}_{x\in X}$ defines a \emph{frame
transform} (operator), i.e.\ the linear operator
\begin{equation} \label{fr-tr}
\frop \colon \Hilb\ni\phi\mapsto
\Phi:=\langle\psic,\phi\rangle\in\lx,
\end{equation}
which is bounded
\begin{equation}
\|\frop\,\phi\|_\lx^2\le \beta\,\|\phi\|_\Hilb^2, \ \ \
\forall\hspace{0.3mm} \phi\in\Hilb,
\end{equation}
injective, and admits a (in general, non unique) bounded left
inverse:
\begin{equation}
\alpha\,\|\phi\|_\Hilb^2\le \|\frop\,\phi\|_\lx^2 ,\ \ \
\forall\hspace{0.3mm} \phi\in\Hilb.
\end{equation}
For every $\phi\in\Hilb$, the $\ccc$-valued function $\frop\,\phi$
will be called the \emph{frame transform} of $\phi$.

Notice that the existence of a bounded left inverse of $\frop$
implies that the range of the frame transform ---
$\mathrm{Ran}(\frop)$
--- is closed in $\lx$:
\begin{equation}
\mathrm{Ran}(\frop)=\overline{\mathrm{Ran}(\frop)}.
\end{equation}
Specifically, $\frop$ admits a (unique) bounded
\emph{pseudo-inverse} $\infrop \hspace{-0.5mm}: \lx\rightarrow
\Hilb$, which is  the linear operator determined by the conditions
\begin{equation} \label{determined}
\infrop\, \frop = I,\ \ \ (\mbox{$\infrop\hspace{-0.5mm}$ is a left
inverse of $\frop$})
\end{equation}
\begin{equation} \label{determined-bis}
\infrop\, \Theta =0,\ \
\forall\hspace{0.3mm}\Theta\in\mathrm{Ran}(\frop)^{\perp},\ \ \
(\mbox{i.e.\ $\mathrm{Ker}(\infrop)=\mathrm{Ran}(\frop)^{\perp}$})
\end{equation}
with $I$ denoting the identity in $\Hilb$ and
$\mathrm{Ran}(\frop)^{\perp}$ the orthogonal complement of the
subspace $\mathrm{Ran}(\frop)$ of $\lx$. Obviously, in the case
where $\mathrm{Ran}(\frop)=\lx$, the pseudo-inverse $\infrop$ is
nothing but the (bounded) inverse $\frop^{-1}$. However, we stress
that the case where $\mathrm{Ran}(\frop)=\lx$ does not occur in
several important examples; typically, $\mathrm{Ran}(\frop)$ is a
proper subspace of $\lx$ consisting of functions with some
regularity property (this happens, for instance, in the case where
$X$ is a topological space and the frame map $x\mapsto\psix$ is
weakly continuous).

It is clear that for the adjoint $\frop^\ast\colon \lx\rightarrow
\Hilb$ of $\frop$ the following formula holds:
\begin{equation} \label{fr-adj}
\frop^\ast \Phi = \int_X \Phi (x)\,\psix \; \de\mu(x),\ \ \
\forall\hspace{0.3mm}\Phi\in\lx,
\end{equation}
where the integral (as all the vector-valued or operator-valued
integrals henceforth) has to be understood `in the weak sense'.

By means of the frame operator $\frop$ and of its adjoint
$\frop^\ast$, one can define the \emph{metric operator} of the frame
$\fra$, i.e.\ the map
\begin{equation}
\met := \frop^\ast\hspace{0.3mm}\frop \colon \Hilb\rightarrow\Hilb,
\end{equation}
which is a bounded, definite positive linear operator (with a
bounded definite positive inverse $\met^{-1}$):
\begin{equation}
\alpha\, I\le \met\le\beta\, I.
\end{equation}
It is easy to verify, using the defining
conditions~{(\ref{determined})\hspace{0.3mm}--\hspace{0.3mm}(\ref{determined-bis})},
that the following relation holds:
\begin{equation} \label{rel-infrop}
\infrop = \met^{-1}\hspace{0.6mm}\frop^\ast.
\end{equation}
The metric operator allows to define the \emph{dual frame} of the
frame $\fra$, namely, the family of operators
\begin{equation}
\frad := \{\psixd\in\Hilb :\  \psixd=\met^{-1}\psix,\ \psix\in
\fra\}.
\end{equation}
We stress that the term `dual frame' is coherent: one can easily
show that $\frad$ is indeed a frame (in $\Hilb$, based on $\mes$).
Notice that, if the frame $\fra$ is tight, then $\frop$ is ---
possibly up to a positive factor --- an isometry, the positive
operator $\met$ is a multiple of the identity, and $\fra$ coincides
with its dual frame $\frad$ up to, possibly, an irrelevant overall
normalization factor; i.e., there is a strictly positive number $r$
such that $\psixd=r\,\psix$, $\forall \hspace{0.4mm} x\in X$. In
particular, we will say that the tight frame $\fra$ is
\emph{normalized} if $r=1$.

Moreover, it is clear that denoting by $\fropd$ the frame transform
associated with the frame $\frad$, we have:
\begin{equation} \label{rel-fropd}
\fropd=\frop\,\met^{-1};
\end{equation}
hence, the metric operator associated with the frame $\frad$ is
$\met^{-1}$ and the dual frame of $\frad$ is $\fra$. From
relations~{(\ref{rel-infrop})} and~{(\ref{rel-fropd})} it follows
that
\begin{equation} \label{for-infrop}
\infrop\hspace{0.3mm} \Phi = \fropd^\ast\hspace{0.3mm} \Phi = \int_X
\Phi (x)\,\psixd \; \de\mu(x),\ \ \ \forall\hspace{0.3mm}\Phi\in\lx.
\end{equation}
If, in particular, the frame $\fra$ is tight, then the
pseudo-inverse $\infrop$ coincides --- possibly up to a positive
factor
--- with $\frop^\ast$.

By means of a couple of mutually dual frames $\fra$ and $\frad$, one
can write some remarkable formulae. In fact, taking into account
formula~{(\ref{for-infrop})}, and using the Dirac notation
$|\phi\rangle\langle\psi|\hspace{0.6mm}\eta\equiv\langle\psi,\eta\rangle\hspace{0.5mm}\phi$,
$\psi,\phi,\eta\in\Hilb$, we can write the following resolutions of
the identity:
\begin{eqnarray}
I  =  \infrop\hspace{0.3mm}\frop \spa & = & \spa \int_X
|\psixd\rangle\langle\psix| \; \de\mu(x)
\nonumber \\
\label{resid} & = & \spa \met \int_X |\psixd\rangle\langle\psix| \;
\de\mu(x) \; \met^{-1} = \int_X |\psix\rangle\langle\psixd| \;
\de\mu(x);
\end{eqnarray}
thus, we have a `reconstruction formula' for the frame transform
$\frop\,\phi$, i.e.
\begin{equation} \label{re-con}
\phi = \int_X \big(\frop\,\phi\big)(x)\; \psixd \; \de\mu(x),\ \ \
\forall\hspace{0.3mm}\phi\in\lx,
\end{equation}
and an analogous formula for the (dual) frame transform
$\fropd\,\phi$. From relations~{(\ref{resid})} we get immediately:
\begin{equation}
\met = \met \int_X |\psixd\rangle\langle\psix| \; \de\mu(x)= \int_X
|\psix\rangle\langle\psix| \; \de\mu(x)
\end{equation}
and
\begin{equation}
\met^{-1} = \met^{-1} \int_X |\psix\rangle\langle\psixd| \;
\de\mu(x)= \int_X |\psixd\rangle\langle\psixd| \; \de\mu(x).
\end{equation}
Moreover, observe that for the orthogonal projection $\pr$ onto the
subspace $\mathrm{Ran}(\frop)$ of $\lx$ we have the following
remarkable expression:
\begin{equation} \label{remexp}
\big(\pr \Phi\big)(x) = \big(\frop\,\infrop\Phi\big)(x) = \int_X
\varkappa (x,x^\prime)\, \Phi(x^\prime) \; \de\mu(x^\prime),\ \ \
\forall\hspace{0.3mm}\Phi\in\lx,
\end{equation}
for $\mu$-almost all (in short, for $\mu$-a.a.) $x\in X$, where
$\varkappa (\cdot,\cdot)$ is the $\ccc$-valued function on $X\times
X$ defined by
\begin{equation} \label{r-ker}
\varkappa (x,x^\prime):=\langle\psix, \psi^{x^\prime}\rangle,\ \ \
\forall\hspace{0.5mm} x,x^\prime\in X.
\end{equation}
Therefore, the range of the frame operator is a \emph{reproducing
kernel Hilbert space} (in short,
r.k.H.s.)~{\cite{Aronszajn,Hille,Saitoh}}.
\begin{remark} \label{RKHS} {\rm
Strictly speaking, $\mathrm{Ran}(\frop)$ is a `r.k.H.s.\ embedded in
$\lx$'. The `true' r.k.H.s.\ is the vector space composed of every
$\ccc$-valued function $\Phi$ on $X$ of the form
$\Phi=\langle\psic,\phi\rangle$, $\phi\in\Hilb$. Embedding this
r.k.H.s.\ in $\lx$ amounts to identifying such a function $\Phi$
with the equivalence class of $\mu$-measurable $\ccc$-valued
functions on $X$ that coincide with $\Phi$ for $\mu$-a.a.\ $x\in X$,
as it is tacitly done usually (e.g., in
definition~{(\ref{fr-tr})}).~$\blacksquare$ }
\end{remark}
It is an interesting fact that every bounded operator in the
r.k.H.s.\ $\mathrm{Ran}(\frop)$ is an integral operator. Precisely,
as the reader may check using formula~{(\ref{for-infrop})}, for
every operator $\opa$ in $\mathcal{B}(\Hilb)$ (the Banach space of
bounded linear operators in $\Hilb$), we have:
\begin{equation} \label{bop}
\Big(\big(\frop\,\opa\,\infrop\big)\Phi\Big)(x) = \int_X \varkappa
(\opa; x,x^\prime)\, \Phi(x^\prime) \; \de\mu(x^\prime),\ \ \
\forall\hspace{0.3mm}\Phi\in\lx,
\end{equation}
for $\mu$-a.a.\ $x\in X$, where
\begin{equation} \label{bop-ker}
\varkappa (\opa; x,x^\prime):=\langle\psix,
\opa\,\psi^{x^\prime}\rangle,\ \ \ \forall\hspace{0.5mm}
x,x^\prime\in X;
\end{equation}
thus:
\begin{equation}
\varkappa (x,x^\prime)= \varkappa (I; x,x^\prime).
\end{equation}

Denoting by $\mathcal{B}_1(\Hilb)$ the Banach space of trace class
operators in $\Hilb$, we now prove the following important result:
\begin{proposition}[the `trace formula for frames'] \label{trforfr}
With the previous notations and assumptions,  for every operator
$\opa$ in $\mathcal{B}_1(\Hilb)$, the following formula holds:
\begin{equation} \label{trace-for0}
\tr(\opa)= \int_X \varkappa (\opa; x,x)\; \de\mu(x).
\end{equation}
Assume now that the frame $\{\psix\}_{x\in X}$ is tight. Then, for
every positive bounded operator $\opb$ in $\Hilb$, $\varkappa (\opb;
x,x)\ge 0$, and
\begin{equation} \label{trace-for0-bis}
\int_X \varkappa (\opb; x,x)\; \de\mu(x)< +\infty
\end{equation}
if and only if $\opb$ is contained in $\mathcal{B}_1(\Hilb)$.
\end{proposition}

\noindent \emph{Proof:} Since, as it is well known, every trace
class operator $\op$ admits a decomposition of the form
\begin{equation}
\op = \op_1 -\op_2 + \ima (\op_3-\op_4),
\end{equation}
where $\op_1,\op_2,\op_3,\op_4$ are \emph{positive} trace class
operators, by linearity of the trace we can prove
relation~{(\ref{trace-for0})}
--- with no loss of generality --- for a generic \emph{positive} trace
class operator $\opa$ in $\Hilb$.

Let us suppose, for the moment, that the frame $\{\psix\}_{x\in X}$
is tight; we can assume that it is normalized (i.e.\ $\fra=\frad$).
Then, choosing an arbitrary orthonormal basis
$\{\eta_n\}_{n\in\mathcal{N}}$ in $\Hilb$ and denoting by
$\opa^{\frac{1}{2}}$ the (positive) square root of $\opa$, we have:
\begin{eqnarray}
\tr(\opa) %\spa & = & \spa
=\sum_{n\in\mathcal{N}}\langle
\opa^{\frac{1}{2}}\,\eta_n,\opa^{\frac{1}{2}}\,\eta_n\rangle
%\nonumber\\
& = & \spa \sum_{n\in\mathcal{N}}\,\int_X \langle
\opa^{\frac{1}{2}}\,\eta_n,\psix\rangle\langle
\psix,\opa^{\frac{1}{2}}\,\eta_n\rangle \; \de\mu(x)
\nonumber\\
& = & \spa \int_X \sum_{n\in\mathcal{N}} \langle
\psix,\opa^{\frac{1}{2}}\,\eta_n\rangle \langle
\opa^{\frac{1}{2}}\,\eta_n,\psix\rangle \; \de\mu(x)
\nonumber\\
& = & \spa \int_X \sum_{n\in\mathcal{N}} \langle
\opa^{\frac{1}{2}}\,\psix,\eta_n\rangle \langle
\eta_n,\opa^{\frac{1}{2}}\,\psix\rangle \; \de\mu(x)
\nonumber\\
& = & \spa \int_X \langle
\opa^{\frac{1}{2}}\,\psix,\opa^{\frac{1}{2}}\,\psix\rangle \;
\de\mu(x) ,
\end{eqnarray}
where the permutation of the (possibly infinite) sum  with the
integral is allowed by the positivity of the integrand functions.
Hence, we obtain:
\begin{equation}
\tr(\opa) =  \int_X  \langle
\opa^{\frac{1}{2}}\,\psix,\opa^{\frac{1}{2}}\,\psix\rangle \;
\de\mu(x) =  \int_X  \langle \psix,\opa\,\psix\rangle \; \de\mu(x) .
\end{equation}
This proves the first assertion of the statement in the case of a
tight frame.

For a \emph{generic} frame $\{\psix\}_{x\in X}$ in $\Hilb$ one can
argue as follows. First observe that --- denoting, as above, by
$\met$ the metric operator of this frame ---  the set
\begin{equation}
\big\{\tpsix=\met^{-\frac{1}{2}}\,\psix=\met^{\frac{1}{2}}\,\psixd\big\}_{x\in
X}
\end{equation}
is a normalized tight frame (exploiting relations~{(\ref{resid})},
the proof of this assertion is straightforward). Next, consider
that, for every $\opa\in\mathcal{B}_1(\Hilb)$,
\begin{eqnarray}
\tr (\opa)= \tr
\big(\met^{\frac{1}{2}}\,\opa\,\met^{-\frac{1}{2}}\big)\spa & = &
\spa  \int_X \langle
\tpsix,\met^{\frac{1}{2}}\,\opa\,\met^{-\frac{1}{2}}\,\tpsix\rangle
\; \de\mu(x)
\nonumber\\
& = & \spa \int_X \langle \psix,\opa\,\psixd\rangle \; \de\mu(x),
\end{eqnarray}
where we have used the cyclic property of the trace and the result
of the first part of the proof.

Let us prove the second assertion of the statement. Assume that the
frame $\{\psix\}_{x\in X}$ is tight (we can suppose that it is
normalized), and let $\opb$ be a positive bounded operator in
$\Hilb$ which is \emph{not} contained in $\mathcal{B}_1(\Hilb)$.
Then, arguing as above, we have:
\begin{eqnarray}
+\infty= \sum_{n\in\mathcal{N}}\langle \eta_n,\opb\,\eta_n\rangle
=\sum_{n\in\mathcal{N}}\langle
\opb^{\frac{1}{2}}\,\eta_n,\opb^{\frac{1}{2}}\,\eta_n\rangle \spa &
= & \spa \int_X \langle
\opb^{\frac{1}{2}}\,\psix,\opb^{\frac{1}{2}}\,\psix\rangle \;
\de\mu(x)
\nonumber\\
& = & \spa \int_X  \langle
\psix,\opb\,\psix\rangle \; \de\mu(x),
\end{eqnarray}
where $\{\eta_n\}_{n\in\mathcal{N}}$ is an arbitrary orthonormal
basis in $\Hilb$.

The proof is now complete.~{$\square$}

At this point, we proceed to the second part of the section, where
we will specialize the scheme outlined above to the case where
$\Hilb=\hs$, with $\hs$ denoting the space of Hilbert-Schmidt
operators in a (separable complex) Hilbert space $\mathcal{H}$ (we
will adopt the symbol $\langle\cdot,\cdot\ranglehs$ for denoting the
scalar product in $\hs$:
$\langle\opa,\opb\ranglehs:=\tr(\opa^\ast\opb)$, $\opa,\opb\in\hs$).
We recall the fact that the Hilbert space $\hs$ is a
$\mathrm{H}^\ast$-algebra~{\cite{Ambrose}}, and a two-sided
$\ast$-ideal in the $\mathrm{C}^\ast$-algebra of bounded operators
$\mathcal{B}(\mathcal{H})$ (see e.g.\ ref.~{\cite{Reed}}).

Then, let $\{\tx\}_{y\in Y}$ be a frame in $\hs$, based on a measure
space $\mesy = (Y,\nu)$, and let $\{\txd\}_{y\in Y}$ be the dual
frame. In order to avoid confusion, we will now denote by $\dequa$
the frame transform associated with the frame $\{\tx\}_{y\in Y}$ and
by $\quant$ its pseudo-inverse; thus, we will set:
\begin{equation}
\dequa\equiv\hspace{0.3mm}\frop\hspace{0.3mm}\colon\hspace{0.3mm}\hs\rightarrow\ly\equiv\lyc
\ \ \ \mbox{and}\ \ \
\quant\equiv\hspace{0.3mm}\infrop\hspace{-0.6mm}\colon\hspace{0.3mm}\ly\rightarrow\hs.
\end{equation}
It is natural to wonder if, in addition to the formulae recalled
above, one can suitably express the product of operators in $\hs$ in
terms of the frame transforms associated with these operators.
Denoting by $A,B$ the frame transforms of $\opa,\opb\in\hs$,
respectively, i.e.\
\begin{equation}
A=\dequa\opa:=\langle\hat{T}_{(\cdot)},\opa\ranglehs \in \ly,\ \ \ B
=\dequa\opb \in \ly,
\end{equation}
we can set:
\begin{equation} \label{stpr}
\big(A \star B\big) (y)  :=  \big(\dequa
\opa\hspace{0.3mm}\opb\big)(y).
\end{equation}
Therefore, the product of operators induces, through the frame
transform $\dequa$, a bilinear map
$(\cdot)\star(\cdot)\colon\rande\times \rande\rightarrow \rande$. As
we are going to show, exploiting the reconstruction formulae
\begin{equation} \label{recfor}
\opa = \int_Y A(y)\; \txd \; \de\nu(y),\ \ \ \opb = \int_Y B(y)\;
\txd \; \de\nu(y),
\end{equation}
one can obtain a suitable expression for this bilinear map.
\begin{remark} \label{weakintegral}
{\rm The integrals in the reconstruction
formulae~{(\ref{recfor})} are weak integrals of vector-valued
functions with respect to the scalar product of $\hs$. Then, \emph{a
fortiori}, they are weak integrals of bounded-operator-valued
functions; indeed:
\begin{eqnarray}
\langle \phi,\Big(\int_Y \hspace{-0.6mm} A(y)\; \txd \;
\de\nu(y)\Big)\psi\rangle \spa & = & \spa \langle \hspace{0.3mm}
|\psi\rangle\langle\phi|,\int_Y  \hspace{-0.6mm} A(y)\; \txd \;
\de\nu(y)\hspace{0.5mm}\ranglehs
\\
& = & \spa \int_Y \hspace{-0.6mm} A(y)\; \langle
|\psi\rangle\langle\phi|, \txd \ranglehs \; \de\nu(y) = \int_Y
\hspace{-0.6mm} A(y)\, \langle\phi,\txd\psi\rangle \; \de\nu(y),
\nonumber
\end{eqnarray}
for any couple of vectors $\phi,\psi\in\mathcal{H}$.~$\blacksquare$
}
\end{remark}

It turns out that the bilinear map $(\cdot)\star(\cdot)$, induced
through the frame transform by  the product of operators in $\hs$,
can be expressed as a `non-local' --- i.e.\ non-pointwise ---
product of functions defined on the range of $\dequa$; in fact, we
have the following result:

\begin{proposition} \label{propo}
With the previous notations and assumptions, for any
$\opa,\opb\in\hs$, the following formula holds:
\begin{eqnarray}
\big(A \star B\big) (y) \spa & = & \spa \int_{Y}\de\nu(y_1)
\int_{Y}\de\nu(y_2)\ \kappa (y, y_1, y_2)\, A(y_1)\, B(y_2)
\nonumber \\
\label{star-pr} & = & \spa \int_{Y}\de\nu(y_2) \int_{Y}\de\nu(y_1)\
\kappa (y, y_1, y_2)\, A(y_1)\, B(y_2),
\end{eqnarray}
for $\nu$-a.a.\ $y\in Y$, where the integral kernel $\kappa \colon
Y\times Y\times Y\rightarrow \ccc$ is defined by
\begin{equation}
\kappa (y, y_1, y_2):= \langle \tx,
\txdu\hspace{0.3mm}\txdd\ranglehs=\tr(\tx^\ast\hspace{0.6mm}\txdu\hspace{0.3mm}\txdd).
\end{equation}
\end{proposition}

\noindent \emph{Proof:} As anticipated, we will exploit the
reconstruction formulae~{(\ref{recfor})}. Let us prove the second of
relations~{(\ref{star-pr})} first. Observe that, for any
$\opa,\opb\in\hs$, we have:
\begin{equation}
\big(A \star B\big) (y) := \langle \tx,
\opa\hspace{0.3mm}\opb\ranglehs =
\tr(\tx^\ast\hspace{0.4mm}\opa\hspace{0.3mm}\opb)=
\tr((\opa^\ast\hspace{0.3mm}\tx)^\ast\hspace{0.3mm}\opb)=\langle
\opa^\ast\hspace{0.3mm}\tx, \opb\ranglehs.
\end{equation}
Hence, using the reconstruction formula for $\opb$, we find that
\begin{eqnarray}
\big(A \star B\big) (y) \spa & = & \spa \langle
\opa^\ast\hspace{0.3mm}\tx, \opb\ranglehs
\nonumber\\
& = & \spa \int_Y \de\nu(y_2)\ \langle
\opa^\ast\hspace{0.3mm}\tx,\txdd \ranglehs\,B(y_2)
\nonumber\\
& = & \spa \int_Y \de\nu(y_2)\ \langle
\tx\hspace{0.3mm}(\txdd)^\ast,\opa \ranglehs\,B(y_2),
\end{eqnarray}
where we have used the cyclic property of the trace:
\begin{equation}
\langle \opa^\ast\hspace{0.3mm}\tx,\txdd \ranglehs =
\tr(\tx^\ast\hspace{0.3mm}\opa\hspace{0.6mm}\txdd)=
\tr(\txdd\hspace{0.6mm}\tx^\ast\hspace{0.3mm}\opa) =\langle
\tx\hspace{0.3mm}(\txdd)^\ast,\opa \ranglehs.
\end{equation}
Next, using the reconstruction formula for $\opa$, we obtain:
\begin{eqnarray}
\big(A \star B\big) (y) \spa & = & \spa  \int_Y \de\nu(y_2)\ \langle
\tx\hspace{0.3mm}(\txdd)^\ast,\opa \ranglehs\,B(y_2)
\nonumber\\
& = & \spa \int_Y \de\nu(y_2)\int_Y \de\nu(y_1)\ \langle
\tx\hspace{0.3mm}(\txdd)^\ast,\txdu \ranglehs\,A(y_1)\,B(y_2)
\nonumber\\
& = & \spa \int_Y \de\nu(y_2)\int_Y \de\nu(y_1)\ \langle
\tx,\txdu\hspace{0.3mm}\txdd \ranglehs\,A(y_1)\,B(y_2).
\end{eqnarray}

On the other hand, we have:
\begin{eqnarray}
\big(A \star B\big) (y) \spa & = & \spa \langle
\tx\hspace{0.3mm}\opb^\ast, \opa\ranglehs
\nonumber\\
& = & \spa \int_Y \de\nu(y_1)\ \langle \tx\hspace{0.3mm}\opb^\ast,
\txdu\ranglehs\,A(y_1)
\nonumber\\
& = & \spa \int_Y \de\nu(y_1)\ \langle
(\txdu)^{\ast}\hspace{0.6mm}\tx, \opb\ranglehs\,A(y_1)
\nonumber\\
& = & \spa \int_Y \de\nu(y_1)\int_Y \de\nu(y_2)\ \langle
(\txdu)^{\ast}\hspace{0.6mm}\tx, \txdd\ranglehs\,A(y_1)\, B(y_2)
\nonumber\\
& = & \spa \int_Y \de\nu(y_1)\int_Y \de\nu(y_2)\ \langle \tx,
\txdu\hspace{0.3mm}\txdd\ranglehs\,A(y_1)\, B(y_2).
\end{eqnarray}

The proof is complete.~{$\square$}

We will call the non-local product of functions~{(\ref{stpr}})
\emph{star product}\footnote{We recall that the notion of star
product of functions on phase space has been extensively studied in
the literature; see, e.g., the classical
papers~{\cite{GLP,Berezin,Fronsdal}} and the recent
contributions~{\cite{Manko1,Manko3}}. Here we show how a notion of
this kind arises in a natural way considering frames of
Hilbert-Schmidt operators.} associated with the frame $\{\tx\}_{y\in
Y}$. Let us observe that the definition of the star product of
functions in $\rande$ can be extended, in a natural way, to all
functions in $\ly$ by setting:
\begin{equation} \label{def-star-pr}
\Phi_1 \star \Phi_2 :=\dequa\big((\quant\Phi_1)\,(\quant
\Phi_2)\big),\ \ \ \forall\hspace{0.5mm}\Phi_1,\Phi_2\in\ly.
\end{equation}
Notice that, since $\quant=\quant\hspace{0.5mm}\prd$, with $\prd$
denoting the orthogonal projection onto $\rande$, we have:
\begin{equation}
\Phi_1 \star \Phi_2 = \big(\prd\Phi_1\big)\star
\big(\prd\Phi_2\big).
\end{equation}
One can easily prove that the `extended star product' --- namely,
the bilinear map $(\cdot)\star(\cdot): \ly\times\ly\rightarrow\ly$
defined by formula~{(\ref{def-star-pr})} --- can be still expressed
as a non-local product of functions; indeed:
\begin{proposition} \label{propo-bis}
With the previous notations and assumptions, for any
$\Phi_1,\Phi_2\in\ly$, the following formula holds:
\begin{eqnarray}
\big(\Phi_1 \star \Phi_2\big)(y) \spa & = & \spa \int_{Y}\deyu
\int_{Y}\deyd\ \kappa (y, y_1, y_2)\, \Phi_1(y_1)\, \Phi_2(y_2)
\nonumber\\
& = & \spa \int_{Y}\deyd \int_{Y}\deyu\ \kappa (y, y_1, y_2)\,
\Phi_1(y_1)\, \Phi_2(y_2),
\end{eqnarray}
for $\nu$-a.a.\ $y\in Y$.
\end{proposition}
\noindent \emph{Proof:} Just recall that
\begin{equation} \label{for-infrop-2}
\quant \Phi  = \int_Y  \Phi (y)\; \txd \; \de\nu(y),\ \ \
\forall\hspace{0.3mm}\Phi\in\ly,
\end{equation}
apply definition~{(\ref{def-star-pr})}, and argue as in the proof of
Proposition~{\ref{propo}}.~$\square$

Since $\hs$ is a two-sided $\ast$-ideal in the
$\mathrm{C}^\ast$-algebra $\mathcal{B}(\mathcal{H})$ of bounded
operators in $\mathcal{H}$,  for every
$\opa\in\mathcal{B}(\mathcal{H})$ one can define the linear maps
\begin{equation} \label{defi-lefrig}
\lef : \hs\ni\opb\mapsto\opa\,\opb\in\hs\ \ \ \mbox{and}\ \ \ \rig :
\hs\ni\opb\mapsto\opb\hspace{0.3mm}\opa\in\hs.
\end{equation}
The maps $\lef$ and $\rig$ are \emph{bounded} linear operators.
Indeed, as it is well known~{\cite{Reed}}, we have:
\begin{equation}
\|\lef\hspace{0.3mm}\opb\norhs\le\|\opa\|\,\|\opb\norhs\ \ \
\mbox{and}\ \ \
\|\rig\hspace{0.3mm}\opb\norhs\le\|\opa\|\,\|\opb\norhs;
\end{equation}
from this relation follows in particular that $\|\lef\|\le\|\opa\|$
and $\|\rig\|\le\|\opa\|$. On the other hand, since
\begin{equation}
\|\lef\hspace{0.3mm}|\psi\rangle\langle\psi|\hspace{0.4mm}\norhs^2=\|\opa\,\psi\|^2
\hspace{0.5mm}\|\psi\|^2
=\|\rig\hspace{0.3mm}|\psi\rangle\langle\psi|\hspace{0.4mm}\norhs^2,\
\ \ \forall\hspace{0.4mm}\psi\in\mathcal{H},
\end{equation}
and
$\|\hspace{0.4mm}|\psi\rangle\langle\psi|\hspace{0.4mm}\norhs=\|\psi\|^2$,
we also have:
\begin{eqnarray}
\|\lef\|:=\sup\{\|\lef\hspace{0.3mm}\opb\norhs\colon
\|\opb\norhs=1\}\spa & \ge & \spa
\sup_{\|\psi\|=1}\|\lef\hspace{0.3mm}|\psi\rangle\langle\psi|\hspace{0.4mm}\norhs
\\ & = & \spa \sup_{\|\psi\|=1}\|\opa\,\psi\|=:\|\opa\|,\ \ \|\rig\|\ge
\|\opa\|. \nonumber
\end{eqnarray}
Hence, we conclude that
\begin{equation} \label{norme}
\|\lef\|=\|\rig\|= \|\opa\|.
\end{equation}
Notice that, if $\opa\in\mathcal{B}(\mathcal{H})$ is selfadjoint,
then the bounded operators $\lef$ and $\rig$ in $\hs$ are
selfadjoint too. The operators $\lef$ and $\rig$  are suitably
represented in the space of frame transforms $\rande$; i.e.
\begin{proposition} \label{prop-bop}
For every bounded operator $\opa\in\mathcal{B}(\mathcal{H})$ and
every Hilbert-Schmidt operator $\opb\in\mathcal{B}_2(\mathcal{H})$,
the following formulae hold:
\begin{equation} \label{forart}
\big((\dequa\lef\quant) B\big) (y)=\big(\dequa\opa\,\opb\big) (y) =
\int_{Y}\de\nu(y^\prime) \ \chil(\opa; y,y^\prime)\,B(y^\prime),
\end{equation}
\begin{equation} \label{forart2}
\big((\dequa\rig\quant)
B\big)(y)=\big(\dequa\opb\hspace{0.4mm}\opa\big) (y) = \int_{Y}\de
\nu(y^\prime) \ \chir(\opa; y,y^\prime)\,B(y^\prime),
\end{equation}
for $\nu$-a.a.\ $y\in Y$, where $B=\dequa\opb$ and
\begin{equation} \label{fortrar}
\chil(\opa; y,y^\prime):= \langle
\hat{T}_{y},\opa\,\hat{T}^{y^\prime}\ranglehs,\ \ \ \chir(\opa;
y,y^\prime):= \langle
\hat{T}_{y},\hat{T}^{y^\prime}\hspace{-0.8mm}\opa\ranglehs;
\end{equation}
hence:
\begin{equation} \label{prodab}
\opa\,\opb=\int_{Y}\deyu \int_{Y}\deyd\ \chil(\opa;
y_1,y_2)\,B(y_2)\;\txdu,
\end{equation}
\begin{equation}
\opb\hspace{0.4mm}\opa=\int_{Y}\deyu \int_{Y}\deyd\ \chir(\opa;
y_1,y_2)\,B(y_2)\;\txdu.
\end{equation}
Moreover, if the frame $\{\tx\}_{y\in Y}$ is tight, then, for every
operator $\opa\in\mathcal{B}(\mathcal{H})$, we have:
\begin{equation} \label{self}
\chil(\opa; y,y^\prime)=\chil(\opa^\ast; y^\prime,y)^\ast\ \ \
\mbox{and}\ \ \ \chir(\opa; y,y^\prime)=\chir(\opa^\ast;
y^\prime,y)^\ast.
\end{equation}
\end{proposition}

\noindent \emph{Proof:} Let us prove formula~{(\ref{forart})}. By
definition we have:
\begin{equation}
\big(\dequa\opa\,\opb\big) (y) =  \langle
\hat{T}_{y},\opa\,\opb\ranglehs.
\end{equation}
Then, exploiting the reconstruction formula for the Hilbert-Schmidt
operator $\opb$, we get:
\begin{eqnarray}
\big(\dequa\opa\,\opb\big) (y) \spa & = & \langle
\opa^\ast\hspace{0.3mm}\hat{T}_{y},\opb\ranglehs \nonumber\\
& = & \spa \int_{Y}\de \nu(y^\prime) \  \langle
\opa^\ast\hspace{0.3mm}\hat{T}_{y},\hat{T}^{y^\prime}\ranglehs\,
B(y^\prime)
\nonumber\\
& = & \spa \int_{Y}\de \nu(y^\prime) \  \langle
\hat{T}_{y},\opa\,\hat{T}^{y^\prime}\ranglehs\, B(y^\prime) ,
\end{eqnarray}
which is what we wanted to prove. The proof of
formula~{(\ref{forart2})} is analogous.

Let us suppose, now, that the frame $\{\tx\}_{y\in Y}$ is tight.
Then, we have:
\begin{eqnarray}
\hspace{-4mm} \chil(\opa; y^\prime,y)^\ast\!=\tr(
\hat{T}_{y^\prime}^\ast\opa\hspace{0.6mm}\hat{T}^{y})^\ast\!=\tr((\hat{T}^{y})^\ast\opa^\ast\hspace{0.3mm}
\hspace{0.7mm}\hat{T}_{y^\prime}) \spa & = & \spa
\tr(\hat{T}_{y}^\ast\hspace{0.4mm}\opa^\ast\hspace{0.4mm}\hat{T}^{y^\prime})
\nonumber\\
& = & \spa \chil(\opa^\ast; y,y^\prime).
\end{eqnarray}
In a similar way, one proves the analogous relation for the function
$\chir(\opa;\cdot,\cdot)$.

The proof is complete.~{$\square$}\\
It is worth stressing that, for every bounded operator
$\opa\in\mathcal{B}(\mathcal{H})$, both the functions
$y^\prime\mapsto\chil(\opa^\ast; y^\prime\hspace{-0.8mm},y)$ and
$y^\prime\mapsto\chir(\opa^\ast; y^\prime\hspace{-0.8mm},y)$ are
contained in $\rande$. If the frame $\{\tx\}_{y\in Y}$ is tight, due
to this fact and to the first of relations~{(\ref{self})}, for every
$\Phi\in\ly$ we have:
\begin{eqnarray}
\int_{Y}\de\nu(y^\prime) \ \chil(\opa; y,y^\prime)\,\Phi(y^\prime)
\spa & = & \spa \int_{Y}\de\nu(y^\prime) \ \chil(\opa^\ast;
y^\prime\hspace{-0.8mm},y)^\ast\hspace{0.6mm}\Phi(y^\prime)
\nonumber\\
\spa & = & \spa \int_{Y}\de\nu(y^\prime) \ \chil(\opa^\ast;
y^\prime\hspace{-0.8mm},y)^\ast\hspace{0.5mm}\big(\prd\Phi\big)(y^\prime)
\nonumber\\
& = & \spa \int_{Y}\de\nu(y^\prime) \ \chil(\opa;
y,y^\prime)\hspace{0.8mm}\big(\prd\Phi\big)(y^\prime).
\end{eqnarray}
Assume that the frame $\{\tx\}_{y\in Y}$ is tight and normalized (so
that $\dequa$ is an isometry). Then, since $\prd=\dequa\quant$, from
the previous relation and from formula~{(\ref{forart})} we obtain:
\begin{equation}
\int_{Y}\de\nu(y^\prime) \ \chil(\opa;
y,y^\prime)\,\Phi(y^\prime)=\big(\dequa(\opa\,\quant\Phi))(y)=
\big(\dequa(\opa\,\quant\hspace{0.5mm}\prd\Phi)\big)(y),
\end{equation}
for all $\Phi\in\ly$; furthermore, for any $\Phi,\Psi\in\ly$ we
have:
\begin{eqnarray}
\hspace{-0.5cm} \int_{Y}\de\nu(y) \int_{Y}\de\nu(y^\prime) \
\chil(\opa; y,y^\prime)\,\Psi(y)^\ast\,\Phi(y^\prime) \spa & = &
\spa \langle \hspace{0.2mm}\prd \Psi,\dequa (\opa\,\quant\Phi) \ranglehs \nonumber\\
& = & \spa \langle\dequa \quant\Psi,\dequa
(\opa\,\quant\Phi)\ranglehs
\nonumber\\
& = & \spa \langle\quant\Psi,\opa\,\quant\Phi\ranglehs
\nonumber\\
& = & \spa \label{forpsiphi}
\langle\quant\hspace{0.5mm}\prd\Psi,\opa\,\quant\hspace{0.5mm}\prd\Phi\ranglehs.
\end{eqnarray}
It is obvious that a completely analogous relation holds for the
integral kernel $\chir(\opa; \cdot,\cdot)$.

\begin{remark} {\rm Notice that the integral kernels $\chil(\opa;\cdot,\cdot)$
and $\chir(\opa;\cdot,\cdot)$ are nothing but the kernels of the
bounded (super-)operators $\lef$ and $\rig$ with respect to the
frame $\{\tx\}_{y\in Y}$ (see formula~{(\ref{bop-ker})}). The `left'
and `right' integral kernels form vector spaces that can be endowed
with the structure of a $\mathrm{C}^\ast$-algebra isomorphic to the
algebra of bounded operators $\mathcal{B}(\mathcal{H})$. Differently
from the case of $\rande\equiv\mathrm{Ran}(\frop)$, we will assume
that these vector spaces are composed of functions rather than of
equivalence classes of functions (see Remark~{\ref{RKHS}}). Observe,
moreover, that for any
$\hat{A}_1,\hat{A}_2\in\mathcal{B}(\mathcal{H})$ we have:
\begin{equation}
\chil(\hat{A}_1\hat{A}_2; y_1,y_2)=\int_{Y}\de\nu(y)\
\chil(\hat{A}_1; y_1,y)\, \chil(\hat{A}_2; y,y_2),
\end{equation}
for all $y_1\in Y$ and all $y_2\in Y$; indeed, exploiting the
resolution of the identity generated by the frame $\{\tx\}_{y\in
Y}$, we get:
\begin{eqnarray}
\chil(\hat{A}_1\hat{A}_2; y_1,y_2) =\langle
\hat{T}_{y_1},\hat{A}_1\hat{A}_2\,\hat{T}^{y_2}\ranglehs \spa & = &
\spa \langle
\hat{A}_1^\ast\,\hat{T}_{y_1},\hat{A}_2\,\hat{T}^{y_2}\ranglehs
\nonumber\\
& = & \spa \int_{Y}\de\nu(y)\ \langle \hat{A}_1^\ast\,
\hat{T}_{y_1},\hat{T}^{y}\ranglehs\, \langle
\hat{T}_{y},\hat{A}_2\,\hat{T}^{y_2}\ranglehs
\nonumber\\
& = & \spa \int_{Y}\de\nu(y)\ \chil(\hat{A}_1; y_1,y)\,
\chil(\hat{A}_2; y,y_2).
\end{eqnarray}
Clearly, an analogous expression holds for the integral kernel
$\chir(\hat{A}_1\hat{A}_2; \cdot,\cdot)$, i.e.
\begin{equation}
\chir(\hat{A}_1\hat{A}_2; y_1,y_2) = \int_{Y}\de\nu(y)\
\chir(\hat{A}_2; y_1,y)\, \chir(\hat{A}_1; y,y_2).
\end{equation}
Therefore --- denoting by $\selfb$ the Jordan-Lie
algebra~{\cite{Emch}} of bounded selfadjoint operators in
$\mathcal{H}$, endowed with the \emph{Jordan product}
$
\hat{A}_1\circ\hat{A}_2:=\frac{1}{2}\hspace{0.4mm}\big(\hat{A}_1\hat{A}_2+\hat{A}_2\hat{A}_1\big)
$
and with the \emph{Lie bracket}
$
\{\hat{A}_1,\hat{A}_2\}:=\frac{1}{i}\hspace{0.4mm}\big[\hat{A}_1,\hat{A}_2\big]
$
--- we find:
\begin{eqnarray}
\chil(\hat{A}_1\hspace{-0.3mm}\circ\hspace{-0.3mm}\hat{A}_2;
y_1,y_2)\spa & = & \spa \frac{1}{2}\int_{Y}\de\nu(y)\;
\Big(\chil(\hat{A}_1; y_1,y)\, \chil(\hat{A}_2; y,y_2)
\nonumber\\
\label{j-prod}
& + & \spa \chil(\hat{A}_2; y_1,y)\, \chil(\hat{A}_1;
y,y_2)\Big),
\end{eqnarray}
\begin{eqnarray}
\chil(\{\hat{A}_1,\hat{A}_2\}; y_1,y_2)\spa & = & \spa
\frac{1}{i}\int_{Y}\de\nu(y)\; \Big(\chil(\hat{A}_1; y_1,y)\,
\chil(\hat{A}_2; y,y_2)
\nonumber\\
\label{l-brac}
& - & \spa
\chil(\hat{A}_2; y_1,y)\, \chil(\hat{A}_1; y,y_2)\Big),
\end{eqnarray}
for any $\hat{A}_1,\hat{A}_2\in\selfb$. Analogous relations hold for
the integral kernels
$\chir(\hat{A}_1\hspace{-0.3mm}\circ\hspace{-0.3mm}\hat{A}_2;\cdot,\cdot)$
and $\chir(\{\hat{A}_1,\hat{A}_2\};\cdot,\cdot)$.~$\blacksquare$}
\end{remark}

It is natural to wonder what is the relation between the functions
$\chil(\opb;\cdot,\cdot)$, $\chir(\opb;\cdot,\cdot)$ --- in the
special case where $\opb\in\hs$ --- and the frame transform
$B\equiv\dequa\opb$. A first half of the answer is contained in the
following:

\begin{proposition} \label{ques} With the previous notations and assumptions,
for every Hilbert-Schmidt operator $\opb\in\hs$, denoting by $B$ the
function $\dequa\opb$, the following formulae hold:
\begin{equation} \label{forchil}
\chil(\opb;y_1,y_2)= \int_{Y}\deyt\ \kappa(y_1,y_3,y_2)\, B(y_3),
\end{equation}
\begin{equation} \label{forchil2}
\chir(\opb;y_1,y_2)= \int_{Y}\deyt\ \kappa(y_1,y_2,y_3)\, B(y_3),
\end{equation}
for $\nu$-a.a.\ $y_1\in Y$ and $\nu$-a.a.\ $y_2\in Y$.
\end{proposition}

\noindent \emph{Proof:} Let us prove formula~{(\ref{forchil})}.
Observe that we have:
\begin{eqnarray}
\chil(\opb;y_1,y_2) :=  \langle
\hat{T}_{y_1},\opb\,\hat{T}^{y_2}\ranglehs \spa & = & \spa \langle
\hat{T}_{y_1}(\hat{T}^{y_2})^\ast,\opb\ranglehs
\nonumber\\
& = & \spa \int_{Y}\deyt\ \langle \hat{T}_{y_1},
\hat{T}^{y_3}\hspace{0.3mm}\txdd\ranglehs\, B(y_3),
\end{eqnarray}
which is what we wanted to prove. The proof of
formula~{(\ref{forchil2})} is analogous.~{$\square$}

Let us now suppose to have, simultaneously, a \emph{couple} of
frames: the frame $\{\tx\}_{y\in Y}$ in the space of Hilbert Schmidt
operators $\hs$ and a frame $\{\psi_x\}_{x\in X}$ in the Hilbert
space $\mathcal{H}$, based on a measure space $\mathcal{X}=(X,\mu)$.
A situation of this kind will be considered in
Sect.~{\ref{quantum}}. Then, in addition to the collection of
formulae previously obtained, we have the following result:
\begin{proposition} \label{prop-simul}
For every bounded operator $\opa\in\mathcal{B}(\mathcal{H})$, every
Hilbert-Schmidt operator $\opb\in\hs$ and every trace class operator
$\hrho\in\mathcal{B}_1(\mathcal{H})$, the following formulae hold:
\begin{equation} \label{kappakappa}
\varkappa (\opb; x,x^\prime):=\langle\psix,
\opb\,\psi^{x^\prime}\rangle= \int_{Y} \de\nu(y)\
\Gamma(x,x^\prime\hspace{-0.8mm},y)\, B(y),
\end{equation}
\begin{equation} \label{trrho}
\tr(\hrho)= \int_{X} \de\mu(x)\int_{Y} \de\nu(y)\ \gamma(x,y)\,
\rho(y),
\end{equation}
\begin{eqnarray} \label{trarho}
\tr(\opa\,\hrho) \spa & = & \spa \int_{X}\de\mu(x) \int_{Y}\deyu
\int_{Y}\deyd\ \gamma (x,y_1)\, \chil(\opa; y_1,y_2)\,\rho(y_2)
\nonumber\\
& = & \spa \int_{X}\de\mu(x) \int_{Y}\deyu \int_{Y}\deyd\ \gamma
(x,y_1)\, \chir(\opa; y_1,y_2)\,\rho(y_2),
\end{eqnarray}
where $B=\dequa\opb$, $\rho=\dequa\hrho$, and
\begin{equation}
\Gamma(x,x^\prime\hspace{-0.8mm},y):=\langle\psi_{x},
\hat{T}^{y}\hspace{0.4mm}\psi^{x^\prime}\rangle,\ \ \ \gamma
(x,y):=\langle\psi_{x},
\hat{T}^{y}\hspace{0.4mm}\psi^{x}\rangle=\Gamma(x,x,y).
\end{equation}
Assume now that the frame $\{\psix\}_{x\in X}$ is tight. Then, for
every \emph{positive} Hilbert-Schmidt operator $\opb$ in
$\mathcal{H}$, $\int_{Y} \de\nu(y)\ \gamma(x,y)\,
\big(\dequa\opb\big)(y)=\langle\psix, \opb\,\psi^{x}\rangle\propto
\langle\psix, \opb\,\psix\rangle\ge 0$, and
\begin{equation} \label{trace-for0-tris}
\int_{X} \de\mu(x)\int_{Y} \de\nu(y)\ \gamma(x,y)\,
\big(\dequa\opb\big)(y)< +\infty
\end{equation}
if and only if $\opb$ is contained in $\mathcal{B}_1(\mathcal{H})$.
\end{proposition}

\noindent \emph{Proof:} Taking into account
Remark~{\ref{weakintegral}}, formula~{(\ref{kappakappa})} follows
from the reconstruction formula for the operator $\opb$.

Let us prove formula~{(\ref{trrho})}. Applying the trace
formula~{(\ref{trace-for0})} to $\hrho$, and using
formula~{(\ref{kappakappa})} for the integral kernel $\varkappa
(\hrho\hspace{0.3mm};\cdot,\cdot)$, we get:
\begin{equation}
\tr(\hrho)= \int_{X} \de\mu(x)\; \varkappa (\hrho\hspace{0.3mm};
x,x) = \int_{X} \de\mu(x)\int_{Y} \de\nu(y)\ \Gamma(x,x,y)\,
\rho(y).
\end{equation}

Let us now prove the first of relations~{(\ref{trarho})}. Applying
formula~{(\ref{trrho})} to the trace class operator $\opa\,\hrho$,
we get:
\begin{equation}
\tr(\opa\,\hrho)  =   \int_{X} \de\mu(x)\int_{Y} \de\nu(y_1)\
\gamma(x,y_1)\,\big(\dequa\opa\,\hrho\big)(y_1)\,.
\end{equation}
Next, by virtue of formula~{(\ref{forart})}, we obtain:
\begin{equation}
\tr(\opa\,\hrho)  =   \int_{X}\de\mu(x) \int_{Y} \de\nu(y_1)\
\gamma(x,y_1)\int_{Y}\deyd\  \chil(\opa; y_1,y_2)\,\rho(y_2)
\end{equation}
where $\rho=\dequa\hrho$.  The proof of the second of
relations~{(\ref{trarho})} is analogous.

The proof of the second assertion of the statement follows from the
second assertion of Proposition~{\ref{trforfr}}.~{$\square$}

We can now show how the frame transform $B\equiv\dequa\opb$ of a
Hilbert-Schmidt operator $\opb\in\hs$ can be recovered from the
functions $\chil(\opb;\cdot,\cdot)$ and $\chir(\opb;\cdot,\cdot)$.
Thus, we have the second part of the answer to question addressed
before Proposition~{\ref{ques}}.
\begin{proposition}
With the previous notations and assumptions, for every
Hilbert-Schmidt operator $\opb\in\hs$, denoting by $B$ the function
$\dequa\opb$, the following formula holds:
\begin{eqnarray} \label{bix}
B(y) \spa & = & \spa \int_{X}\de\mu (x) \int_{Y}\deyu\int_{Y}\deyd\
\gamma (x,y_1)\, \chil(\opb; y_1,y_2)\,\delta(y_2,y)
\nonumber\\
& = & \spa \int_{X}\de\mu (x) \int_{Y}\deyu\int_{Y}\deyd\ \gamma
(x,y_1)\, \chir(\opb; y_1,y_2)\,\delta(y_2,y),
\end{eqnarray}
where
\begin{equation}
\delta(y_1,y_2):=\langle\hat{T}_{y_1}^{\phantom{\ast}},\hat{T}_{y_2}^\ast\ranglehs=
\big(\dequa \hat{T}_{y_2}^\ast\big)(y_1).
\end{equation}
\end{proposition}

\noindent \emph{Proof:} Let us prove the first of
relations~{(\ref{bix})}. Observe that we have:
\begin{eqnarray}
B(y)\spa & = & \spa \tr(\hat{T}_{y}^\ast\hspace{0.3mm}\opb)
\nonumber\\
& = & \spa \tr(\opb\hspace{0.5mm}\hat{T}_{y}^\ast)
\nonumber\\
& = & \spa \int_{X}\de\mu (x) \int_{Y}\deyu\int_{X}\deyd\ \gamma
(x,y_1)\, \chil(\opb; y_1,y_2)\,\big(\dequa
\hat{T}_{y}^\ast\big)(y_2),
\end{eqnarray}
where we have used the first of relations~{(\ref{trarho})}. Using
the second of relations~{(\ref{trarho})} one proves the second of
relations~{(\ref{bix})}.~{$\square$}

The frame transform $\dequa\equiv\frop$ associated with a frame in
$\hs$ may be regarded as a `dequantization map', which associates
with any operator in $\hs$ a square integrable function. Conversely,
the pseudo-inverse $\quant\equiv\infrop$ may be regarded as a
`quantization map' which suitably associates an operator with a
$\ccc$-valued function. In this context, the counterpart of the
product of operators is given by the star product of functions. At
this point, the reader will have recognized the typical scheme
underlying the subject which is usually called `quantum mechanics on
phase space': the Wigner transform (dequantization), the Weyl map
(quantization) and the Gr\"onewold-Moyal product of functions (star
product), see ref.~{\cite{Zachos}}. In the following, we will show
that there is a precise link between the `frame formalism' discussed
in the present section and the Weyl-Wigner-Gr\"onewold-Moyal
formalism for quantum mechanics.

%------------------------------------------------------------------------------
\section{Quantum mechanics on phase space: the Wigner distribution}
\label{wigdis}
%------------------------------------------------------------------------------

As it is well known, due to the indetermination relations, the
notion of phase space is not straightforward in the
quantum-mechanical setting as it is in the classical setting. Since
particles cannot have, simultaneously, a well defined position $q$
and momentum $p$, it is not possible to define a genuine phase space
probability distribution for a quantum particle as it happens in
classical statistical mechanics; in other words, quantum mechanics
is not a statistical theory in the classical sense. It is, however,
possible to introduce a notion of `quasi-probability distribution'
or `quasi-distribution' that allows one to express quantum averages
in a way analogous to classical averages.

In the following, for the sake of notational simplicity, we will
consider the case of a $(1+1)$-dimensional phase space (with
coordinates denoted by $q,p$); the extension to the ordinary
$(3+3)$-dimensional case is straightforward. In the classical
setting, a particle can be described by a classical probability
distribution on phase space $(q,p)\mapsto\mathcal{P}(q,p)$ (or, more
generally, by a probability measure). The average (at a certain
time) of a function of position and momentum $(q,p)\mapsto A(q,p)$
--- namely, of a classical observable --- is given by the expression
\begin{equation} \label{for1}
\langle A\rangle_{\mathcal{P}} = \intrr A(q,p)\, \mathcal{P}(q,p)
\dqdp\;.
\end{equation}
On the other hand, a quantum-mechanical state is described by a
density operator $\dens$ --- a positive trace class operator of unit
trace
--- and the mean value of a quantum observable $\opa$, which (by
virtue of the spectral decomposition of a selfadjoint operator) can
always be assumed to be a \emph{bounded} selfadjoint operator, is
given by the well known `trace formula'
\begin{equation} \label{for2}
\langle \opa\rangle_{\hrho} =  \tr (\opa\,\dens)\,.
\end{equation}
If one tries to establish a link between the classical
formula~{(\ref{for1})} and the quantum one~{(\ref{for2})}, one has
to face the following problem: how one can \emph{set a suitable
correspondence between a quantum observable} $\opa$ (i.e.\ a
selfadjoint operator, in the standard formulation of quantum
mechanics) \emph{and a `corresponding classical-like observable'}
$(q,p)\mapsto A(q,p)$ (a numerical function), \emph{and between a
density operator $\dens$ and a suitable `quantum quasi-distribution
function'} $(q,p)\mapsto \qua_{\dens} (q,p)$, in such a way that it
is then possible to \emph{express the expectation value of a quantum
observable in a `formally classical fashion'}, i.e.\ as a phase
space average of the type~{(\ref{for1})}:
\begin{equation} \label{for3}
\langle \opa\rangle_{\dens} = \intrr A(q,p)\, \qua_{\dens}(q,p)
\dqdp\,.
\end{equation}
It is a remarkable fact that this problem can be solved --- at least
partially --- within a theoretical scheme usually called
`Weyl-Wigner formulation of quantum mechanics', or, in a slightly
more general sense, `phase space formulation of quantum mechanics'.
It turns out that the correspondence \emph{operator}
$\leftrightarrow$ \emph{numerical function} is of the same kind
(i.e.\ it is obtained using the same formulae) both for the density
operator $\dens$ and the observable $\opa$ (at least for a suitable
class of observables).

As it is well known, the notion of quasi-distribution function has
been introduced by E.~{Wigner} in his celebrated
paper~\cite{Wigner}, with the aim of exploring the quantum
corrections to classical statistical mechanics. The
quasi-distribution introduced by Wigner
--- which is still regarded nowadays as the `standard'
quasi-distribution function (other quasi-distributions, with
remarkable applications in quantum optics, can also be defined,
see~{\cite{Cahill,Aniello-quasi,Schleich}}; see also the recent
proposals~{\cite{ Ventriglia,Klauder3}})
--- is universally known as the \emph{Wigner distribution}. In the
following, we will recall a few basic results; for the proofs, the
reader may consult standard references on the subject
like~{\cite{Folland-bis}} and~\cite{Wong-bis}. As above, in order to
simplify notation, we will consider the case of a quantum particle
with a single degree of freedom (hence, we will deal with a
$(1+1)$-dimensional phase space). Then, let us denote by $\psi$ a
vector in the Hilbert space $\lr$ and, using the Dirac notation, let
us set $\hpsi\equiv\psipsi$. With the vector $\psi$  --- or, more
precisely, with the operator $\hpsi$ --- one can associate the
function
\begin{equation}
\quapsi :\, \rr\longrightarrow \mathbb{C}\, ,
\end{equation}
defined by ($\hbar = 1$):
\begin{equation}
\label{defwig}
\quapsi (q,p) := \fac\intr e^{-\ima p x}\,
\psi\!\left(q-\frac{x}{2}\right)^*
\psi\!\left(q+\frac{x}{2}\right)\dx \, .
\end{equation}
If $\psi\in\lr$ is, in particular, a normalized nonzero vector
(i.e.\ $\|\psi\|=1$), then $\quapsi$ is called the ``Wigner
distribution associated with the pure state $\hpsi\,$''. Notice
that, for almost all $q\in\mathbb{R}$, the function
\begin{equation}
\mathbb{R}\ni x \mapsto \psi\!\left(q-\frac{x}{2}\right)^*
\psi\!\left(q+\frac{x}{2}\right)\!\in\mathbb{C}
\end{equation}
is contained in $\lru$; hence the Fourier integral in
definition~{(\ref{defwig})} is indeed an ordinary integral.
Moreover, this integral can be regarded as $1/\pi$ times the scalar
product of the normalized functions
\begin{equation}
\mathbb{R}\ni x \mapsto \frac{e^{\ima p x}}{\sqrt{2}}\;
\psi\!\left(q-\frac{x}{2}\right)\!\in\mathbb{C} \, \ \ \mbox{and}\ \
\mathbb{R}\ni x\mapsto
\frac{1}{\sqrt{2}}\;\psi\!\left(q+\frac{x}{2}\right)\!\in\mathbb{C}\,
;
\end{equation}
hence, according to the Cauchy-Schwarz inequality, we have:
\begin{equation}
|\quapsi (q,p)|\le \frac{1}{\pi}\; \|\psi\|^2, \ \
\forall\,\psi\in\lr\, ,\ \forall\, q,p\in\mathbb{R}\, .
\end{equation}
Actually, one can prove that, for any $\psi\in\lr$, the function
$\quapsi$ belongs to the space of continuous functions on $\rr$
`vanishing at infinity'; i.e.\ $\quapsi\in\cfvi (\rr)$, where:
\begin{eqnarray}
\cfvi (\rr) \spa & := & \spa \,\Big\{ f\in\mathrm{C}(\rr):\
\forall\,\epsilon
>0\, ,\ \mbox{the set} \nonumber\\
& & \hspace{1.4mm}\{(q,p)\in\rr: |f((q,p)|\ge \epsilon\}\ \mbox{is
compact in $\rr$} \Big\}\,.
\end{eqnarray}
One can easily prove, moreover, that the function $\quapsi$
\emph{assumes only real values}.

As far as we know, it is not completely clear in what way Wigner
obtained formula~{(\ref{defwig})}. It seems that he achieved this
expression by requiring that some general properties were satisfied
in a `simple way' (see~\cite{Hillery} and references therein); in
particular:
\begin{enumerate}
\item As already mentioned, the function $\quapsi$ assumes only real
values.

\item The \emph{marginal sub-distributions}
\begin{equation}
\quapsi (q,\cdot) :\, \mathbb{R}\ni p \mapsto\quapsi (q,p)\, ,\
q\in\mathbb{R}\,,\ \ \quapsi (\cdot,p) :\, \mathbb{R}\ni q
\mapsto\quapsi (q,p)\, ,\ p\in\mathbb{R}\,,
\end{equation}
satisfy the following relations:
\begin{equation}
\label{marg1}
\intr \quapsi (q,p) \dep = |\psi(q)|^2\, ,\ \
\mbox{for a.a.\ $q\in\mathbb{R}\, ,$}
\end{equation}
\begin{equation}
\label{marg2} \intr \quapsi (q,p) \deq =
|\big(\mathcal{F}\psi\big)(p)|^2\, ,\ \ \mbox{for a.a.\
$p\in\mathbb{R}\, ,$}
\end{equation}
where $\mathcal{F}:\, \lr\rightarrow\lr$ is the Fourier-Plancherel
operator. We remark that, rigorously, the function $\quapsi$ and the
associated marginal sub-distributions are not integrable, in
general. However, one can easily prove that, if $\mathcal{F}\psi$
belongs to $\lru$ (hence, $\mathcal{F}\psi\in\lru\cap\lr$), then the
marginal sub-distribution $\quapsi (q,\cdot)$ is contained in $\lru$
too and relation~{(\ref{marg1})} holds true. Analogously, if $\psi$
belongs to $\lru$ ($\cap\;\lr$), then $\quapsi (\cdot,p)$ is
contained in $\lru$ and relation~{(\ref{marg2})} is satisfied as
well. Notice that, if relation~{(\ref{marg1})} holds (in particular,
if $\mathcal{F}\psi\in\lru$), then
\begin{equation}
\intr\left(\intr \quapsi (q,p) \dep\right)\!\deq = \|\psi\|^2\, ;
\end{equation}
similarly, if relation~{(\ref{marg2})} holds (in particular, if
$\psi\in\lru$), then
\begin{equation}
\intr\left(\intr \quapsi (q,p) \deq\right)\!\dep = \|\psi\|^2\, .
\end{equation}
Moreover, it is possible to prove that if $\psi$ belongs to the
Schwartz space $\mathcal{S}(\mathbb{R})$, then $\quapsi$ belongs to
$\mathcal{S}(\rr)$; thus, both relations~{(\ref{marg1})}
and~{(\ref{marg2})} hold true, and we have that
\begin{equation}
\label{integr} \intrr \quapsi (q,p) \dqdp = \|\psi\|^2\, .
\end{equation}
However, we stress that, for $\|\psi\|=1$, the Wigner distribution
associated with the pure state $\hpsi$ cannot be regarded as a
genuine probability distribution as it assumes, in general, both
positive and negative values (this fact is already explicitly
observed in Wigner's original paper~\cite{Wigner}).

\item The function $\quapsi$ behaves in an `elementary way' with
respect to position and momentum translations; namely:
\begin{eqnarray} \label{euno}
\hspace{-0.5cm} \psi(q)\mapsto\psi(q-q^\prime)=\left(e^{-\ima
q^\prime\hat{p}}\,\psi\right)(q)\;  & \Longrightarrow & \;
\quapsi(q,p)\mapsto\quapsi(q-q^\prime, p)\,,
\\ \label{edue} \hspace{-0.5cm}
\psi(q)\mapsto e^{\ima p^\prime q}\,\psi(q)=\left(e^{\ima
p^\prime\hat{q}}\,\psi\right)(q)\; & \Longrightarrow & \;
\quapsi(q,p)\mapsto\quapsi(q, p-p^\prime)\,,
\end{eqnarray}
where we have denoted by $\hq$ and $\hp$ the standard position and
momentum operators in $\lr$, respectively.

\end{enumerate}

However, we point out that it is the peculiar property of satisfying
a relation of the type~{(\ref{for3})} for the expectation values of
observables the salient feature of the Wigner distribution. As it
will be shown later on, one can actually associate with \emph{any}
trace class operator in $\lr$ (in particular, with any physical
state, i.e.\ not only with a pure state) a suitable (generalized)
Wigner distribution; this association will then allow to obtain an
expression of the type~{(\ref{for3})}. The first step of this
generalization is to associate with any finite-rank operator a
Wigner distribution (we will not attempt at establishing
formula~{(\ref{for3})} itself, for the moment). To this aim, for any
couple of vectors $\phi, \psi$ in $\lr$, let us set:
\begin{equation}
\label{defwig2} \quafp (q,p) := \fac\intr e^{-\ima p x}\,
\psi\!\left(q-\frac{x}{2}\right)^*
\phi\!\left(q+\frac{x}{2}\right)\dx \, ;
\end{equation}
this expression is a straightforward generalization of
formula~{(\ref{defwig})}, relating a \emph{generic} rank-one
operator $\hfp\equiv\phipsi$ with a $\mathbb{C}$-valued function.
Notice that, as in the case of
$\quapsi\equiv\qua_{\widehat{\psi\psi}}$, the function $\quafp$ is
well defined since that map $x\mapsto
\phi\!\left(q-\frac{x}{2}\right)^* \psi\!\left(q+\frac{x}{2}\right)$
belongs to $\lr$ for all $q\in\mathbb{R}$. It is also immediate to
observe that, for any $q,p\in\mathbb{R}$, $|\quafp (q,p)|\le
\frac{1}{\pi}\; \|\phi\|\, \|\psi\|$, and
\begin{eqnarray}
\quafp (q,p)^* \spa & = & \spa \fac\intr e^{\ima p x}\,
\psi\!\left(q-\frac{x}{2}\right)\,
\phi\!\left(q+\frac{x}{2}\right)^*\dx
\nonumber\\
& = & \spa \fac\intr e^{-\ima p x}\,
\phi\!\left(q-\frac{x}{2}\right)^*
\psi\!\left(q+\frac{x}{2}\right)\dx \, ;
\end{eqnarray}
hence:
\begin{equation}
\quafp (q,p)^*=\quapf (q,p)\, ,\ \ \forall\, \phi,\psi\in\lr\, .
\end{equation}
One can prove, moreover, that for any $\phi,\psi\in\lr$ the function
$\quafp$ is contained in $\lrr\cap\cfvi (\rr)$, and the following
important relation --- the \emph{Moyal identity} --- holds true:
\begin{eqnarray}
\intrr \quafpa (q,p)^*\, \quafpb (q,p) \dqdp  \spa & = & \spa
\frac{1}{2\pi}\,\langle\phi_1,\phi_2\rangle\,\langle\psi_2,\psi_1\rangle
\nonumber\\
& = & \spa \frac{1}{2\pi}\,\tr\! \left(\hfpa^*\,\hfpb\right),
\label{Moya}
\end{eqnarray}
for all $\phi_1,\psi_1,\phi_2,\psi_2\in\lr$; in particular, for
$\phi_1=\psi_1=\phi_2=\psi_2\equiv\psi$, and recalling that $\quapsi
(q,p)\in\mathbb{R}$, we have:
\begin{equation}
\label{integr2} \intrr \quapsi (q,p)^2 \dqdp = \fac\, \|\psi\|^4
\end{equation}
(compare with formula~{(\ref{integr})}; notice, however, that
formula~{(\ref{integr2})} holds for \emph{every} vector $\psi$ in
$\lr$).

Consider now the family of unitary operators
\begin{equation}
\{U\qp\}_{q,p\in\mathbb{R}}\subset \mathcal{U}(\lr)
\end{equation}
(given a Hilbert space $\mathcal{H}$, we denote by
$\mathcal{U}(\mathcal{H})$ the unitary group of $\mathcal{H}$),
defined by
\begin{eqnarray}
U\qp \spa & := & \spa \disp
\nonumber\\
& = & \spa e^{-\frac{\ima}{2}\, qp}\, \exp(\ima
p\hspace{0.5mm}\hq)\, \exp(-\ima q\hspace{0.3mm}\hp) \nonumber\\
\label{Weyl-sys}
& = & \spa   e^{\frac{\ima}{2}\, qp}\, \exp(-\ima
q\hspace{0.3mm}\hp)\, \exp(\ima p\hspace{0.5mm}\hq)\, , \ \
q,p\in\mathbb{R}\, .
\end{eqnarray}
One can prove (see ref.~\cite{Ali2}) that the function
$\tr(U(\cdot,\cdot)^\ast \hfp)\big)\colon (q,p)\mapsto
\tr(U(q,p)^\ast \hfp)$ belongs to $\lrr$ and the following relation
holds:
\begin{equation} \label{for-quafp}
\quafp (q,p)=\frac{1}{2\pi}\big(\fs \tr(U(\cdot,\cdot)^\ast
\hfp)\big)(q,p),
\end{equation}
where $\fs \colon \lrr \rightarrow\lrr$ is the \emph{symplectic
Fourier transform}, i.e.\ the unitary operator determined by
\begin{equation}
\big(\fsy f\big)(q,p)=\frac{1}{2\pi}\intrr f(q^\prime,p^\prime)\,
e^{\ima (qp^\prime - pq^\prime)}\; \de q^\prime \de p^\prime,\ \ \
\forall\hspace{0.4mm} f\in\lurr\cap\lrr.
\end{equation}
Recall that $\fsy$ enjoys the remarkable property of being both
unitary and selfadjoint:
\begin{equation}
\fs=\fsy^\ast,\ \ \ \fsy^2=I.
\end{equation}
Thus, for any $\phi,\psi\in\lr$, the Wigner distribution is the
symplectic Fourier transform of the function
\begin{equation} \label{F-Wig}
\fwfp \colon \rr\ni(q,p)\mapsto (2\pi)^{-1}\,\tr(U(q,p)^\ast
\hfp)\in\ccc,
\end{equation}
which is usually called \emph{Fourier-Wigner distribution}
associated with the rank-one operator $\hfp$. It is a peculiar fact
that the  Fourier-Wigner distribution can be cast in a form similar
to the standard Wigner distribution (compare with
formula~{(\ref{defwig2})}):
\begin{eqnarray}
\fwfp (q,p) \spa & = & \spa \fac\intr e^{\frac{\ima}{2}\, q p}\,
e^{-\ima p y}\, \psi(y-q)^*
\phi(y)\; \de y  \nonumber\\
& = & \spa \fac\intr e^{-\ima p x}\,
\psi\!\left(x-\frac{q}{2}\right)^*
\phi\!\left(x+\frac{q}{2}\right)\dx .
\end{eqnarray}
It is clear that, since $\fs$ is unitary, the function
$\fwfp=\fs\quafp$ satisfies a relation completely analogous to the
Moyal identity~{(\ref{Moya})}.

As it is well known, the map
$\mathbb{R}\times\mathbb{R}\ni\qp\mapsto U\qp$ that appears in the
definition of the Wigner and Fourier-Wigner distributions is an
\emph{irreducible projective representation} of the group
$\mathbb{R}\times\mathbb{R}$ in $\lr$; with a slight abuse of
terminology, we will call it \emph{Weyl system}.\footnote{Strictly
speaking, it is the pair of unitary representations
($p\mapsto\exp(\ima p\hspace{0.5mm}\hq)$, $q\mapsto\exp(-\ima
q\hspace{0.3mm}\hp)$) that it is customary to call `Weyl system',
see~\cite{Marmo}; however, the irreducible projective representation
$U$ has the same physical meaning since it `codifies' the canonical
commutation relations (in integrated form), as shown
in~{(\ref{Weyl-sys})}. The representation $U$ is strictly related to
a \emph{Schr\"odinger representation} of the Heisenberg-Weyl group,
see ref.~\cite{Folland-bis}.} The Moyal identity~{(\ref{Moya})} is a
manifestation of the fact that the representation $U$ is
\emph{square integrable}. This property, whose main technical
aspects will be recalled in the next section, allows to extend the
notion of Wigner distribution defining a \emph{Wigner transform}
which associates with any Hilbert-Schmidt operator in $\lr$ a
suitable numerical function; furthermore, as it will be shown in
Sect.~{\ref{revisited}}, one can actually define a (generalized)
Wigner transform for \emph{every} square integrable representation.

%-------------------------------------------------------------------------------
\section{A technical interlude: square integrable representations}
\label{interlude}
%-------------------------------------------------------------------------------

In this section, we will use some basic facts of the theory of
topological groups and their representations; standard references on
the subject are~{\cite{Raja,Folland}}.

Let $G$ be a locally compact second countable Hausdorff topological
group (in short, l.c.s.c.\ group). We will denote by $\mG$ and
$\Delta_G$ respectively a {\it left Haar measure} (of course
uniquely defined up to a multiplicative constant) and the {\it
modular function} on $G$. The symbol $e$ will indicate the unit
element in $G$.

Given a separable complex Hilbert space $\mathcal{H}$, the symbol
$\mathcal{U}(\mathcal{H})$ will denote, as in Sect.~{\ref{wigdis}},
the {\it unitary group} of $\mathcal{H}$ --- i.e.\ the group of all
unitary operators in $\mathcal{H}$, endowed with the strong operator
topology --- which is a metrizable second countable Hausdorff
topological group.

We will mean by the term {\it projective representation} of a
l.c.s.c.\ group $G$ a Borel projective representation of $G$ in a
separable complex Hilbert space $\mathcal{H}$ (see, for instance,
ref.~\cite{Raja}, chapter~{VII}), namely a map of $G$ into
$\mathcal{U}(\mathcal{H})$ such that
\begin{itemize}
\item
$U$ is a weakly Borel map, i.e.\ $G\ni g\mapsto
\langle\phi,U(g)\,\psi\rangle\in\mathbb{C}$ is a Borel
function,\footnote{The terms {\it Borel function} (or map) and {\it
Borel measure} will be always used with reference to the natural
Borel structures on the topological spaces involved, namely to the
smallest $\sigma$-algebras containing all open subsets.} for any
couple of vectors $\phi,\psi\in\mathcal{H}$;
\item
$U(e)=I$, where $I$ the identity operator in $\mathcal{H}$;
\item
denoted by $\mathbb{T}$ the circle group, namely the group of
complex numbers of modulus one, there exists a Borel function $\mm
\colon G\times G\rightarrow\mathbb{T}$ such that
\[
U(gh)=\mm (g,h)\,U(g)\,U(h),\ \ \ \forall\, g,h\in G.
\]
\end{itemize}
The function $\mm$, which is called the {\it multiplier associated
with} $U$, satisfies the following conditions:
\begin{equation}
\mm(g,e)=\mm(e,g)=1,\ \ \ \ \forall g\in G,
\end{equation}
and
\begin{equation}
\mm(g_1,g_2g_3)\, \mm(g_2,g_3)= \mm(g_1 g_2,g_3)\, \mm(g_1,g_2),\ \
\ \ \forall\, g_1,g_2,g_3\in G.
\end{equation}
In particular, in the case where $\mm\equiv 1$, $U$ is a standard
unitary representation; in this case, according to a well known
result, the hypothesis that the map $U$ is weakly Borel implies that
it is, actually, strongly continuous. The notion of irreducibility
is defined for projective representations as for unitary
representations.

Let $\ut \colon G\rightarrow\mathcal{U}(\widetilde{\mathcal{H}})$ be
a projective representation of $G$ in a (separable complex) Hilbert
space $\widetilde{\mathcal{H}}$. We say that $\ut$ is
\emph{physically equivalent} to $U$ if there exist a Borel function
$\beta \colon G\rightarrow\mathbb{T}$ and a unitary or antiunitary
operator $W \colon \mathcal{H}\rightarrow\widetilde{\mathcal{H}}$
such that
\begin{equation}
\ut(g)=\beta(g)\, W\,U(g)\,W^\ast,\ \ \ \forall\hspace{0.3mm} g\in
G.
\end{equation}
Notice that the notion of physical equivalence is coherent with
Wigner's theorem on symmetry actions. It is clear that a projective
representation, physically equivalent to an irreducible projective
representation, is irreducible too.

Let $U$ be an \emph{irreducible} projective representation of the
l.c.s.c.\ group $G$ in the Hilbert space $\mathcal{H}$. Then, given
two vectors $\psi,\phi\in\mathcal{H}$, we define the function
(usually called `coefficient')
\begin{equation}
c_{\psi ,\phi}^U \colon G\ni g\mapsto \langle U(g)\,\psi
,\phi\rangle\in\ccc\, ,
\end{equation}
and we consider the set (of `admissible vectors for $U$')
\begin{equation}
\mathcal{A}(U):=\left\{\psi\in\mathcal{H}\,|\
\exists\phi\in\mathcal{H}:\, \phi\neq 0,\, c_{\psi,\phi}^U \in
\ldg\right\},
\end{equation}
where $\ldg\equiv\mathrm{L}^2(G,\mu_G;\ccc)$. The representation $U$
is said to be {\it square integrable} if
\begin{equation}
\mathcal{A}(U)\neq\{0\}.
\end{equation}
Square integrable projective representations are characterized by
the following result --- see ref.~{\cite{Aniello}} --- which is a
generalization of a classical theorem of Duflo and
Moore~\cite{Duflo} concerning unitary representations:
\begin{theorem} \label{Duflo-Moore}
Let the projective representation $U\colon\,G\rightarrow
\mathcal{U}(\mathcal{H})$ be square integrable. Then, the set
$\mathcal{A}(U)$ is a dense linear span\footnote{Throughout the
paper, we call a nonempty subset of a vector space $\mathsf{V}$
`linear span' if it is a linear space itself (with respect to the
operations of $\mathsf{V}$), {\it with no extra requirement of
closedness with respect to any topology on} $\mathsf{V}$; we prefer
to use the term `(vector) subspace' of $\mathsf{V}$ for indicating a
\emph{closed} linear span (with respect to a given topology on
$\mathsf{V}$).} in $\mathcal{H}$, stable under the action of $U$,
and, for any couple of vectors $\phi\in\mathcal{H}$ and
$\psi\in\mathcal{A}(U)$, the coefficient $c_{\psi,\phi}^U$ is square
integrable with respect to the left Haar measure $\mu_G$ on $G$.
Moreover, there exists a unique positive selfadjoint injective
linear operator $\duu$ in $\mathcal{H}$
--- {\em which we will call the `Duflo-Moore operator associated with} $U$' ---
such that
\begin{equation}
\mathcal{A}(U)=\mathrm{Dom}\big(\duu\big)
\end{equation}
and the following `orthogonality relations' hold:
\begin{eqnarray}
\int_G c_{\psi_1,\phi_1}(g)^\ast \,c_{\psi_2,\phi_2}(g)\ \de\mu_G(g)
\spa  & = & \spa \int_G \langle\phi_1,U(g)\,\psi_1\rangle\, \langle
U(g)\,\psi_2,\phi_2 \rangle\ \de\mu_G (g)
\nonumber \\
\label{orthrel} & = & \spa \langle\phi_1 ,\phi_2\rangle\, \langle
\duu\,\psi_2, \duu\,\psi_1\rangle ,
\end{eqnarray}
for all $\phi_1,\phi_2\in\mathcal{H}$ and all $\psi_1,\psi_2\in
\mathcal{A}(U)$. The Duflo-Moore operator $\duu$ is
\emph{semi-invariant}, with respect to $U$, \emph{with weight}
$\Delta_G^{1/2}$, i.e.
\begin{equation} \label{semi-invariance}
U(g)\,\duu = \Delta_G(g)^{\frac{1}{2}}\,\duu\, U(g),\ \ \
\forall\hspace{0.4mm} g\in G;
\end{equation}
it is bounded if and only if $G$ is unimodular \emph{(i.e.\
$\Delta_G\equiv 1$)} and, in such case, it is a multiple of the
identity.
\end{theorem}
\begin{remark} \label{dufmoo}
{\rm If $U$ is square integrable, the associated Duflo-Moore
operator $\duu$, being injective and selfadjoint, has a selfadjoint
densely defined inverse. Duflo and Moore call (for historical
reasons) the square of $\du^{-1}$ the {\it formal degree} of the
representation $U$. Notice that the operator $\duu$ is linked to the
normalization of the Haar measure $\mu_G$. Indeed, if $\mu_G$ is
rescaled by a positive constant, then $\duu$ is rescaled by the
square root of the same constant. Keeping this fact in mind, we will
say that $\duu$ is {\it normalized according to} $\mu_G$. On the
other hand, if a normalization of the left Haar measure on $G$ is
not fixed, $\duu$ is defined up to a positive factor and we will
call a specific choice a {\it normalization of the Duflo-Moore
operator}. In particular, if $G$ is unimodular, then $\duu=I$ is a
normalization of the Duflo-Moore operator; the corresponding Haar
measure will be said to be \emph{normalized in agreement with the
representation} $U$. Moreover, observe that, according to
relation~{(\ref{semi-invariance})}, the dense linear span
$\mathrm{Dom}\big(\du^{-1}\big)=\mathrm{Ran}\big(\duu\big)$ (like
the linear span $\mathcal{A}(U)=\mathrm{Dom}\big(\duu\big)$) is
stable under the action of $U$ and
\begin{equation} \label{semi-invariance-bis}
U(g)^\ast\hspace{0.3mm}\du^{-1} =
\Delta_G(g)^{\frac{1}{2}}\,\du^{-1}\hspace{0.4mm} U(g)^\ast,\ \ \
\forall\hspace{0.4mm} g\in G.
\end{equation}
Finally, we notice that the orthogonality
relations~{(\ref{orthrel})} can also be written replacing the
positive selfadjoint operator $\duu$ with a closed injective
operator $\Kuu$ which is only required to be selfadjoint. Such an
operator $\Kuu$ is not unique (e.g., trivially, one can set
$\Kuu=-\duu$), and it is characterized by a polar decomposition of
the form $\Kuu= V\hspace{0.5mm}\duu$, where $V$ is a suitable
unitary operator in $\mathcal{H}$. A selfadjoint operator satisfying
the orthogonality relations will be called a \emph{variant of the
Duflo-Moore operator}.~$\blacksquare$ }
\end{remark}

Let us list a few basic facts about square integrable
representations:
\begin{enumerate}

\item The square-integrability of a representation extends to all its
physical equivalence class. Thus, we can say consistently that a
certain physical equivalence class of representations is square
integrable.

\item In the case where the l.c.s.c.\ group $G$ is compact (hence, unimodular),
every irreducible projective representation of $G$ is square
integrable (since, in this case, the Haar measure on $G$ is finite)
and, in the case of a unitary representation,
Theorem~{\ref{Duflo-Moore}} coincides with the celebrated Peter-Weyl
theorem. The trivial representation of $G$ in $\mathbb{C}$ is square
integrable if and only if $G$ is compact.

\item If the representation $U$ of $G$ is square integrable, then the
orthogonality relations~{(\ref{orthrel})} imply that, for any
nonzero admissible vector $\psi\in\mathcal{A}(U)$, one can define
the linear operator
\begin{equation}
\wt \colon \mathcal{H}\ni \phi\mapsto \|\duu\,\psi\|^{-1}\, c_{\psi
,\phi}^U\in \ldg
\end{equation}
--- sometimes called {\it (generalized)
wavelet transform} generated by $U$, with {\it analyzing} or {\it
fiducial vector} $\psi$ --- which is an isometry. Notice that $\wt$
is the frame transform associated with the \emph{normalized tight
frame} $\{\|\duu\,\psi\|^{-1}\,U(g)\,\psi\}_{g\in G}$ in
$\mathcal{H}$ based on $(G,\mu_G)$. For the adjoint
$\wta\colon\ldg\rightarrow\mathcal{H}$ of the isometry $\wt$ the
following formula holds (compare with the reconstruction
formula~{(\ref{re-con})}):
\begin{equation} \label{re-con-wt}
\wta\hspace{0.5mm} f = \| \duu\,\psi\|^{-1} \int_G
f(g)\hspace{0.5mm}\big(U(g)\,\psi\big)\,\de\mu_G(g),\ \ \
\forall\hspace{0.4mm} f\in\ldg.
\end{equation}
The ordinary wavelet transform arises in the special case where $G$
is the 1-dimensional affine group $\mathbb{R}\sdp\mathbb{R}^+_\ast$
(see~\cite{Grossmann}).

\item The isometry $\wt$ intertwines the square integrable
representation $U$ with the {\it left-regular $\mm$-representation}
$R_\mm\hspace{-0.3mm}$ of $G$ in $\ldg$, see ref.~{\cite{Aniello}},
which is the projective representation (with multiplier $\mm$)
defined by:
\begin{equation}
\big(R_\mm\hspace{-0.3mm}(g) f\big)(g^\prime
)=\overset{\rightarrow}{\mm}(g,g^\prime)\, f(g^{-1}g^\prime ),\ \ \
g,g^\prime\in G,
\end{equation}
\begin{equation}
\overset{\rightarrow}{\mm}(g,g^\prime):=\mm(g,g^{-1})^\ast\,\mm(g^{-1},g^\prime),
\end{equation}
for every $f\in \ldg$; namely:
\begin{equation}
\wt\, U(g) = R_\mm\hspace{-0.3mm}(g)\; \wt,\ \ \
\forall\hspace{0.4mm} g\in G.
\end{equation}
Hence, $U$ is (unitarily) equivalent to a subrepresentation of
$R_\mm$. Notice that, for $\mm\equiv 1$, $R\equiv
R_\mm\hspace{-0.3mm}$ is the standard left regular representation of
$G$.

\item Since $\wt$ is a frame transform, the range
$\rpsi\equiv\mathrm{Ran}\big(\wt\big)$ --- which, by Schwarz
inequality, consists of (equivalence classes of $\mu_G$-almost
everywhere) \emph{bounded} square integrable functions
--- is a r.k.H.s.\ (embedded in $\ldg$; see Remark~{\ref{RKHS}}), and the reproducing kernel
is given explicitly by:
\begin{equation}
\varkappa_\psi^U(g,g^\prime) := \| \duu\,\psi\|^{-2}\, \langle
U(g)\,\psi ,U(g^\prime)\, \psi\rangle, \ \ \ g,g^\prime\in G.
\end{equation}
Namely, for every function $f$ in $\rpsi$, we have:
\begin{equation}
f(g)=\int_G \varkappa_\psi^U(g,g^\prime)\, f(g^\prime )\
\de\mu_G(g^\prime),\ \ \ \mbox{for $\mu_G$-a.a.\ $g\in G$}.
\end{equation}

\item The wavelet transform $\wt$ intertwines a bounded
operator $\opa$ in $\mathcal{H}$ with an integral operator in
$\ldg$:
\begin{equation} \label{relwtapsi}
\wt\, \opa = \Apsi\;\wt,\ \ \ \opa\in\mathcal{B}(\mathcal{H}),
\end{equation}
where
\begin{equation} \label{forapsi}
\big(\Apsi\hspace{0.3mm} f\big)(g):=\int_G \varkappa_\psi^U(\opa;
g,g^\prime)\, f(g^\prime )\ \de\mu_G(g^\prime),\ \ \ f\in\ldg,
\end{equation}
\begin{equation}
\varkappa_\psi^U(\opa; g,g^\prime) := \| \duu\,\psi\|^{-2}\, \langle
U(g)\,\psi ,\opa\,U(g^\prime)\, \psi\rangle, \ \ \ g,g^\prime\in G;
\end{equation}
in particular: $\varkappa_\psi^U(I;
g,g^\prime)=\varkappa_\psi^U(g,g^\prime)$. Since
\begin{equation}
\varkappa_\psi^U(\opa;g,\cdot)= \| \duu\,\psi\|^{-2}\, \langle
U(\cdot)\,\psi ,\opa^\ast\,U(g)\, \psi\rangle^\ast
\end{equation}
and the function $\| \duu\,\psi\|^{-2}\, \langle U(\cdot)\,\psi
,\opa^\ast\,U(g)\, \psi\rangle$ belongs to $\rpsi$, denoting by
$\rpsip$ the orthogonal complement in $\ldg$ of $\rpsi$, the
operator $\Apsi$ satisfies:
\begin{equation}
\Apsi\hspace{0.3mm} f = 0,\ \ \ \forall\hspace{0.4mm}f\in\rpsip;
\end{equation}
therefore, we have (compare with relation~{(\ref{bop})}):
\begin{equation} \label{relapsi}
\Apsi=\hspace{0.4mm}\wt\,\opa\; \wta\,.
\end{equation}
Moreover, relation~{(\ref{relwtapsi})} implies that
$\opa=\hspace{0.4mm}\wta\,\Apsi\hspace{0.8mm} \wt$; hence, by means
of formulae~{(\ref{re-con-wt})} and~{(\ref{forapsi})}, we get the
following (weak integral) formula:
\begin{equation} \label{w-int-for}
\opa= \| \duu\,\psi\|^{-2} \int_G \de\mu_G(g) \int_G
\de\mu_G(g^\prime)\ \varkappa_\psi^U(\opa; g,g^\prime)\,
|U(g)\,\psi\rangle\langle U(g^\prime)\,\psi|\, .
\end{equation}

\item Since for the Fourier-Wigner transform a relation analogous to the Moyal
identity holds true, namely,
\begin{eqnarray}
\intrr \fwfpu (q,p)^*\, \fwfpd (q,p) \dqdp  \spa & = & \spa \intrr
\langle\phi_1,U(q,p)\,\psi_1\rangle\, \langle U(q,p)\,\psi_2,\phi_2
\rangle \frac{\dqdp}{(2\pi)^2}
\nonumber \\
& = & \spa
\frac{1}{2\pi}\,\langle\phi_1,\phi_2\rangle\,\langle\psi_2,\psi_1\rangle,
\end{eqnarray}
we conclude that the projective representation
\begin{equation}
U\colon\rr\ni(q,p)\mapsto\disp\in\mathcal{U}(\lr)
\end{equation}
is square integrable and, fixing $(2\pi)^{-1}\de q\de p$ as the Haar
measure on $\rr$, we have that $\du=I$. Therefore, the Haar measure
$(2\pi)^{-1}\de q\de p$ is normalized in agreement with $U$. If
$\psi\in\lr$ is the ground state of the quantum harmonic oscillator,
then $\{ U(q,p)\,\psi\}_{q,p\in\mathbb{R}}$ is the family of
standard \emph{coherent states}~{\cite{Perelomov,Klauder}}, which is
a normalized tight frame in $\lr$ based on
$(\mathbb{R}\times\mathbb{R},(2\pi)^{-1}\de q\de p)$.

\end{enumerate}

As a consequence of the `trace formula for frames' --- see
Proposition~{\ref{trforfr}} --- we have the following further
remarkable property of square integrable representations:
\begin{proposition}[the `first trace formula for sq.\ int.\ reps.']
Let $U\colon G\rightarrow \mathcal{U}(\mathcal{H})$ be a square
integrable projective representation and $\duu$ the associated
Duflo-Moore operator {\rm (normalized according to the left Haar
measure $\mu_G$)}. Then, for any couple of admissible vectors
$\psi,\phi\in\mathcal{A}(U)$ and any trace class operator $\opa$ in
$\mathcal{H}$, the following formula holds:
\begin{equation} \label{trace-for}
\tr(\opa)\,\langle \duu\, \psi, \duu\,\phi\rangle = \int_G \langle
U(g)\,\psi ,\opa\, U(g)\, \phi\rangle \deggi.
\end{equation}
\end{proposition}

\noindent \emph{Proof:} We will assume that $\psi\neq 0 \neq \phi$,
otherwise the statement is trivial; we will further assume, for the
moment, that $\phi=\psi\in\mathcal{A}(U)$. Then, as already
observed, the set of vectors
$\{\|\duu\,\psi\|^{-1}\,U(g)\,\psi\}_{g\in G}$ is a normalized tight
frame in $\mathcal{H}$ based on $(G,\mu_G)$, and
formula~{(\ref{trace-for})}
--- for every $\opa\in\mathcal{B}_1(\mathcal{H})$ and with
$\phi=\psi$
--- follows from formula~{(\ref{trace-for0})} applied to this frame.

In order to extend the proof to the case where $\phi\neq\psi$, we
can use the result just obtained and a standard `polarization
argument'.  Let $\opa$ be a trace class operator in $\mathcal{H}$
and $\psi,\phi$ arbitrary vectors in $\mathcal{A}(U)$. Notice that
we have:
\begin{eqnarray}
\tr(\opa)\, \langle \duu\, \psi, \duu\,\phi\rangle  \spa & = & \spa
\tr(\opa)\,\frac{1}{4}\Big( \langle \duu (\psi+\phi),
\duu(\psi+\phi)\rangle - \langle \duu(\psi - \phi), \duu(\psi -
\phi)\rangle
\nonumber\\
& - & \spa \ima \big(\langle \duu(\psi+\ima\phi),
\duu(\psi+\ima\phi)\rangle - \langle \duu(\psi - \ima\phi),
\duu(\psi - \ima\phi)\rangle\big) \Big)
\nonumber\\
& = & \spa \frac{1}{4}\intg\Big( \langle U(g)\, (\psi+\phi),
\opa\,U(g)\,(\psi+\phi)\rangle
\nonumber\\
& - & \spa \langle U(g)\,(\psi - \phi), \opa\,U(g)\,(\psi -
\phi)\rangle - \ima \big(U(g)\,(\psi+\ima\phi),
\opa\,U(g)\,(\psi+\ima\phi)\rangle
\nonumber\\
& - & \spa \langle U(g)\,(\psi - \ima\phi), \opa\,U(g)\,(\psi -
\ima\phi)\rangle\big) \Big) \deggi
\nonumber\\
& = & \spa \intg \langle U(g)\,\psi,\opa\,U(g)\,\phi\rangle \deggi .
\end{eqnarray}

The proof is complete.~$\square$

One can furthermore prove that, in the case where the l.c.s.c.\
group $G$ is unimodular, the first trace formula for square
integrable representations is a particular case of the following
result:
\begin{proposition}[the `second trace formula for sq.\ int.\ reps.']
Let $U \colon G\rightarrow \mathcal{U}(\mathcal{H})$ be a square
integrable projective representation of a unimodular l.c.s.c.\ group
$G$ and let $\duu=\ddu\,I$, $\ddu>0$, be the associated Duflo-Moore
operator {\rm (normalized according to the Haar measure $\mu_G$)}.
Then, for any couple of trace class operators $\opa,\op$ in
$\mathcal{H}$, the following formula holds:
\begin{equation} \label{trace-for-bis}
\tr(\opa)\,\tr(\op) = d_U^{-2}\int_G \tr( U(g)\,\op\,U(g)^\ast \opa)
\deggi.
\end{equation}
\end{proposition}

\noindent \emph{Proof:} As in the proof of
Proposition~{\ref{trforfr}}, we can exploit the fact that every
trace class operator can be expressed as a linear combination of
four \emph{positive} trace class operators, and we can restrict the
proof of relation~{(\ref{trace-for-bis})}
--- with no loss of generality
--- to the case where $\opa,\op$ are generic nonzero
\emph{positive} trace class operator in $\mathcal{H}$. Then, let us
consider the canonical decomposition of $\op$ as a nonzero
(positive) compact operator
\begin{equation} \label{suma}
\op=\sum_{n\in\mathcal{N}} \tau_n\,|\psi_n\rangle\langle\psi_n|,\ \
\ \psi_n,\phi_n\in\mathcal{H},
\end{equation}
where $\mathcal{N}$ is a finite or countably infinite index set,
$\{\psi_n\}_{n\in\mathcal{N}}$ is an orthonormal system and
$\{\tau_n\}_{n\in\mathcal{N}}$ is a set of strictly positive numbers
--- the nonzero singular values of $\op$ (which, being $\op$ positive,
coincide with the nonzero eigenvalues of $\op$) --- such that
\begin{equation}
\sum_{n\in\mathcal{N}} \tau_n =\tr(\op) ;
\end{equation}
the sum~{(\ref{suma})} converges with respect to the trace norm.
Observe that the map
\begin{equation}
\mathcal{B}_1(\mathcal{H})\ni\hat{S}\mapsto\tr(
U(g)\,\hat{S}\,U(g)^\ast \opa)=\tr(\hat{S}\,U(g)^\ast \opa\,
U(g))\in\ccc
\end{equation}
is a bounded linear functional; hence:
\begin{equation}
\tr(U(g)\,\op\,U(g)^\ast \opa)= \sum_{n\in\mathcal{N}} \tau_n\,
\tr(U(g)\,|\psi_n\rangle\langle\psi_n|\,U(g)^\ast \opa).
\end{equation}
Therefore, we have:
\begin{eqnarray}
\int_G \tr( U(g)\,\op\,U(g)^\ast \opa) \deggi \spa & = & \spa \int_G
\sum_{n\in\mathcal{N}} \tau_n\hspace{0.5mm} \langle
U(g)\,\psi_n,\opa\,U(g)\,\psi_n\rangle \deggi
\nonumber\\
& = & \spa \sum_{n\in\mathcal{N}} \tau_n\hspace{0.3mm} \int_G
\langle U(g)\,\psi_n,\opa\,U(g)\,\psi_n\rangle \deggi
\nonumber\\
& = & \spa d_U^2\,\tr(\opa)\,\tr(\op),
\end{eqnarray}
where the permutation of the (possibly infinite) sum  with the
integral is allowed by the positivity of the integrand functions and
we have used the first trace formula~{(\ref{trace-for})}.~$\square$

In the next section, it will be shown that the notion of square
integrable representation allows to give a rigorous definition of
the Wigner transform, and to generalize this definition in a
straightforward way: with every square integrable projective
representation one can associate a suitable isometry, i.e.\ a
(generalized) Wigner transform.

%------------------------------------------------------------------------------
\section{Wigner transforms associated with square integrable
representations and the Wigner distribution (revisited)}
\label{revisited}
%------------------------------------------------------------------------------

The (generalized) wavelet transform defined in the previous section
is not the only remarkable linear map that one can construct, in a
natural way, by means of a square integrable representation. Indeed,
following ref.~{\cite{Ali2}}, we will show that --- given a square
integrable projective representation $U \colon
G\rightarrow\mathcal{U}(\mathcal{H})$ (with miltiplier $\mm$)
--- with every Hilbert-Schmidt operator $\opa\in\hs$ one can suitably
associate a function
\begin{equation}
G\ni g \mapsto \big(\wig \opa\big)(g)\in\ccc
\end{equation}
contained in $\ldg\equiv\mathrm{L}^2(G,\mu_G;\ccc)$. Denoting by
$\duu$, as in Sect.~{\ref{interlude}}, the Duflo-Moore operator
associated with $U$ (normalized according to a left Haar measure
$\mu_G$ on $G$), \emph{formally} we set:
\begin{equation} \label{formal}
\big(\wig \opa\big)(g):=\tr\big(U(g)^\ast\opa\,\du^{-1}\big).
\end{equation}
Since the operator $U(g)^\ast\opa\,\du^{-1}$ (or, possibly, its
closure) is not, in general, a trace class operator,
definition~{(\ref{formal})} is meaningless unless we provide a
rigorous interpretation. To this aim, we will exploit the fact that
finite rank operators form a dense linear span $\fr$ in $\hs$.
Precisely, consider those rank one operators in $\mathcal{H}$ that
are of the type
\begin{equation}
\hfp=|\phi\rangle\langle\psi |,\ \ \ \phi\in\mathcal{H},\
\psi\in\mathrm{Dom}(\du^{-1}).
\end{equation}
The linear span generated by the operators of this form, namely the
set
\begin{equation}
\fri := \{ \hat{F}\in\fr:\
\mathrm{Ran}(\hat{F^\ast})\subset\mathrm{Dom}(\du^{-1})\},
\end{equation}
is dense in $\fr$, hence, in $\hs$:
\begin{equation}
\overline{\fri}=\hs.
\end{equation}
Observe, moreover, that if we set
\begin{equation}
\big(\wig\,
\hfp\big)(g):=\tr(U(g)^\ast|\phi\rangle\langle\du^{-1}\psi|)=\langle
U(g)\,\du^{-1}\psi,\phi\rangle,\ \ \
\forall\hspace{0.4mm}\hfp\in\fri,
\end{equation}
then,  by virtue of the orthogonality relations~{(\ref{orthrel})},
for any $\hfpa ,\hfpb\in\fri$ we have:
\begin{eqnarray}
\int_G \big(\wig\, \hfpa\big)(g)^\ast\,\big(\wig\, \hfpb\big)(g)\;
\de\mu_G(g) \spa & = & \spa \int_G
\langle\phi_1,U(g)\,\du^{-1}\psi_1\rangle\langle
U(g)\,\du^{-1}\psi_2,\phi_2\rangle\; \de\mu_G(g) \nonumber\\
& = & \spa \langle\phi_1,\phi_2\rangle\,\langle\psi_2,\psi_1\rangle
= \big\langle \hfpa ,\hfpb\big\rangle_{\hs}.
\end{eqnarray}
Therefore, extending the map $\wig$ to all $\fri$ by linearity, and
then to the whole Hilbert space $\hs$ by continuity, we obtain an
\emph{isometry}
\begin{equation}
\wig \colon\hs\rightarrow\ldg
\end{equation}
called the \emph{(generalized) Wigner transform} generated by $U$.
As the reader may check, if the group $G$ is unimodular
($\Rightarrow \duu=\ddu\hspace{0.4mm}I$, with $\ddu>0$), then for
every trace class operator $\hrho\in\mathcal{B}_1(\mathcal{H})$
--- in particular, for every density operator in $\mathcal{H}$ ---
we have simply:
\begin{equation} \label{non-formal}
\big(\wig\hspace{0.3mm} \hrho\big)(g)=d_U^{-1}\,\tr(U(g)^\ast\hrho).
\end{equation}

Let us now investigate the intertwining property of the isometry
$\wig$ with respect to the natural action of the group $G$ in $\hs$.
Precisely, let us consider the map
\begin{equation}
\rep \colon G\rightarrow\mathcal{U}(\hs)
\end{equation}
defined by
\begin{equation} \label{definrep}
\rep(g)\hspace{0.3mm} \opa := U(g)\, \opa\, U(g)^\ast, \ \ \ \forall
\hspace{0.5mm}g\in G,\ \ \opa\in\hs.
\end{equation}
The map $\rep$ is a (strongly continuous) \emph{unitary}
representation
--- even if, in general, the representation $U$ has only been
assumed to be \emph{projective}
---  which can be regarded as the standard action of the `symmetry
group' $G$ on the quantum `observables' (or on the `states'). Next,
let us consider the map
\begin{equation}
\two \colon G\rightarrow \mathcal{U}(\ldg)
\end{equation}
defined by
\begin{equation} \label{two-sided}
\big(\two(g) f)(g^\prime)= \Delta_G(g)^{\frac{1}{2}}\hspace{0.9mm}
\mmm(g,g^\prime)\hspace{0.8mm} f(g^{-1}g^\prime g),
\end{equation}
where the function $\mmm :G\times G\rightarrow \mathbb{T}$ has the
following expression:
\begin{equation}
\mmm(g,g^\prime):= \mm(g,g^{-1}g^\prime)^\ast\hspace{0.5mm}
\mm(g^{-1}g^\prime,g).
\end{equation}
As the reader may check by means of a direct calculation involving
multipliers, the map $\two$ is a unitary representation; the
presence of the square root of the modular function $\Delta_G$ in
formula~{(\ref{two-sided})} takes into account the right action of
$G$ on itself. Notice that, for $\mm\equiv 1$, it coincides with the
restriction to the `diagonal subgroup' of the \emph{two-sided
regular representation} of the direct product group $G\times G$;
see~\cite{Folland,Segal}. As the reader may check using
relation~{(\ref{semi-invariance-bis})}, the Wigner transform $\wig$
intertwines the representations $\rep$ and $\two$:
\begin{equation}
\wig\hspace{0.8mm} \rep (g)= \two (g)\, \wig,\ \ \ \forall
\hspace{0.5mm}g\in G.
\end{equation}

Since the generalized Wigner transform $\wig$ is an isometry, the
adjoint map
\begin{equation}
\wigg^\ast \colon \ldg\rightarrow\hs
\end{equation}
is a partial isometry such that
\begin{equation}
\wigg^\ast\,\wig =I,\ \ \ \wig\,\wigg^\ast=\proi,
\end{equation}
where $\proi$ is the orthogonal projection onto the subspace
$\ru\equiv\mathrm{Ran}(\wig)=\mathrm{Ker}(\wigg^\ast)$ of $\ldg$.
Thus, the partial isometry $\wigg^\ast$ is the pseudo-inverse of
$\wig$ and we will call it \emph{(generalized) Weyl map} associated
with the representation $U$. It is remarkable that the Weyl map
$\wigg^\ast$ admits the following weak integral expression
(see~{\cite{Ali2}}):
\begin{equation}
\wigg^\ast\hspace{0.5mm} f=\int_G f(g)\, U(g)\,\du^{-1}\;
\de\mu_G(g),\ \ \ \forall\hspace{0.3mm}f\in\ldg.
\end{equation}
Observe that, in the case where the group $G$ is unimodular, with
the Haar measure $\mu_G$ normalized in agreement with $U$, we have
simply:
\begin{equation}
\wigg^\ast\hspace{0.5mm} f=\int_G f(g)\, U(g)\; \de\mu_G(g),\ \ \
\forall\hspace{0.3mm}f\in\ldg.
\end{equation}

Let us now focus on the case where $G=\rrr\times\rrr$ and $U$ is the
square integrable projective representation
\begin{equation} \label{schrep}
U\colon\rr\ni(q,p)\mapsto\disp\in\mathcal{U}(\lr).
\end{equation}
We recall from Sect.~{\ref{interlude}} that $(2\pi)^{-1}\de q\de p$
is the Haar measure on $\rrr\times\rrr$ normalized in agreement with
$U$. Then, in this case, the generalized Wigner transform $\wig$ is
the isometry from $\mathcal{B}_2(\lr)$ into
$\lrr\equiv\mathrm{L}^2\big(\rrr\times\rrr, (2\pi)^{-1}\de q\de
p\hspace{0.3mm};\ccc\big)$ determined by:
\begin{equation} \label{deter}
\big(\wig\hspace{0.3mm} \hrho\big)\qp =\tr(U\qp^\ast\hrho),\ \ \
\forall\hspace{0.4mm}\hrho\in\mathcal{B}_1(\lr).
\end{equation}
For a pure state
$\hpsi\equiv|\psi\rangle\langle\psi|\in\mathcal{B}_2(\lr)$,
$\|\psi\|=1$, the function $\wig\hspace{0.3mm}\hpsi$ coincides, up
to an irrelevant normalization factor, with the Fourier-Wigner
distribution associated with $\hpsi$ (compare with
definition~{(\ref{F-Wig})}). The multiplier
$\mm\colon(\rrr\times\rrr)\times(\rrr\times\rrr)\rightarrow\mathbb{T}$
associated with $U$ is given by
\begin{equation}
\mm(q,p\hspace{0.6mm};q^\prime,p^\prime)=\exp\hspace{-0.5mm}\Big(\frac{\ima}{2}(qp^\prime-pq^\prime)\Big).
\end{equation}
Hence, for the function $\mmm$ we find, in this case, the following
expression:
\begin{equation}
\mmm(q,p\hspace{0.6mm};q^\prime,p^\prime)=\mm(q,p\hspace{0.6mm};q^\prime-q,p^\prime-p)^\ast\,
\mm(q^\prime -q,p^\prime
-p\hspace{0.6mm};q,p)=\exp\hspace{-0.5mm}\big(\hspace{-0.7mm}-\ima(qp^\prime-pq^\prime)\big).
\end{equation}
Recalling formula~{(\ref{two-sided})}, we conclude that the
generalized Wigner transform $\wig$ intertwines the unitary
representation
\begin{equation}
\rep \colon \rrr\times\rrr\rightarrow\mathcal{U}(\mathcal{B}_2(\lr))
\end{equation}
with the representation $\two \colon
\rrr\times\rrr\rightarrow\mathcal{U}(\lrr)$ defined by
\begin{equation}
\big(\two(q,p)\hspace{0.4mm} f\big)(q^\prime,p^\prime)=
e^{-\ima(qp^\prime - pq^\prime)}\hspace{0.8mm} f(q^\prime,
p^\prime),\ \ \ \forall\hspace{0.3mm}f\in\lrr.
\end{equation}
The \emph{standard Wigner transform} --- we will denote it by
$\wigt$ --- is the isometry obtained composing the isometry $\wig$
determined by~{(\ref{deter})} with the symplectic Fourier transform:
\begin{equation}
\wigt := \fsy \,\wig\colon \mathcal{B}_2(\lr)\rightarrow \lrr.
\end{equation}
In particular, for a pure state $\hpsi\in\mathcal{B}_2(\lr)$ the
function $\wigt\,\hpsi$ coincides, up to an irrelevant normalization
factor, with the Wigner distribution associated with $\hpsi$
(compare with formula~{(\ref{for-quafp})}):
\begin{equation}
\big(\wigt\,\hpsi\big)\qp = 2\pi\,\quapsi\qp.
\end{equation}
It is clear that the isometry $\wigt$ intertwines the representation
$\rep$ with the unitary representation
$\mathcal{T}\colon\rrr\times\rrr\rightarrow\mathcal{U}(\lrr)$
defined by
\begin{equation}
\mathcal{T}\qp=\hspace{0.3mm}\fsy\hspace{0.6mm}\two\qp\hspace{0.7mm}\fsy,
\ \ \ \forall\hspace{0.5mm}\qp\in\rrr\times\rrr;
\end{equation}
as the reader may easily check, explicitly, we have:
\begin{equation}
\big(\mathcal{T}(q,p)\hspace{0.4mm} f\big)(q^\prime,p^\prime)=
 f(q^\prime-q, p^\prime-p),\ \ \ \forall\hspace{0.3mm}f\in\lrr.
\end{equation}
Notice that this result is consistent with relations~{(\ref{euno})}
and~{(\ref{edue})}. It is also a remarkable result --- see
ref.~{\cite{Pool}} --- that
\begin{equation}
\mathrm{Ran}\big(\wig\big)=\mathrm{Ran}(\wigt)=\lrr.
\end{equation}
Therefore, the standard Wigner transform $\wigt$ --- and its adjoint
$\wigt^\ast$, the \emph{standard Weyl map} --- are both unitary
operators.

Notice that, according to the definition of the map $\wig$, the
Wigner transform associated with a square integrable representation
is not --- in general --- a frame transform. For instance, in the
case where $U$ is the Weyl system~{(\ref{schrep})}, it is not. This
is coherent with the fact that, in the mentioned case,
$\mathrm{Ran}\big(\wig\big)=\lrr$ and hence
$\mathrm{Ran}\big(\wig\big)$ is not a r.k.H.s.\ as it should be if
$\wig$ were a frame transform. For the same reason, the standard
Wigner transform $\wigt$ is not a frame transform. It is then
natural to address the following problem: given a square integrable
projective representation $U$, is it possible to associate with $U$,
in a straightforward way, a frame transform in $\hs$? We will give
an (affirmative) answer to this question in the subsequent section.

%------------------------------------------------------------------------------
\section{Frames in Hilbert-Schmidt spaces from square integrable representations}
\label{transforms}
%------------------------------------------------------------------------------

In this section, we will show that it is possible to obtain from a
square integrable representation --- in a natural way ---  frame
transforms having as domain the space of Hilbert-Schmidt operators
in the Hilbert space where the representation acts. In the
following, we will assume that $G$ is a l.c.s.c.\ group and $U :
G\rightarrow\unig$ a square integrable projective representation of
$G$ in the Hilbert space $\mathcal{H}$. For the sake of simplicity,
we will suppose that the group $G$ is \emph{unimodular}, but the
results that we are going to prove actually extend to the general
case (see Remark~{\ref{generalcase}} below). We will denote by
$\mu_G$ the Haar measure on $G$ \emph{normalized in agreement with
the representation} $U$ (see Remark~{\ref{dufmoo}}).

Now, for any couple of Hilbert-Schmidt operators $\opa,\op\in\hs$,
we can define the function
\begin{equation} \label{defa}
A\colon G\times G \ni (\gi,\hi)\mapsto \langle \tgh ,
\opa\ranglehs\in\mathbb{C},
\end{equation}
where:
\begin{equation}
\tgh := \ug\,\op\hspace{0.4mm}\uh^\ast,\ \ \ \gi,\hi\in G.
\end{equation}

At this point, we have the following result:
\begin{theorem} \label{main-th}
With the previous notations and assumptions, for any
$\opa,\op\in\hs$, the map
\begin{equation} \langle \txx ,
\opa\ranglehs \colon G\times G\ni
(g_1,g_2)\mapsto\langle\tgh,\opa\ranglehs \in \mathbb{C}
\end{equation}
is a Borel function contained in $\lgg\equiv\lggi$, and the linear
application
\begin{equation} \label{dequa}
\dequat \colon \hs\ni\opa\mapsto A=\langle \txx , \opa\ranglehs\in
\lgg,
\end{equation}
for $\op$ nonzero and normalized {\rm (i.e.\ $\|\op\norhs=1$)}, is
an isometry {\rm (the `dequantization map' associated with the
representation $U$, with `analyzing operator' $\op$)}; namely, for
$\op$ normalized, the family of operators $\{ \hspace{0.4mm}
\tgh\colon\, (\gi,\hi)\in G\times G \hspace{0.5mm} \}$ is a
normalized tight frame in $\hs$, based on the measure
space $(G\times G,\mu_G\otimes\mu_G)$.\\
Moreover, for any
$\opa,\opb,\hat{S},\op\in\hs$, the following relation holds:
\begin{equation} \label{ortho-ortho}
\hspace{-1.6mm} \intgg \hspace{-1.2mm}
\big(\dequat\hspace{0.3mm}\opa\big)\hspace{-0.2mm}(\gi,\hi)^\ast
\big(\dequas\hspace{0.3mm}\opb\big)\hspace{-0.2mm}(\gi,\hi)\degh =
\langle \opa , \opb\ranglehs \hspace{0.3mm} \langle \hat{S} ,
\op\ranglehs.
\end{equation}
\end{theorem}

\noindent\emph{Proof:} Let $\op$ be a nonzero operator in $\hs$. As
an Hilbert-Schmidt operator, $\op$ will admit a canonical
decomposition of the form
\begin{equation} \label{sum}
\op=\sum_{n\in\mathcal{N}} \tau_n\,|\phi_n\rangle\langle\psi_n|,\ \
\ \psi_n,\phi_n\in\mathcal{H},
\end{equation}
where $\mathcal{N}$ is a finite or countably infinite index set,
$\{\psi_n\}_{n\in\mathcal{N}}$, $\{\phi_n\}_{n\in\mathcal{N}}$ are
orthonormal systems and $\{\tau_n\}_{n\in\mathcal{N}}$ is a set of
strictly positive numbers (the nonzero singular values of $\op$)
such that
\begin{equation}
\sum_{n\in\mathcal{N}} \tau_n^2=\|\op\norhs^2 ;
\end{equation}
the sum~{(\ref{sum})} converges with respect to the Hilbert-Schmidt
norm.

The fact that the representation $U$ is a weakly Borel map implies
that the function $\langle \txx , \opa\ranglehs$ --- for any
$\opa,\op\in\hs$ --- is Borel; namely, that the  application
$G\times G \ni (\gi,\hi)\mapsto \tgh\in\hs$ is weakly Borel. In
fact, by means of the canonical decompositions of the operators
$\opa$ and $\op$, one can express the function $\langle \txx ,
\opa\ranglehs$ as a finite --- or countably infinite and pointwise
converging --- sum of Borel functions; we leave the details to the
reader (recall that, given Borel functions $f_j\colon G
\rightarrow\ccc$, $j=1,2$, the function $f\colon G\times G\ni
(g_1,g_2)\mapsto f_1(g_1)\,f_2(g_2)\in\ccc$ is Borel too).

Assume, now, that $\op\neq 0$ and $\|\op\norhs\hspace{-0.6mm} =1$,
and let $\opa$ be an arbitrary operator in $\hs$. Consider the
associated Borel complex-valued function $A\equiv \langle \txx ,
\opa\ranglehs$ on $G\times G$. We will prove that this function
belongs to $\lgg$ and, simultaneously, that the dequantization
map~{(\ref{dequa})} is an isometry. To this aim, it will be
convenient to assume for the moment that $\op$ is a \emph{finite
rank operator}; this is equivalent to suppose that the index set
$\mathcal{N}$ is \emph{finite}. Then, by Tonelli's theorem and the
(finite) canonical decomposition of $\op$, we have:
\begin{eqnarray}
\hspace{-8.5mm} \intgg \hspace{-2mm} |A(\gi,\hi)|^2
\hspace{-0.6mm}\degh \spa \hspace{-0.3mm}& = & \spa \intg\left(\intg
|A(\gi,\hi)|^2 \degi\! \right)\!\deh
\nonumber\\
& = & \spa \sum_{n,k\in\mathcal{N}} \tau_n\tau_k \intg\bigg(\intg
\langle \opa ,
(|\phi_n(\gi)\rangle\langle\psi_n(\hi)|) \ranglehs \nonumber\\
\label{alal} & \times & \spa \hspace{-1mm} \langle
(|\phi_k(\gi)\rangle\langle\psi_k(\hi)|) , \opa\ranglehs\! \degi
\!\bigg)\!\!\deh ,
\end{eqnarray}
where, for the sake of notational conciseness, we have set
\begin{equation}
\phi_n(g):=U(g)\, \phi_n ,\ \ \psi_n(g) =U(g)\, \psi_n,\ \ \ \forall
\hspace{0.5mm}g\in G,\ \ \forall\hspace{0.4mm} n\in\mathcal{N}.
\end{equation}
Next, observe that
\begin{equation} \label{elel}
\langle \opa , (|\phi_n(\gi)\rangle\langle\psi_n(\hi)|) \ranglehs
=\tr(|\psi_n(\hi)\rangle\langle\phi_n(\gi)|\opa)^\ast = \langle
\opa\, \psi_n(\hi) , \phi_n(\gi) \rangle;
\end{equation}
hence, from relations~{(\ref{alal})} and~{(\ref{elel})}, we obtain:
\begin{eqnarray}
\hspace{-8.5mm} \intgg\hspace{-2mm} |A(\gi,\hi)|^2 \degh
\hspace{-0.3mm}\spa & = & \spa \sum_{n,k\in\mathcal{N}} \tau_n\tau_k
\intg\bigg(\intg
\langle \opa \,\psi_n(\hi),\phi_n(\gi) \rangle \nonumber\\
\label{blbl} &\times& \spa \langle \phi_k(\gi),\opa\,\psi_k(\hi)
\rangle \degi \!\bigg)\!\deh
\nonumber\\
& = & \spa \sum_{n\in\mathcal{N}} \tau_n^2 \intg \langle \opa\,
U(\hi) \,\psi_n,\opa\, U(\hi)\,\psi_n \rangle \deh,
\end{eqnarray}
where we have used the orthogonality relations for the square
integrable representation $U$ ($G$ unimodular, $\mu_G$ normalized in
agreement with $U$). At this point, using the trace
formula~{(\ref{trace-for})}, we get:
\begin{equation}
\intgg |A(\gi,\hi)|^2 \degh = \tr(\opa^\ast\opa)
\sum_{n\in\mathcal{N}} \tau_n^2 = \|\opa \norhs^2 \, \|\op \norhs^2,
\end{equation}
with $\|\op \norhs^2\hspace{-1.2mm}=1$. Thus, in the case where the
index set $\mathcal{N}$ is finite, the proof is complete.

Suppose now that $\dim(\mathcal{H})=\infty$ and
$\mathcal{N}=\mathbb{N}$. In this case, we can consider a sequence
$\{\op_\enne\}_{\enne\in\mathbb{N}}\subset\hs$ of \emph{finite rank
operators} converging to $\op$:
\begin{equation}
\|\op - \op_\enne \norhs \xrightarrow{\enne\rightarrow\infty} 0;
\end{equation}
in particular, we can consider the sequence of finite truncations of
the canonical decomposition of $\op$, i.e.
\begin{equation}
\op_\enne := \sum_{n=1}^\enne \tau_n\,|\psi_n\rangle\langle\phi_n|.
\end{equation}
Then, setting $\op_\enne(\gi,\hi):=
\ug\,\op_\enne\hspace{0.5mm}\uh^\ast$, we get:
\begin{equation}
\|\op(\gi,\hi) - \op_\enne (\gi,\hi) \norhs \hspace{-0.5mm} = \|\op
- \op_\enne \norhs \xrightarrow{\enne\rightarrow\infty} 0,
\end{equation}
and
\begin{equation}
A(\gi,\hi):= \langle \tgh , \opa\ranglehs
=\lim_{\enne\rightarrow\infty} \langle \op_\enne(\gi,\hi) ,
\opa\ranglehs,\ \ \ \forall \hspace{0.5mm}\gi,\hi\in G.
\end{equation}
Next, observe that for every $\mbox{\small
$\mathsf{N}$}\in\mathbb{N}$ the function $A_{\enne}:=\langle
\op_\enne(\cdot,\cdot) , \opa\ranglehs\colon G\times
G\rightarrow\mathbb{C}$ belongs to $\lgg$, and
$\{A_{\enne}\}_{\enne\in\mathbb{N}}$ is a Cauchy sequence in $\lgg$.
Indeed
--- according to the first segment of the proof --- one finds out that,
for any $\mbox{\small $\mathsf{N},\mathsf{N}^\prime$}\in\mathbb{N}$,
\begin{eqnarray}
\hspace{-2mm}\intgg |A_{\enne^\prime}(g_1,g_2)-A_{\enne}(g_1,g_2)|^2
\degh \spa & = & \spa  \|\langle
\op_{\enne,\enne^\prime}(\cdot,\cdot) , \opa\ranglehs\|_{\lgg}^2
\nonumber\\
%& = & \spa
%\|\opa\norhs^2\, \| \op_{\enne,\enne^\prime}\norhs^2
%\nonumber\\
& = & \spa \|\opa\norhs^2\, \| \op_{\enne^\prime}-\op_\enne\norhs^2,
\end{eqnarray}
where we have set
\begin{equation}
\op_{\enne,\enne^\prime}(g_1,g_2)\equiv
\ug\,\op_{\enne,\enne^\prime}\hspace{0.4mm}\uh^\ast,\ \ \
\op_{\enne,\enne^\prime}\equiv \op_{\enne^\prime}-\op_\enne,
\end{equation}
and we have exploited the fact that $\op_{\enne,\enne^\prime}$ is a
finite rank operator.

Therefore, the function $A \colon G\times G\rightarrow\mathbb{C}$ is
the pointwise limit of a Cauchy sequence of functions  in $\lgg$, so
that --- according to a well known result --- it belongs to $\lgg$
too and
\begin{equation}
\|A - A_\enne \|_{\lgg} \xrightarrow{\enne\rightarrow\infty} 0.
\end{equation}
Hence, taking into account that $\|A_\enne
\|_{\lgg}\hspace{-0.5mm}=\|\opa\norhs\| \op_\enne\norhs$, we have:
\begin{eqnarray}
\|A \|_{\lgg} \spa & = & \spa \lim_{\enne\rightarrow\infty}
\|A_\enne \|_{\lgg}
\nonumber\\
& = & \spa \|\opa\norhs\, \lim_{\enne\rightarrow\infty}\|
\op_\enne\norhs= \|\opa\norhs\, \|\op\norhs,
\end{eqnarray}
with $\|\op\norhs\hspace{-0.6mm}=1$. Thus, the first part of the
proof is complete.

We will now prove relation~{(\ref{ortho-ortho})}. This second part
of the proof goes along lines similar to the ones already traced in
the first part, so we will be rather sketchy.

Let $\opa,\opb,\hat{S},\op$ be operators in $\hs$, with $\hat{S}\neq
0\neq\op$ (otherwise relation~{(\ref{ortho-ortho})} is trivial), and
consider the canonical decompositions
\begin{equation} \label{sum2}
\hat{S}=\sum_{m\in\mathcal{M}}
\sigma_m\,|\eta_m\rangle\langle\chi_m|,\ \
\op=\sum_{n\in\mathcal{N}} \tau_n\,|\phi_n\rangle\langle\psi_n|,\ \
\ \eta_m,\chi_m,\psi_n,\phi_n\in\mathcal{H},
\end{equation}
where $\mathcal{M},\mathcal{N}$ are finite or countably infinite
index sets, $\{\eta_m\}_{n\in\mathcal{M}}$,
$\{\chi_m\}_{n\in\mathcal{M}}$, $\{\psi_n\}_{n\in\mathcal{N}}$,
$\{\phi_n\}_{n\in\mathcal{N}}$ are orthonormal systems,
$\{\sigma_m\}_{m\in\mathcal{M}}$, $\{\tau_n\}_{n\in\mathcal{N}}$ are
sets of strictly positive numbers such that $\sum_{m\in\mathcal{M}}
\sigma_m^2=\|\hat{S}\norhs^2$, $\sum_{n\in\mathcal{N}}
\tau_n^2=\|\op\norhs^2$, and we have:
\begin{equation}
\langle \hat{S},\op\ranglehs  = \sum_{\substack{m\in\mathcal{M} \\
n\in\mathcal{N}}} \sigma_m \tau_n \, \langle \eta_m,\phi_n\rangle\,
\langle \psi_n,\chi_m\rangle\, . \vspace{-1.8mm}
\end{equation}
The sums~{(\ref{sum2})} converge with respect to the Hilbert-Schmidt
norm.

Suppose first that the index sets $\mathcal{M},\mathcal{N}$ are both
\emph{finite}. For notational conciseness, we define the function
$\Phi\colon G\times G\rightarrow \ccc$,
\begin{equation}
\Phi
(\gi,\hi):=\big(\dequat\hspace{0.3mm}\opa\big)\hspace{-0.2mm}(\gi,\hi)^\ast
\big(\dequas\hspace{0.3mm}\opb\big)\hspace{-0.2mm}(\gi,\hi),
\end{equation}
and we set
\begin{equation}
\eta_m(g):=U(g)\, \eta_m ,\ \ \chi_m(g):=U(g)\, \chi_m ,\ \
\phi_n(g):=U(g)\, \phi_n ,\ \ \psi_n(g) =U(g)\, \psi_n .
\end{equation}
Then, since the function $\Phi$ belongs to $\lugg$ (according to the
first part of the proof), we can apply Fubini's theorem thus
getting:
\begin{eqnarray}
\hspace{-8mm} \intgg\! \Phi (\gi,\hi) \degh \spa & = & \spa
\sum_{m\in\mathcal{M}} \sum_{n\in\mathcal{N}} \sigma_m \tau_n
\intg\bigg(\intg \langle \opa \,\psi_n(\hi),\phi_n(\gi) \rangle
\nonumber\\
& \times & \spa \langle \eta_m(\gi),\opb\,\chi_m(\hi) \rangle \degi
\!\bigg)\!\deh
\nonumber\\
& = & \spa \sum_{m\in\mathcal{M}} \sum_{n\in\mathcal{N}} \sigma_m
\tau_n\, \langle \eta_m,\phi_n\rangle
\nonumber\\
& \times & \spa \intg \langle \opa\, U(\hi) \,\psi_n,\opb\,
U(\hi)\,\chi_m \rangle \deh,
\end{eqnarray}
where we have used the orthogonality relations for $U$. Next, use
the trace formula~{(\ref{trace-for})}:
\begin{eqnarray}
\hspace{-8mm} \intgg\! \Phi (\gi,\hi) \degh \spa & = & \spa
\sum_{m\in\mathcal{M}} \sum_{n\in\mathcal{N}} \sigma_m \tau_n\,
\langle \eta_m,\phi_n\rangle
\nonumber\\
& \times & \spa \intg \langle  U(\hi) \,\psi_n,\opa^\ast\opb \,
U(\hi)\,\chi_m \rangle \deh \nonumber \\
& = & \spa \sum_{m\in\mathcal{M}} \sum_{n\in\mathcal{N}} \sigma_m
\tau_n\, \langle \eta_m,\phi_n\rangle\, \langle
\psi_n,\chi_m\rangle\,\tr(\opa^\ast\opb)
\nonumber\\
& = & \spa \langle \opa , \opb\ranglehs \hspace{0.3mm} \langle
\hat{S} , \op\ranglehs.
\end{eqnarray}

Suppose now that $\dim(\mathcal{H})=\infty$, and that
$\mathcal{M}=\mathbb{N}$ and/or $\mathcal{N}=\mathbb{N}$. Then, one
can adopt a reasoning similar to the one used in the second half of
the first part of the proof: consider sequences
$\{\hat{S}_\mme\}_{\mme\in\mathbb{N}}$ and/or
$\{\op_\enne\}_{\enne\in\mathbb{N}}$ of finite rank operators ---
converging to $\hat{S}$ and/or to $\op$, respectively --- and
exploit the continuity (in both arguments) of the scalar products in
$\lgg$ and $\hs$, for proving relation~{(\ref{ortho-ortho})} also in
this case.

The proof of the theorem is complete.~{$\square$}

\begin{remark} \label{irrep}
{\rm In order to prove Theorem~{\ref{main-th}}, we could have shown
that the map
\begin{equation}
\mathbf{U} \colon G\times G\rightarrow\mathcal{U}(\hs),
\end{equation}
defined by
\begin{equation}
\mathbf{U}(g_1,g_2)\hspace{0.8mm}\op:=
\ug\,\op\hspace{0.4mm}\uh^\ast=:\tgh,\ \ \ \gi,\hi\in G,\ \op\in\hs,
\end{equation}
is an \emph{irreducible} projective representation of the
(unimodular) direct product group $G\times G$, and that, moreover,
it is square integrable. Then, formula~{(\ref{ortho-ortho})} can be
regarded as the `orthogonality relations' of the square integrable
representation $\mathbf{U}$. The advantage of the above proof is
that of `explicitly illustrating' what happens for finite rank
operators. In the general case where $G$ is not assumed to be
unimodular
--- see Remark~{\ref{generalcase}} below --- this kind of proof
allows to provide an explicit expression for (a variant of) the
Duflo-Moore operator associated with the representation $\mathbf{U}$
in terms of the Duflo-Moore operator associated with
$U$.~$\blacksquare$ }
\end{remark}

\begin{remark} \label{finite-rank}
{\rm Assume that the analyzing operator $\op\in\hs$ is a nonzero
\emph{finite rank operator} (with $\|\op\norhs=1$). Then, arguing as
in the proof of Theorem~{\ref{main-th}}, one shows that for every
trace class operator $\opa\in\mathcal{B}_1(\mathcal{H})$ and every
bounded operator $\opb\in\mathcal{B}(\mathcal{H})$ --- setting:
$B(\gi,\hi):=\tr(\tgh^\ast\hspace{0.3mm}\opb)$ --- the function
\begin{equation}
G \ni \hi\mapsto\big(\dequat\hspace{0.3mm}\opa\big)
\hspace{-0.2mm}(\gi,\hi)^\ast\, B(\gi,\hi)\in\ccc,\ \ \
\forall\hspace{0.4mm}\gi\in G,
\end{equation}
is contained in $\lug$, as well as the function $ \gi\mapsto\int_G
\big(\dequat\hspace{0.3mm}\opa\big) \hspace{-0.2mm}(\gi,\hi)^\ast
B(\gi,\hi)\,\de\mu_G(\hi)$, and the following formula holds:
\begin{equation} \label{new-new}
\intg\hspace{-0.8mm}\de\mu_G(\gi)
\intg\hspace{-0.8mm}\de\mu_G(\hi)\hspace{0.5mm}
\big(\dequat\hspace{0.3mm}\opa\big)\hspace{-0.2mm}(\gi,\hi)^\ast\hspace{0.3mm}
B(\gi,\hi)\degh = \tr (\opa^\ast \opb) ,
\end{equation}
where one can interchange the order of the integrals. Furthermore
--- taking into account the fact that, for any
$\phi,\eta\in\mathcal{H}$,
$\langle\phi,\tgh\,\eta\rangle=\big(\dequat|\phi\rangle\langle\eta|\big)(\gi,\hi)^\ast$
--- the following weak integral reconstruction formula holds:
\begin{equation} \label{new-dec}
\opb= \intg\hspace{-0.8mm}\de\mu_G(\gi)
\intg\hspace{-0.8mm}\de\mu_G(\hi)\hspace{0.8mm} B(\gi,\hi)\; \tgh \,
;
\end{equation}
in particular, for $\op= |\psi\rangle\langle\psi|$, $\|\psi\|=1$, we
re-obtain relation~{(\ref{w-int-for})}.~{$\blacksquare$} }
\end{remark}

Let us now investigate the intertwining property of the isometry
$\dequat$ with respect to the natural action of the group $G$ in
$\hs$. Precisely, let us consider the representation
\begin{equation}
\rep \colon G\rightarrow\mathcal{U}(\hs)
\end{equation}
defined in Sect.~{\ref{revisited}}; see formula~{(\ref{definrep})}.
As already observed, $\rep$ is a \emph{unitary} representation, even
in the case where the representation $U$ is genuinely
\emph{projective}. Consider, now, the map
\begin{equation}
\repr \colon G\rightarrow \mathcal{U}(\lgg)
\end{equation}
defined by
\begin{equation}
\left(\repr(g) f\right)(\gi,\hi):= \MM(g;\gi,\hi)\,
f(g^{-1}\gi,g^{-1}\hi),
\end{equation}
where the function $\MM \colon G\times G\times
G\rightarrow\mathbb{T}$ is given by
\begin{equation} \label{asso-fun}
\MM(g;\gi,\hi):= \mm (g^{-1},\gi)\,\mm (g^{-1},\hi)^\ast ,
\end{equation}
with $\mm$ denoting the multiplier of $U$. The map $\repr$ is a
unitary representation too, as the reader may verify by checking
that the following relation holds:
\begin{equation}
\MM(gg^\prime;\gi,\hi)=
\MM(g;\gi,\hi)\,\MM(g^\prime;g^{-1}\gi,g^{-1}\hi).
\end{equation}
It is clear that the unitary representation $\repr$ is weakly Borel;
hence, according to a well known result, it is strongly continuous.
Between the representations $\rep$ and $\repr$ there is a precise
relation: $\rep$ is unitarily equivalent to a sub-representation of
$\repr$; indeed, we have:
\begin{proposition} With the previous notations and assumptions, for every
normalized Hilbert-Schmidt operator $\op\in\hs$, the isometry
$\dequat$ intertwines the unitary representation $\rep\colon
G\rightarrow\mathcal{U}(\hs)$ with the unitary representation $\repr
\colon G\rightarrow \mathcal{U}(\lgg)$; namely:
\begin{equation} \label{fund-intert}
\dequat\, \rep (g)= \repr (g)\, \dequat,\ \ \ \forall
\hspace{0.5mm}g\in G.
\end{equation}
\end{proposition}

\noindent\emph{Proof:} Let $\opa$ an arbitrary operator in $\hs$. We
want to prove that
\begin{equation} \label{toshow}
\big(\dequat\hspace{0.4mm} (U(g)\hspace{0.3mm}\opa\hspace{0.5mm}
U(g)^\ast)\big)(g_1,g_2)=
\MM(g;\gi,\hi)\,\big(\dequat\hspace{0.4mm}\opa\big)(g^{-1}\gi,g^{-1}\hi).
\end{equation}
In fact, the l.h.s.\ of eq.~{(\ref{toshow})} is equal to
\begin{eqnarray}
\langle\tgh, U(g)\hspace{0.3mm}\opa\hspace{0.5mm} U(g)^\ast\ranglehs
\spa & = & \spa \tr\big(\uh\,\op^\ast\hspace{0.3mm}\ug^\ast\,
U(g)\hspace{0.3mm}\opa\hspace{0.5mm} U(g)^\ast\big)
\nonumber\\
& = & \spa
\tr\big((U(g)^\ast\hspace{0.3mm}\uh)\,\op^\ast(U(g)^\ast\hspace{0.3mm}\ug)^\ast\opa\big)
\nonumber\\
& = & \spa \tr\big((\mm
(g,g^{-1})\hspace{0.3mm}U(g^{-1})\hspace{0.3mm}\uh)\,\op^\ast
\nonumber\\
& \times & \spa(\mm
(g,g^{-1})\hspace{0.3mm}U(g^{-1})\hspace{0.3mm}\ug)^\ast\opa\big)
\nonumber\\
& = & \spa
\tr\big((U(g^{-1})\hspace{0.3mm}\uh)\,\op^\ast(U(g^{-1})\hspace{0.3mm}\ug)^\ast\opa\big).
\end{eqnarray}
Hence, we have that
\begin{eqnarray}
\langle\tgh, U(g)\hspace{0.3mm}\opa\hspace{0.5mm} U(g)^\ast\ranglehs
\spa & = & \spa \mm (g^{-1},g_1)\,\mm (g^{-1},g_2)^\ast\,
\tr\big(U(g^{-1}g_2)\,\op^\ast\hspace{0.3mm}
U(g^{-1}g_1)^\ast\opa\big)
\nonumber\\
& = & \spa \MM(g;g_1,g_2)\, \langle\op(g^{-1}g_1,g^{-1}g_2), \opa
\ranglehs.
\end{eqnarray}
We have thus obtained the r.h.s.\ of eq.~{(\ref{toshow})} and the
proof is complete.~$\square$

We conclude this section with a few remarks.

\begin{remark}{\rm Let $\ut \colon G\rightarrow\mathcal{U}(\widetilde{\mathcal{H}})$ be a
projective representation physically equivalent to $U$ (hence,
square integrable too):
\begin{equation}
\ut(g)=\beta(g)\, W\,U(g)\,W^\ast,\ \ \ \forall\hspace{0.4mm} g\in
G,
\end{equation}
where $\beta \colon G\rightarrow\mathbb{T}$ is a Borel function and
$W\colon \mathcal{H}\rightarrow\widetilde{\mathcal{H}}$ a unitary or
antiunitary operator. Then, the unitary representations $\rep$ and
$\rept$ are unitarily or antiunitarily equivalent (indeed, the
operator $\hs\ni\opa\mapsto W\opa\hspace{0.7mm}W^\ast\in\hs$ is
unitary if $W$ is unitary, antiunitary if $W$ is antiunitary).
Moreover, denoting by $\mmt$ the multiplier of $\ut$ and by $\MMt
\colon G\times G\times G\rightarrow\mathbb{T}$ the associated
function defined as in~{(\ref{asso-fun})}, it turns out that the
unitary representations $\repr(g)$ and $\reprt(g)$ are, accordingly,
unitarily or antiunitarily equivalent. Indeed --- using the fact
that for $W$ unitary or antiunitary we have, respectively:
\begin{equation}
\mmt(\gi,\hi)=\frac{\beta(\gi\hi)}{\beta(\gi)\beta(\hi)}\,\mm(\gi,\hi)\
\ \ \mbox{or}\ \ \
\mmt(\gi,\hi)=\frac{\beta(\gi\hi)}{\beta(\gi)\beta(\hi)}\,\mm(\gi,\hi)^\ast
\end{equation}
--- one can easily check the following relations:
\begin{equation}
\MMt(g;\gi,\hi)= \beta(g_1)^\ast \beta(g_2)\,\beta(g^{-1}\gi)\,\beta
(g^{-1}\hi)^\ast\, \MM(g;\gi,\hi),\ \ \mbox{for $W$ unitary},
\end{equation}
\begin{equation}
\MMt(g;\gi,\hi)= \beta(g_1)^\ast \beta(g_2)\,\beta(g^{-1}\gi)\,\beta
(g^{-1}\hi)^\ast\, \MM(g;\gi,\hi)^\ast,\ \ \mbox{for $W$
antiunitary}.
\end{equation}
Hence --- denoting by $J$ the standard complex conjugation in
$\lgg$, i.e.\ the antiunitary operator
\begin{equation}
J\colon \lgg\ni f\mapsto f^\ast\in\lgg,\ \ \ J=J^\ast,
\end{equation}
and by $\hat{\beta}$ the multiplication operator in $\lgg$ by the
$\mathbb{T}$-valued Borel function $(g_1,g_2)\mapsto\beta(g_1)^\ast
\beta(g_2)$ (operator which is obviously unitary) --- for every
$g\in G$ we have:
\begin{equation}
\reprt (g)=\hat{\beta}\,\repr(g)\,\hat{\beta}^\ast \ \ \mbox{($W$
unitary)},\ \ \ \reprt (g)=
\hat{\beta}\hspace{0.5mm}J\,\repr(g)\,J\hspace{0.5mm}\hat{\beta}^\ast
\ \ \mbox{($W$ antiunitary)}.
\end{equation}
This result is coherent with the fact that, denoting by $\tdequat$
the dequantization operator associated with the representation
$\ut$, with analyzing operator
$\op^\prime\in\mathcal{B}_2(\widetilde{\mathcal{H}})$ --- where
$\op^\prime= W\hspace{0.3mm}\op\hspace{0.3mm} W^\ast$, for some
$\op\in\hs$ such that $\|\op\norhs =1$ --- for every $\opa\in\hs$ we
have:
\begin{equation} \label{checkrel}
\tdequat (W\opa\hspace{0.7mm} W^\ast)=
\big(\hat{\beta}_W\hspace{0.3mm}\dequat\big) \opa,
\end{equation}
with $\hat{\beta}_W\equiv\hat{\beta}$, for $W$ unitary, and
$\hat{\beta}_W\equiv \hat{\beta}\hspace{0.5mm} J$, for $W$
antiunitary. We leave the simple check of
relation~{(\ref{checkrel})} to the reader.~$\blacksquare$ }
\end{remark}

\begin{remark} {\rm
We stress that, excluding the trivial case where
$\dim(\mathcal{H})=1$, $\mathrm{Ran}(\dequat)$ is a \emph{proper}
subspace of $\lgg$. In fact, if $\dim(\mathcal{H})\ge 2$, according
to relation~{(\ref{ortho-ortho})} we have:
\begin{equation}
\mathrm{Ran}\big(\dequatu\big)\perp\mathrm{Ran}\big(\dequatd\big),\hspace{2mm}
\mbox{for all}\hspace{1.5mm} \op_1,\op_2\in\hs\hspace{1.5mm}
\mbox{such that} \hspace{1.5mm} \langle\op_1,\op_2\ranglehs=0;
\end{equation}
hence, the ranges of a couple of dequantization maps, with mutually
orthogonal analyzing operators, are mutually orthogonal subspaces of
$\lgg$. Therefore, the ranges of dequantization maps must be proper
subspaces of $\lgg$.~$\blacksquare$ }
\end{remark}

\begin{remark} \label{rem-invo}
{\rm With every function $f\in\lgg$ one can associate a function $f^\diamond$, contained in $\lgg$
too, defined by
\begin{equation}
f^\diamond(g_1,g_2):=f(g_2,g_1)^\ast,\ \ \
\forall\hspace{0.8mm}(g_1,g_2)\in G\times G.
\end{equation}
Clearly, the antilinear application
\begin{equation} \label{invol}
\invo : \lgg\ni f\mapsto f^\diamond\in\lgg
\end{equation}
is a complex conjugation ($\invo=\invo^\ast$ and $\invo^2=I$).
Observe that, for every Hilbert-Schmidt operator $\opa\in\hs$, the
following relation holds
\begin{equation}
\dequat(\opa^\ast)=\big(\dequatt\opa\big)^\diamond ;
\end{equation}
indeed, we have:
\begin{eqnarray}
\dequat(\opa^\ast)(g_1,g_2)=\tr(U(g_2)\,\op^\ast\hspace{0.3mm}U(g_1)^\ast\opa^\ast)
\spa & = & \spa \tr(\opa\hspace{0.8mm}
U(g_1)\,\op\,U(g_2)^\ast)^\ast
\\
& = & \spa
\tr(U(g_1)\,\op\,U(g_2)^\ast\opa)^\ast=\big(\dequatt\opa\big)^\diamond
(g_1,g_2). \nonumber
\end{eqnarray}
Suppose that the analyzing operator $\op\in\hs$ is selfadjoint.
Then, the isometry $\dequat$ intertwines the standard complex
conjugation $\opa\mapsto\opa^\ast$ in $\hs$ with the complex
conjugation $\invo$ in $\lgg$, i.e.\
$\dequat(\opa^\ast)=\big(\dequat\opa\big)^\diamond$. Therefore,
taking into account the injectivity of the map $\dequat$, a function
$\Psi$ belonging to $\mathrm{Ran}\big(\dequat\big)$ is the image of
a selfadjoint operator \emph{if and only if}
$\Psi=\Psi^\diamond$.~$\blacksquare$ }
\end{remark}

\begin{remark} \label{generalcase}
{\rm Up to this point, we have focused on the case where the group
$G$ is unimodular. We stress that a suitable dequantization map can
be defined even if $G$ is \emph{not} unimodular (we denote by
$\mu_G$, as usual, a left Haar measure on $G$ and by $\duu$ the
Duflo-Moore operator normalized according to $\mu_G$), though in
this case the construction is slightly more complicate. Here we will
sketch the main points of this construction; further details (and
suitable examples) will be contained in a forthcoming paper. Let us
denote by $\fr$ the linear span of finite rank operators and let us
consider the set
\begin{equation}
\fru := \{ \hat{F}\in\fr:\ \mathrm{Ran}(\hat{F}),
\mathrm{Ran}(\hat{F^\ast})\subset\mathrm{Dom}\big(\duu\big)\}.
\end{equation}
The set $\fru$ is a dense linear span in $\hs$, and a generic
nonzero vector in $\fru$ is of the form $\sum_{n=1}^\enne
|\psi_n\rangle\langle\phi_n|$, where $\{\psi_n\}_{n=1}^{\enne}$,
$\{\phi_n\}_{n=1}^{\enne}$ are linearly independent sets in
$\mathrm{Dom}\big(\duu\big)$. Let us introduce a linear operator
$\preku$, with domain $\fru$, defined by
\begin{equation}
\preku\Big(\sum_{n=1}^\enne |\psi_n\rangle\langle\phi_n|\Big)=
\sum_{n=1}^\enne |\duu\,\psi_n\rangle\langle\duu\,\phi_n|.
\end{equation}
It is easy to check that, due to the selfadjointness of $\duu$,
$\preku$ is a symmetric operator. It follows that $\preku$ is
closable, and we denote by $\ku$ the closure of $\preku$; hence,
$\ku$ is a closed, symmetric, densely defined operator in $\hs$
whose restriction to $\fru$ coincides with $\preku$.\\
At this point, with every operator $\op$ in the dense linear span
$\mathrm{Dom}(\ku)$ one can associate a linear map $\dequat : \hs
\rightarrow \lgg\equiv\lggi$ defined by
\begin{equation}
\big(\dequat\hspace{0.3mm} \opa\big)(g_1,g_2):=
\langle\hspace{0.2mm} \ug\,\op\hspace{0.4mm}\uh^\ast,\opa\ranglehs,\
\ \ \gi,\hi\in G,
\end{equation}
which --- for $\op$ nonzero and such that
$\|\ku\hspace{0.3mm}\op\norhs\hspace{-0.3mm}= 1$ --- is an isometry.
Moreover, for any $\opa,\opb\in\hs$ and any $\hat{S},\op$ in the
dense linear span $\mathrm{Dom}(\ku)\subset\hs$, the following
orthogonality relations hold:
\begin{equation} \label{ortho-ortho-2}
\big\langle \dequat\hspace{0.3mm}\opa,
\dequas\hspace{0.3mm}\opb\big\rangle_{\lgg} = \langle \opa ,
\opb\ranglehs \hspace{0.6mm} \langle \ku\hspace{0.3mm}\hat{S}
,\ku\hspace{0.3mm}\op\ranglehs \,.
\end{equation}
The proof of these statements goes along lines similar to the ones
traced in the proof of Theorem~{\ref{main-th}}. First one proves the
statements with the operator $\op$ (and $\hat{S}$) belonging to the
dense linear span $\fru$. Then, one extends the result to a generic
$\op$ in $\mathrm{Dom}(\ku)$ by means of a limit argument. This time
the sequence $\{\op_\enne\}_{\enne\in\mathbb{N}}$ converging to
$\op$ should be chosen as follows. It must be a sequence in $\fru$
such that
\begin{equation}
\|\op - \op_\enne \norhs \xrightarrow{\enne\rightarrow\infty} 0\ \ \
\mbox{and}\ \ \ \|\ku\hspace{0.3mm}\op - \ku\hspace{0.3mm}\op_\enne
\norhs \xrightarrow{\enne\rightarrow\infty} 0
\end{equation}
(such a sequence exists since $\ku$ is the closure of $\preku$).\\
One can prove that the operator $\preku$ is essentially selfadjoint;
hence, its closure $\ku$ is the unique selfadjoint extension of
$\preku$. Thus, $\ku$ is a variant (Remark~{\ref{dufmoo}}) of the
Duflo-Moore operator associated with the square integrable
projective representation $\mathbf{U}$, see Remark~{\ref{irrep}}.
Therefore, for $\op\in\mathrm{Dom}(\ku)$ such that
$\|\ku\hspace{0.3mm}\op\norhs\hspace{-0.3mm}= 1$, the linear map
$\dequat$ can be regarded as the generalized wavelet transform
generated by $\mathbf{U}$, with analyzing vector
$\op$.~$\blacksquare$ }
\end{remark}

In the next section, we will exploit the class of frames introduced
above and the results of Sect.~{\ref{frame}} for deriving suitable
expressions of quantum-mechanical formulae in terms of functions on
`phase space'. Although most of the results hold in the general
case, we will assume, for the sake of simplicity, that the l.c.s.c.\
group $G$ is unimodular.

%%---------------------------------------------------------------------
\section{Frame transforms and quantum mechanics}
\label{quantum}
%%---------------------------------------------------------------------

Since we are now equipped with a wide class of tight frames in the
space $\hs$ of Hilbert-Schmidt operators in the Hilbert space
$\mathcal{H}$, we can exploit the results of Sect.~\ref{frame}. It
will be convenient to denote by $\GG$ the direct product group
$G\times G$ ($G$ unimodular), by $\gigi\equiv(g_1,g_2)$ a typical
element of $\GG$, by $\diag$ the `diagonal element' $(g,g)$ of $\GG$
and by $\mugg$ the Haar measure $\mu_G\otimes\mu_G$ on $\GG$ (which
is, obviously, a unimodular l.c.s.c.\ group). Then, according to
Theorem~{\ref{main-th}}, for every nonzero Hilbert-Schmidt operator
$\op\in\hs$ such that $\|\op\norhs=1$ ($\mu_G$ is normalized in
agreement with $U$), the family of operators
\begin{equation} \label{ourframe}
\{\op(\gigi)\equiv U(g_1)\,\op\,
U(g_2)^\ast=\mathbf{U}(\gigi)\,\op\}_{\gigi\in\GG}\,,
\end{equation}
is a normalized tight frame in $\hs$, based on $(\GG,\mugg)$. Thus,
we can identify the measure space $(Y,\nu)$ of Sect.~\ref{frame}
with the measure space $(\GG,\mugg)$. The frame transform associated
with the frame~{(\ref{ourframe})} is the linear map
$\dequat\colon\hs\rightarrow\ldgg\equiv\mathrm{L}^2(\GG,\mugg;\ccc)$
--- $\big(\dequat\opa\big)(\gigi):=\langle\op(\gigi),\opa\ranglehs$,
for every  $\opa\in\hs$ --- which is an isometry (the
`dequantization map'). We will denote by
\begin{equation}
\quat \colon \ldgg\rightarrow\hs
\end{equation}
the adjoint of the isometry $\dequat$; then, $\quat$ (the
`quantization map') is a partial isometry that coincides with the
pseudo-inverse of $\dequat$:
\begin{equation}
\quat\,\dequat=I,\ \ \ \mathrm{Ker}\big(\quat\big)=
\mathrm{Ran}\big(\dequat\big)^{\perp}.
\end{equation}
For the partial isometry $\quat$ we have the following simple
formula (compare with relation~{(\ref{fr-adj})}):
\begin{equation} \label{recfor-bis}
\quat\hspace{0.3mm}\Phi = \int_{\GG}\de\mugg({\gigi})\;
\Phi(\gigi)\,\op(\gigi),\ \ \ \forall\hspace{0.5mm}\Phi\in\ldgg.
\end{equation}
We stress that the integral in formula~{(\ref{recfor-bis})} is a
weak integral of $\hs$-valued functions; hence, \emph{a fortiori},
it can also be regarded as a weak integral of
bounded-operator-valued functions (see Remark~{\ref{weakintegral}).

As observed in Sect.~\ref{frame}, the linear maps $\dequat$ and
$\quat$ induce in $\ldgg$ a star product of functions defined by
(see definition~{(\ref{def-star-pr})}):
\begin{equation}
\Phi_1 \star \Phi_2 :=\dequat\big((\quat\Phi_1)\,(\quat
\Phi_2)\big),\ \ \ \forall\hspace{0.5mm}\Phi_1,\Phi_2\in\ldgg.
\end{equation}
According to Proposition~{\ref{propo-bis}}, we have:
\begin{equation}
\big(\Phi_1 \star \Phi_2\big)(\gigi)=
\int_{\GG}\de\mugg({\gigi}^{\prime})\int_{\GG}\de\mugg({\gigi}^{\prime\prime})
\,\kerst\,\Phi_1(\gigi^{\prime})\,\Phi_2(\gigi^{\prime\prime}),
\end{equation}
where
\begin{equation}
\kerst:=\langle\op(\gigi),\op(\gigi^{\prime})\,\op(\gigi^{\prime\prime})\ranglehs=
\tr\big(\op(\gigi)^\ast\,\op(\gigi^{\prime})\,\op(\gigi^{\prime\prime})\big).
\end{equation}
In particular, the subspace $\mathrm{Ran}\big(\dequat\big)$ of
$\ldgg$ is a r.k.H.s.\ (compare with formulae~{(\ref{remexp})}
and~{(\ref{r-ker})}):
\begin{equation}
\Phi(\gigi)=
\int_{\GG}\de\mugg(\gigi^{\prime})\,\repker\,\Phi(\gigi^{\prime}),\
\ \ \ \forall\hspace{0.4mm}\Phi\in\mathrm{Ran}\big(\dequat\big)
\end{equation}
--- where the reproducing kernel has the following expression:
\begin{equation}
\repker:=\langle \op(\gigi), \op(\gigi^{\prime})\ranglehs
\end{equation}
--- and, for every couple of Hilbert-Schmidt operators
$\opa_1,\opa_2\in\hs$, we have:
\begin{equation} \label{profo}
\big(\dequat \opa_1  \opa_2\big)(\gigi)=
\int_{\GG}\de\mugg({\gigi}^{\prime})\int_{\GG}\de\mugg({\gigi}^{\prime\prime})
\,\kerst\,A_1(\gigi^{\prime})\,A_2(\gigi^{\prime\prime}),
\end{equation}
with $A_1(\gigi)\equiv\big(\dequat\opa_1\big)(\gigi)$,
$A_2(\gigi)\equiv\big(\dequat\opa_2\big)(\gigi)$.

Observe that it is possible to express, within the present
framework, the expectation values of quantum mechanical observables.
Recall, in fact, that the (bounded) left and right multiplication
operators in $\hs$ by a bounded operator $\opa$ --- i.e.,
respectively: $\lef \colon \hs\ni\opb\mapsto\opa\,\opb\in\hs$ and
$\rig \colon \hs\ni\opb\mapsto\opb\hspace{0.3mm}\opa\in\hs$ --- are
represented as suitable integral operators in the Hilbert space of
frame transforms $\mathrm{Ran}\big(\dequat\big)=\dequat(\hs)$.
Precisely, the `left' and `right' integral kernels
\begin{equation}
\chilt\big(\opa;\gigi,\gigi^\prime\big):=\langle
\op(\gigi),\opa\,\op(\gigi^\prime)\ranglehs,\ \ \
\chirt\big(\opa;\gigi,\gigi^\prime\big):=\langle
\op(\gigi),\op(\gigi^\prime)\,\opa\ranglehs
\end{equation}
--- see Proposition~{\ref{prop-bop}}
--- correspond to the `super-operators' $\lef$ and $\rig$, respectively.
In particular, for every
trace class operator $\hrho\in\mathcal{B}_1(\mathcal{H})$, the
following formulae apply:
\begin{equation} \label{leffor}
\big(\dequat\hspace{0.3mm} \opa\hspace{0.4mm}
\hrho\big)(\gigi)=\int_{\GG}\de\mugg({\gigi}^{\prime})\;\chilt\big(\opa;\gigi,\gigi^\prime\big)\,
\rho(\gigi^\prime),\ \ \ \rho\equiv\dequat\hspace{0.3mm}\hrho,
\end{equation}
\begin{equation} \label{rigfor}
\big(\dequat\hspace{0.3mm} \hrho\hspace{0.3mm}\opa\big)(\gigi)
=\int_{\GG}\de\mugg({\gigi}^{\prime})\;\chirt\big(\opa;\gigi,\gigi^\prime\big)\,
\rho(\gigi^\prime).
\end{equation}

Besides, for every normalized non-zero vector $\psi\in\mathcal{H}$
--- more precisely, for every rank one projector
$\hpsi\equiv|\psi\rangle\langle\psi|$ ---  setting
\begin{equation} \label{def-gam}
\gam:=\langle\rep(g)\hspace{0.6mm}\hpsi,\op(\gigi)\ranglehs=\langle
U(g)\,\psi,\op(\gigi)\, U(g)\,\psi\rangle,
\end{equation}
we have (see Proposition~{\ref{prop-simul}}; consider that
$\{U(g)\,\psi\}_{g\in G}$ is a normalized tight frame in
$\mathcal{H}$, based on $(G,\mu_G)$):
\begin{equation} \label{trtrtr}
\tr(\hrho)=  \int_G\de\mu_G(g) \int_{\GG}\de\mugg(\gigi)\, \gam\,
\rho(\gigi)\equiv\tr(\rho).
\end{equation}
According to the second assertion of Proposition~{\ref{prop-simul}},
a \emph{positive} Hilbert-Schmidt operator $\opb\in\hs$ is a trace
class operator if and only if
\begin{equation} \label{posiff}
\int_G\de\mu_G(g) \int_{\GG}\de\mugg(\gigi)\,
\gam\,\big(\dequat\hspace{0.3mm}\opb\big)(\gigi) < +\infty.
\end{equation}
Observe also that, recalling the intertwining
relation~{(\ref{fund-intert})}, from definition~{(\ref{def-gam})} we
get:
\begin{eqnarray}
\gam := \langle\rep(g)\hspace{0.6mm}\hpsi,\op(\gigi)\ranglehs \spa &
= & \spa \langle\op(\gigi), \rep(g)\hspace{0.6mm}\hpsi\ranglehs^\ast
\nonumber\\
& = & \spa
\big(\dequat\hspace{0.5mm}\rep(g)\hspace{0.6mm}\hpsi\big)(\gigi)^\ast
\nonumber\\ \label{relgamma} & = & \spa
\big(\repr(g)\hspace{0.4mm}\dequat\hspace{0.4mm}\hpsi\big)(\gigi)^\ast.
\end{eqnarray}

\begin{remark}
{\rm Formula~{(\ref{trtrtr})} is a special case of a more general
relation. In fact, let $\hat{S}$ be a trace class operator in
$\mathcal{H}$ such that $\tr(\hat{S})=1$; then, extending
definition~{(\ref{def-gam})}, let us set
\begin{equation}
\gams := \langle\rep(g)\hspace{0.4mm}\hat{S},\op(\gigi)\ranglehs=
\tr\big((\rep(g)\hspace{0.4mm}\hat{S}^\ast)\hspace{0.8mm}\op(\gigi)\big).
\end{equation}
At this point, using the `second trace
formula'~{(\ref{trace-for-bis})} and the reconstruction formula for
the operator $\hrho$, we find:
\begin{equation}
\tr(\hrho)= \int_G\de\mu_G(g)\,
\tr\big((\rep(g)\hspace{0.4mm}\hat{S}^\ast)\hspace{0.8mm}\hrho\big)=
\int_G\de\mu_G(g) \int_{\GG}\de\mugg(\gigi)\,\gams\,\rho(\gigi).
\end{equation}
Moreover, arguing as above, we conclude that
\begin{equation}
\gams=\big(\repr(g)\hspace{0.4mm}\dequat\hspace{0.3mm}\hat{S}\big)(\gigi)^\ast.
\end{equation}
This formula shows that the function $\gigi\mapsto\gams^\ast$ is
contained in $\mathrm{Ran}\big(\dequat\big)$.~$\blacksquare$}
\end{remark}

In the special case where $\op\in\mathcal{B}_1(\mathcal{H})$,
exploiting again the second trace formula~{(\ref{trace-for-bis})},
we find also that
\begin{equation} \label{fordiag}
\tr(\hrho)\, \tr(\op)^\ast = \tr(\hrho)\, \tr(\op^\ast) =
\int_G\de\mu_G(g)\; \rho(\diag),\ \ \
\rho\equiv\dequat\hspace{0.3mm}\hrho\,.
\end{equation}
Hence, in particular, $|\tr(\op)|^2 = \int_G\de\mu_G(g)\,
\big(\dequat\hspace{0.3mm} \op\big)(\diag)$, and, if
$\op\in\mathcal{B}_1(\mathcal{H})$ is such that $\tr(\op)\neq 0$, we
have:
\begin{equation} \label{fordiag-bis}
|\tr(\hrho)| = \frac{1}{\sqrt{\mbox{$\int_G\de\mu_G(g)\,
\big(\dequat\hspace{0.3mm} \op\big)(\diag)$}}}\; \Big|
\int_G\de\mu_G(g)\; \rho(\diag)\, \Big|.
\end{equation}

We are now ready to provide a suitable expression for the quantity
$\tr(\opa\hspace{0.5mm}\hrho)$, which --- in the special case where
the bounded operator $\opa$ is selfadjoint, and the trace class
operator $\hrho$ is positive and of unit trace --- can be regarded
as a quantum-mechanical expectation value. From
relations~{(\ref{leffor})}, (\ref{rigfor}) and~{(\ref{trtrtr})} it
follows immediately that
\begin{eqnarray}
\tr(\opa\hspace{0.5mm}\hrho) \spa & = & \spa
\int_G\de\mu_G(g)\int_{\GG}\de\mugg(\gigi)\int_{\GG}\de\mugg({\gigi}^{\prime})
\,\gam\;\chilt\big(\opa;\gigi,\gigi^\prime\big)\, \rho(\gigi^\prime)
\nonumber\\
& = & \spa
\int_G\de\mu_G(g)\int_{\GG}\de\mugg(\gigi)\int_{\GG}\de\mugg({\gigi}^{\prime})
\,\gam\;\chirt\big(\opa;\gigi,\gigi^\prime\big)\,
\rho(\gigi^\prime).
\end{eqnarray}
Of course, analogous formulae involving the more general type of
integral kernel $\gamd$ defined above hold too. Moreover, in the
special case where $\op\in\mathcal{B}_1(\mathcal{H})$, with
$\tr(\op)\neq 0$, formula~{(\ref{fordiag})} implies:
\begin{eqnarray}
\tr(\opa\hspace{0.5mm}\hrho) \spa & = & \spa \tr(\op^\ast)^{-1}
\int_G\de\mu_G(g)\int_{\GG}\de\mugg({\gigi}^{\prime})\;\chilt\big(\opa;\diag,\gigi^\prime\big)\,
\rho(\gigi^\prime)
\nonumber\\
& = & \spa  \tr(\op^\ast)^{-1}
\int_G\de\mu_G(g)\int_{\GG}\de\mugg({\gigi}^{\prime})\;\chirt\big(\opa;\diag,\gigi^\prime\big)\,
\rho(\gigi^\prime) = \tr(\hrho\hspace{0.3mm}\opa).
\end{eqnarray}

In conclusion, having in mind applications to quantum mechanics,
within the framework outlined in the present section we have the
following picture. With \emph{states} (density operators) are
associated functions  --- the frame transforms of the density
operators --- belonging to the r.k.H.s.\
$\mathrm{Ran}\big(\dequat\big)$, which is endowed with a star
product that reproduces the product of the $\mathrm{H}^\ast$-algebra
$\hs$. On the other hand, with \emph{observables} are associated
suitable (left and right) integral kernels. The quantum-mechanical
expectation values are given by integral formulae involving the
frame transforms associated with states and the integral kernels.
Notice that in this picture the norm of a quantum observable can be
defined `intrinsically'. Indeed, for every bounded selfadjoint
operator $\opa$ in $\mathcal{H}$, recalling
definition~{(\ref{defi-lefrig})} and relation~{(\ref{norme})}, and
using the fact that $\lef$ is a bounded \emph{selfadjoint} operator
in $\hs$, we have:
\begin{eqnarray}
\|\opa\| \spa & = & \spa \|\lef\|
\nonumber\\
& = & \spa \hspace{-1mm}\sup_{\opb\in\hs,\, \opb\neq
0}\|\opb\norhs^{-2}\hspace{0.6mm}|\langle\opb,\opa\,\opb\ranglehs|\
\ \ \ \ \ \ \ \ \ \mbox{($\lef$ selfadjoint)}
\nonumber\\
\label{norma-norma0}
& = & \spa \hspace{-1mm} \sup_{\substack{\Phi\in \mathrm{Ran}(\dequat) \\
\Phi\neq 0}}\hspace{-1mm}
\|\Phi\|_{\ldgg}^{-2}\hspace{0.5mm}\Big|\int_{\GG}\hspace{-0.6mm}\de\mugg(\gigi)
\int_{\GG}\hspace{-0.6mm}\de\mugg({\gigi}^{\prime})
\;\chilt\big(\opa;\gigi,\gigi^\prime\big)\,\Phi(\gigi)^\ast\hspace{0.3mm}
\Phi(\gigi^\prime)\hspace{0.5mm}\Big| .
\end{eqnarray}
Moreover, taking into account relation~{(\ref{forpsiphi})}, we find
out that in formula~{(\ref{norma-norma0})} one can relax the
condition that $\Phi\in \mathrm{Ran}\big(\dequat\big)$; i.e.
\begin{eqnarray}
\|\opa\| \spa & = & \spa \hspace{-1mm} \sup_{\substack{\Phi\in \ldgg \\
\Phi\neq 0}}\hspace{-0.8mm}
\|\Phi\|_{\ldgg}^{-2}\hspace{0.5mm}\Big|\int_{\GG}\hspace{-0.5mm}\de\mugg(\gigi)
\int_{\GG}\hspace{-0.5mm}\de\mugg({\gigi}^{\prime})
\;\chilt\big(\opa;\gigi,\gigi^\prime\big)\,\Phi(\gigi)^\ast\hspace{0.3mm}
\Phi(\gigi^\prime)\hspace{0.5mm}\Big| \nonumber\\
\label{norma-norma} & =: & \spa
\big\|\chilt\big(\opa;\boldsymbol{\cdot},\boldsymbol{\cdot}\big)\big\|.
\end{eqnarray}
Of course, using the fact that $\|\opa\| = \|\rig\|$, one obtains a
completely analogous relation involving the right integral kernel
$\chirt\big(\opa;\boldsymbol{\cdot},\boldsymbol{\cdot}\big)$.

Therefore, we can identify the Jordan-Lie algebra of bounded
selfadjoint operators in $\mathcal{H}$ with the vector space of the
associated left integral kernels endowed with the norm defined by
formula~{(\ref{norma-norma})}, and with the Jordan product and the
Lie bracket defined by (compare with formulae~{(\ref{j-prod})}
and~{(\ref{l-brac})}, respectively):
\begin{eqnarray} \label{j-prod-bis}
\chilt\big(\hat{A}_1\hspace{-0.3mm}\circ\hspace{-0.3mm}\hat{A}_2;
\gigigia\big)\spa & = & \spa \frac{1}{2}\int_{\GG} \de\mugg(\gigi)\;
\Big(\chilt\big(\hat{A}_1; \gigigib\big)\, \chilt\big(\hat{A}_2;
\gigigic\big)
\\
& + & \spa \chilt\big(\hat{A}_2; \gigigib\big)\,
\chilt\big(\hat{A}_1;
\gigigic\big)\Big)=:\chilt\big(\opa_1;\boldsymbol{\cdot},\boldsymbol{\cdot}\big)
\circ\,\chilt\big(\opa_2;\boldsymbol{\cdot},\boldsymbol{\cdot}\big),
\nonumber
\end{eqnarray}
\begin{eqnarray} \label{l-brac-bis}
\chilt\big(\{\hat{A}_1,\hat{A}_2\}; \gigigia\big)\spa & = & \spa
\frac{1}{i}\int_{\GG} \de\mugg(\gigi)\; \Big(\chilt\big(\hat{A}_1;
\gigigib\big)\, \chilt\big(\hat{A}_2; \gigigic\big)
\\
& - & \spa \chilt\big(\hat{A}_2; \gigigib\big)\,
\chilt\big(\hat{A}_1;
\gigigic\big)\Big)=:\big\{\hspace{0.5mm}\chilt\big(\opa_1;\boldsymbol{\cdot},\boldsymbol{\cdot}\big)
,\chilt\big(\opa_2;\boldsymbol{\cdot},\boldsymbol{\cdot}\big)\hspace{-0.3mm}\big\},
\nonumber
\end{eqnarray}
for any couple of bounded selfadjoint operators
$\hat{A}_1,\hat{A}_2\in\mathcal{B}(\mathcal{H})$. It is clear that a
similar identification holds for the (suitably equipped) vector
space of right integral kernels.

Assume now that the analyzing operator $\op\in\hs$ is selfadjoint.
Observe that, in this case, the image through $\dequat$ of the set
$\mathsf{P}(\mathcal{H})$ of \emph{pure states} (rank-one
projectors) in the Hilbert space $\mathcal{H}$ is characterized as a
subset of $\mathrm{Ran}\big(\dequat\big)$ in the following way:
\begin{equation}
\dequat(\mathsf{P}(\mathcal{H}))=\{\Psi\in\mathrm{Ran}\big(\dequat\big)\colon\,
\Psi=\Psi^\diamond,\ \Psi\star\Psi=\Psi ,\ \tr(\Psi)=1\},
\end{equation}
where
\begin{equation}
\tr(\Psi) =  \int_G\de\mu_G(g) \int_{\GG}\de\mugg(\gigi)\, \gam\,
\Psi(\gigi).
\end{equation}
Indeed --- recalling Remark~{\ref{rem-invo}}, and
formulae~{(\ref{profo})} and~(\ref{trtrtr}) --- the image through
the isometry $\dequat$ of the set of orthogonal projectors in
$\mathcal{H}$ is characterized by the couple of conditions
\begin{equation}
\Psi=\Psi^\diamond,\ \ \ \Psi\star\Psi=\Psi.
\end{equation}
At this point, the third condition --- $\tr(\Psi)=1$ --- ensures
that $\quat\Psi$ is a trace class operator (notice that $\quat\Psi$
is positive and recall condition~{(\ref{posiff})}), i.e.\ a finite
rank projector, and in particular a rank one projector. This
characterization of the set $\dequat(\mathsf{P}(\mathcal{H}))$
allows to obtain an alternative expression of the norm of an
observable in terms of its left and right integral kernels. In fact,
for every bounded selfadjoint operator $\opa$ in $\mathcal{H}$, we
have that
\begin{equation}
\|\opa\|= \sup_{\psi\in\mathcal{H}:\; \|\psi\|=1}
|\langle\psi,\opa\,\psi\rangle | = \sup_{\hat{\mathrm{P}}\in
\mathsf{P}(\mathcal{H})}|\tr(\opa\,\hat{\mathrm{P}})|.
\end{equation}
Therefore, if the analyzing operator $\op\in\hs$ is selfadjoint, in
terms of the left integral kernel
$\chilt\big(\opa;\boldsymbol{\cdot},\boldsymbol{\cdot}\big)$, the
norm of the operator $\opa$ has the following alternative
expression:
\begin{eqnarray}
\|\opa\|\spa & = & \spa
\hspace{-1.8mm}\sup_{\substack{\Psi\in\mathrm{Ran}(\dequat):\,
\Psi=\Psi^\diamond
\\
\Psi\star\Psi=\Psi ,\, \tr(\Psi)=1
}}\hspace{-1mm}\Big\{\Big|\hspace{-0.5mm}\int_G\hspace{-0.9mm}\de\mu_G(g)\hspace{-0.5mm}\int_{\GG}\hspace{-1.1mm}
\de\mugg(\gigi)\hspace{-0.5mm}
\int_{\GG}\hspace{-1.1mm}\de\mugg({\gigi}^{\prime})
\,\gam\;\chilt\big(\opa;\gigi,\gigi^\prime\big)\, \Psi(\gigi^\prime)
\Big|\Big\} \nonumber \\
\label{norm-norm} & = & \spa
\big\|\chilt\big(\opa;\boldsymbol{\cdot},\boldsymbol{\cdot}\big)\big\|;
\end{eqnarray}
clearly, an analogous expression involving the right integral kernel
$\chirt\big(\opa;\boldsymbol{\cdot},\boldsymbol{\cdot}\big)$ holds
too.

We leave to the reader the simple exercise of deriving how the
natural symmetry action of the group $G$ on bounded operators in
$\mathcal{H}$ is represented in the vector spaces of the associated
left and right integral kernels.

%------------------------------------------------------------------------------
\section{A remarkable example}
\label{example}
%------------------------------------------------------------------------------

In this section, we will focus on the case where the group $G$ is
the additive group $\rrr\times\rrr$ (the group of translations on
the $1+1$-dimensional phase space; the generalization to the
$n+n$-dimensional case is straightforward) and the square integrable
projective representation $U$ is the Weyl system~{(\ref{schrep})}.
We will denote a generic element of $\rrr\times\rrr$ as a complex
variable
--- $z\equiv q+\ima p$ --- and a generic element of the direct
product group $\GG\equiv (\rrr\times\rrr)\times(\rrr\times\rrr)$,
accordingly, as $\zz=(\zu,\zd)$. As in Sect.~{\ref{quantum}}, the
diagonal element $(z,z)$ of $\GG$ will be denoted by $\diagz$. We
recall that the Haar measure $\mu_G$ on $G\equiv\rrr\times\rrr$,
normalized in agreement with $U$, is given by
$\de\mu_G(z)=(2\pi)^{-1}\hspace{0.3mm}\de z\equiv
(2\pi)^{-1}\hspace{0.3mm}\de q\hspace{0.6mm}\de p\,$; hence, the
Haar measure $\mugg$ on $\GG$ is given by
$\de\mugg(\zz)=(2\pi)^{-2}\hspace{0.3mm}\de \zz\equiv
(2\pi)^{-2}\hspace{0.3mm}\de \zu\hspace{0.3mm}\de \zd$. A this
point, as a consequence of Theorem~{\ref{main-th}}, we have that for
every normalized nonzero Hilbert-Schmidt operator $\op$ in $\lr$ the
family of operators
\begin{equation}
\{\op(\zz)\equiv U(z_1)\,\op\,
U(z_2)^\ast=\mathbf{U}(\zz)\,\op\}_{\zz\in\GG}
\end{equation}
is a normalized tight frame in $\mathcal{B}_2(\lr)$, based on
$(\GG,\mugg)$. This frame allows to define the isometry
\begin{equation}
\dequat\colon \BB\equiv\mathcal{B}_2(\lr)\rightarrow\ZZ\equiv\ldgg
\end{equation}
by setting:
\begin{equation}
\big(\dequat\hspace{0.3mm}\opa\big)(\zz):=\langle
\op(\zz),\opa\rangle_{\BB},\ \ \ \forall \hspace{0.4mm}\opa\in\BB.
\end{equation}
The range of the isometry $\dequat$ is a proper subspace of $\ZZ$
and a r.k.H.s.\ (embedded in $\ZZ$), with reproducing kernel
\begin{equation}
\repkert(\zz,\tzz):=\langle \op(\zz),\op(\tzz)\rangle_{\BB};
\end{equation}
taking into account the fact that $U(z)^\ast=U(-z)$, we have:
\begin{eqnarray}
\repkert(\zz,\tzz)\spa & = & \spa \tr\big(U(\zd)\,\op^\ast\,
U(-\zu)\,U(\tzu)\,\op\, U(-\tzd)\big)
\nonumber\\
& = & \spa e^{\frac{1}{4}(z_1^\ast \tzu - \zu
\tilde{z}_1^\ast)}\,e^{-\frac{1}{4}(z_2^\ast \tzd - \zd
\tilde{z}_2^\ast)}\,\tr\big(U(z_2-\tilde{z}_2)\,\op^\ast\,
U(z_1-\tilde{z}_1)^\ast\,\op\big)
\nonumber\\
& = & \spa \exp\Big(\frac{1}{4}(z_1^\ast \tzu - \zu
\tilde{z}_1^\ast- z_2^\ast \tzd + \zd
\tilde{z}_2^\ast)\Big)\big(\dequat\hspace{0.3mm}\op\big)(\zz-\tzz),
\end{eqnarray}
with $\zz\equiv(\zu,\zd)$, $\tzz\equiv(\tzu,\tzd)$. Moreover, the
isometry $\dequat$ intertwines the unitary representation
$\rep\colon G\equiv\rrr\times\rrr\rightarrow\mathcal{U}(\BB)$,
\begin{equation}
\rep(z)\,\opa = U(z)\,\opa\, U(-z),\ \ \ U(-z)=U(z)^\ast,
\end{equation}
with the unitary representation $\repr \colon
G\rightarrow\mathcal{U}(\ZZ)$ defined by
\begin{equation}
\big(\repr(z)\,f\big)(\zz):= \MM(z;\zz)\,
f(\zz-\hspace{-0.4mm}\diagz),\ \ \ \forall\hspace{0.4mm} f\in\ZZ,
\end{equation}
where:
\begin{equation}
\MM(z;\zz):=\exp\Big(\frac{\ima}{2}\big(q\,(p_2-p_1)-p\,(q_2-q_1)\big)
\Big),\ \ z\equiv q+\ima p,\ \zz \equiv (q_1+\ima p_1,q_2+\ima p_2).
\end{equation}
Of course all the formulae obtained in Sect.~{\ref{quantum}} apply
to this case; we will present some detailed calculations  and
examples elsewhere. We want now to highlight, briefly, the relation
between our results and the fundamental seminal papers~\cite{Cahill}
of Cahill and Glauber on quasi-distributions. In the cited papers,
Cahill and Glauber (with aims partially distinct from ours)
introduced and studied a family of normal operators with spectral
decomposition
\begin{equation}
\ops := \frac{2}{1-s} \sum_{n=0}^{\infty}
\left(\frac{s+1}{s-1}\right)^n |n\rangle\langle n|,\ \ \ s\in\ccc,\
s\neq 1,
\end{equation}
where $\{|n\rangle\}_{n=0,1,\ldots}$ are the standard eigenfunctions
of the harmonic oscillator Hamiltonian. From the first of the
papers~\cite{Cahill} we learn, in particular, the following (easily
verifiable) facts:\footnote{We warn the reader that in the mentioned
paper the terminology for indicating the bounded, Hilbert-Schmidt
and trace class operators, as well as the choice of the symbols for
the associated norms, is somewhat unusual.}
\begin{itemize}

\item for $\re(s)\le 0$, the operator $\ops$ is bounded and
\begin{equation}
\|\ops\|=\left| \frac{2}{1-s}\right|;
\end{equation}
moreover: $\ops^\ast=\op_{s^\ast}^{\phantom{\ast}}$;

\item for $\re(s)< 0$, the operator $\ops$ belongs to the Banach
space $\mathcal{B}_1(\lr)$ (hence, in particular, to the Hilbert
space $\BB\equiv\mathcal{B}_2(\lr)$), and
\begin{equation}
\|\ops\|_{1}:=\tr(|\ops|)=\frac{2}{|1-s|}\,\sum_{n=0}^{\infty}\left|\frac{s+1}{s-1}\right|^n
=\frac{2}{|1-s|-|1+s|},
\end{equation}
\begin{equation} \label{norops}
\|\ops\|_{2}:=\sqrt{\langle\ops,\ops\rangle_{\BB}\hspace{-0.6mm}}=\frac{1}{\sqrt{|\re(s)|}}\,
;
\end{equation}
thus, $\|\cdot\|_1$ and $\|\cdot\|_2$ are the trace class and
Hilbert Schmidt norms, respectively; moreover:
\begin{equation}
\tr(\ops)=1; \ \ \ (\re(s)<0)
\end{equation}

\item for $\re(s)= 0$, the operator $\ops$ belongs to the set
$\big(\mathcal{B}(\lr)\smallsetminus\BB\big)$;

\item for $\re(s)> 0$, $s\neq 1$, the operator $\ops$ is unbounded.

\end{itemize}

Cahill and Glauber proposed the following (in general,
\emph{formal}) decomposition of a Hilbert-Schmidt operator
(`bounded', in their terminology) $\opa\in\BB$:
\begin{equation} \label{deco}
\opa=\int_{\doma} \aas(z)\,\ops(z)\ \frac{\de z}{2\pi}
\end{equation}
where:
\begin{equation}
\ops(z):= U(z)\,\ops\, U(-z), \ \ \ s \neq 1,
\end{equation}
and
\begin{equation} \label{defi-quasi}
\aas(z):= \tr\big(\op_{-s}(z)\,\opa\big).
\end{equation}
In particular, one can show that, for $s=0$,
formula~{(\ref{defi-quasi})} --- with the trace suitably interpreted
as in Sect.~{\ref{revisited}} --- defines the Wigner distribution
(notice that $\hat{\Pi}\equiv\frac{1}{2}\op_0$ is the parity
operator in $\lr$: $\big(\hat{\Pi}\,f\big)(x)=f(-x)$). In general,
the mathematically rigorous interpretation of the decomposition
formula~{(\ref{deco})} is problematic since, for $\re(s)\neq 0$, it
involves unbounded operators, either in the formula itself, or in
the definition of the quasi-distribution $\aas$ (i.e.\ the pair
$(\ops,\op_{-s})$ contains an unbounded operator, for $\re(s)\neq
0$). Notice, moreover, that for $s=1$ the decomposition is not
defined at all (the operator $\op_1$ is not defined); therefore,
with the \emph{Husimi-Kano quasi-distribution} $\mathsf{A}_{-1}$
(see~{\cite{Kano,Cahill,Aniello-quasi,Schleich}})
---
\begin{equation}
\mathsf{A}_{-1}(z) := \langle z|\opa|z\rangle,
\end{equation}
where $\{|z\rangle\equiv U(z)\, |0\rangle \}_{z\in\ccc}$ is the
family of \emph{coherent states} of the quantum harmonic oscillator
--- is not associated any (even formal) reconstruction formula.

In our framework, taking into account relation~{(\ref{norops})},
with every Hilbert-Schmidt operator $\ops$
--- with $\re(s)<0$
--- one can associate a \emph{normalized tight frame}
\begin{equation}
\big\{\sqrt{|\re(s)|}\;\ops(\zz)\big\}_{\zz\in \GG}\, ,\ \
\mbox{where}\ \ \ops(\zz):= U(\zu)\,\ops\, U(-\zd), \ \
\zz\equiv(\zu,\zd),
\end{equation}
(thus: $\ops(z)\equiv\ops(\diagz)$), in the Hilbert space $\BB$,
based on $(\GG,\mugg)$. Besides, we have the decomposition formula:
\begin{equation} \label{decomia}
\opa=\frac{|\re(s)|^{\frac{1}{2}}}{(2\pi)^2}\int_{\GG}
\as(\zz)\,\ops(\zz)\ \de \zz,\ \ \ \opa\in\BB,
\end{equation}
where
\begin{eqnarray}
\as(\zz)\spa & := & \spa \sqrt{|\re(s)|}\;\langle \ops(\zz), \opa
\rangle_{\BB}
\nonumber\\
& = & \spa \sqrt{|\re(s)|}\;\tr\big(\ops(\zz)^\ast\opa\big)
=\sqrt{|\re(s)|}\;\tr\big(\op_{s^\ast}(\check{\zz})\hspace{0.5mm}\opa\big),\
\ \ \check{\zz}\equiv (\zd,\zu).
\end{eqnarray}
Therefore, for every $s\in\ccc$ such that $\re(s)<0$, we have that
\begin{equation}
\as(\diagz)= \sqrt{|\re(s)|}\;\mathsf{A}_{s^\ast}(z),
\end{equation}
and, if $\opa$ is a trace class operator,
\begin{equation}
\int_{\doma}\as(\diagz)\hspace{1.2mm}\de z
=\sqrt{|\re(s)|}\hspace{0.9mm}\tr(\opa),\ \ \ \int_{\doma}
\mathsf{A}_{s}(z) \hspace{1.2mm}\de z  = \tr(\opa),
\end{equation}
where we have used formula~{(\ref{fordiag})} and the fact that
$\tr(\ops)=1$. For $s=-1$, we have that $\op_{-1}=|0\rangle\langle
0|$; hence:
\begin{equation}
\op_{-1}(\zz)= |\zu\rangle\langle\zd|,\ \ \ A_{-1}(\zz)=
\langle\zu|\opa|\zd\rangle,\ \ \
\mathsf{A}_{-1}(z)=A_{-1}(\diagz)=\langle z|\opa|z\rangle,
\end{equation}
\begin{equation} \label{decomiapart}
\opa=\frac{1}{(2\pi)^2}\int_{\GG}\,
\langle\zu|\opa|\zd\rangle\,|\zu\rangle\langle\zd|\ \de\zu\de\zd.
\end{equation}
Thus, the Husimi-Kano quasi-distribution $\mathsf{A}_{-1}$ can be
regarded as the `restriction to the diagonal' of the function
$A_{-1}$, and formula~{(\ref{decomiapart})} is the `non-diagonal
coherent state representation of an operator'
(see~{\cite{Klauder}}). Moreover, for every bounded operator
$\opb\in\mathcal{B}(\lr)$, we have the following double integral
decomposition (see relation~{(\ref{w-int-for})} and
Remark~{\ref{finite-rank}}):
\begin{equation}
\opb=\frac{1}{(2\pi)^2}\int_{\doma}\de\zu\int_{\doma}\de\zd
\;\langle\zu|\opb|\zd\rangle\,|\zu\rangle\langle\zd|\, .
\end{equation}

%------------------------------------------------------------------------------
\section{Conclusions and perspectives}
\label{conclusions}
%------------------------------------------------------------------------------

In the present paper, we have reconsidered some fundamental aspects
of the quantization-dequantization theory in the light of the
mathematical notion of \emph{frame}. We have shown (see
Sect.~{\ref{frame}}) that --- in addition to the standard formulae
that play a fundamental role in (generalized) wavelet analysis ---
by considering frames of Hilbert-Schmidt operators one is able to
obtain a remarkable representation of a quantum system. It turns out
that \emph{states} (density operators) are naturally represented by
`phase space functions' belonging to a r.k.H.s.\ which is endowed
with a `star product'; while \emph{observables} are represented by
(left and right) `integral kernels' forming vector spaces endowed
with a structure of Jordan-Lie algebra. Quantum mechanical
expectation values are given by simple integral formulae. We have
then shown (see Sects.~{\ref{wigdis}--{\ref{revisited}}) that the
classical Weyl-Wigner approach to quantization-dequantization,
although not directly related to the notion of frame, relies on the
notion of \emph{square integrable projective representation}. Using
this mathematical tool one can introduce (see
Sect.~{\ref{transforms}}) a class of tight frames of Hilbert Schmidt
operators. A frame of this kind is generated by a square integrable
representation of a group that can be regarded as the `symmetry
group' of a quantum system, and by an `analyzing operator', whose
choice can be adapted to specific applications or requirements (as
it happens in wavelet analysis). Such a frame allows to achieve a
remarkable implementation (see Sect.~{\ref{quantum}}) of the
abstract scheme outlined in Sect.~{\ref{frame}}. In the case where
the square integrable representation is the Weyl system, there is a
link between our approach and the formalism of `$s$-parametrized
quasi-distributions' introduced by Cahill and Glauber (see
Sect.~{\ref{example}}), a link that on our opinion will deserve
further exploration. We plan, moreover, to develop the basic results
established in the present contribution in several directions; in
particular, we will mention the representation --- in our framework
--- of specific quantum systems and of `super-operators' (that play
a fundamental in the theory of open quantum systems), and the study
of the classical limit of quantum mechanics.

\section*{Aknowledgments}

V.I.\ Man'ko wishes to thank the University of Napoli `Federico II'
and the INFN (Sezione di Napoli) for the kind hospitality, and the
Russian Foundation for Basic Research (under Project n.\
07-02-00598) for partial support.

%%%%%%--------------------------------------------------------------------------

%-----------------------------------------------------------------------------
\end{document}